\newcommand{\CLASSINPUTinnersidemargin}{1.015in}
\newcommand{\CLASSINPUToutersidemargin}{1.015in}
\documentclass[12pt,onecolumn]{IEEEtran}
\usepackage{algorithm,amsbsy,amsmath,amssymb,epsfig,bbm,mathrsfs,multirow,amsthm}
\usepackage{color}
\usepackage{algpseudocode}
\usepackage{graphicx,caption,subcaption,array,multirow}
\usepackage{makecell}

\renewcommand{\baselinestretch}{1.39}
\DeclareMathAlphabet{\mathpzc}{OT1}{pzc}{m}{it}

\newtheorem{theorem}{\textbf{\textsc{Theorem}}}

\begin{document}
	
\title{``Jam Me If You Can'': Defeating Jammer with Deep Dueling Neural Network Architecture and Ambient Backscattering Augmented Communications}
	
\author{Nguyen Van Huynh, Diep N. Nguyen, Dinh Thai Hoang, and Eryk Dutkiewicz
\thanks{Nguyen Van Huynh, Diep N. Nguyen, Dinh Thai Hoang, and Eryk Dutkiewicz are with University of Technology Sydney, Australia. E-mails: huynh.nguyenvan@student.uts.edu.au, \{Hoang.Dinh, Diep.Nguyen, and Eryk.Dutkiewicz\}@uts.edu.au.}
\thanks{The authors contributed equally to this work.}
}
\maketitle
\thispagestyle{empty}
\renewcommand{\baselinestretch}{1}	
\begin{abstract}
With conventional anti-jamming solutions like frequency hopping or spread spectrum, legitimate transceivers often tend to ``escape" or ``hide" themselves from jammers. These reactive anti-jamming approaches are constrained by the lack of timely knowledge of jamming attacks (especially from smart jammers). Bringing together the latest advances in neural network architectures and ambient backscattering communications, this work allows wireless nodes to effectively ``face" the jammer (instead of escaping) by first learning its jamming strategy, then adapting the rate or transmitting information right on the jamming signal (i.e., backscattering modulated information on the jamming signal). Specifically, to deal with unknown jamming attacks (e.g., jamming strategies, jamming power levels, and jamming capability), existing work often relies on reinforcement learning algorithms, e.g., Q-learning. However, the Q-learning algorithm is notorious for its slow convergence to the optimal policy, especially when the system state and action spaces are large. This makes the Q-learning algorithm pragmatically inapplicable. To overcome this problem, we design a novel deep reinforcement learning algorithm using the recent dueling neural network architecture. Our proposed algorithm allows the transmitter to effectively learn about the jammer and attain the optimal countermeasures (e.g., adapt the transmission rate or backscatter or harvest energy or stay idle) thousand times faster than that of the conventional Q-learning algorithm. Through extensive simulation results, we show that our design (using ambient backscattering and the deep dueling neural network architecture) can improve the average throughput (under smart and reactive jamming attacks) by up to 426\% and reduce the packet loss by 24\%. By augmenting the ambient backscattering capability on devices and using our algorithm, it is interesting to observe that the (successful) transmission rate increases with the jamming power. Our proposed solution can find its applications in both civil (e.g., ultra-reliable and low-latency communications or URLLC) and military scenarios (to combat both inadvertent and deliberate jamming).
\end{abstract}
	

\section{Introduction}

Due to its broadcast nature, wireless communications are particularly vulnerable to jamming attacks, especially in low-power wireless networks. By injecting interference to the wireless communication channel (i.e., deliberate jamming), a jammer can degrade the effective signal-to-interference-plus-noise ratio (SINR), thereby disrupting or even bringing down legitimate communications links. The jamming attacks can be easily launched by using commercial off-the-shelf products~\cite{Hanawal2016Joint,Xu2005Feasibility} and have a significant detriment to wireless applications, especially for mission-critical systems (e.g., cyber-physical systems in traffic safety, industry automation or military missions). In practice, jamming signals can also come from inadvertent sources, e.g., due to device malfunctioning.
	

\subsection{Related Work and Motivation}
Anti-jamming has a very rich literature, originating from the early days of wireless communications. With most conventional anti-jamming solutions like frequency hopping or spread spectrum, legitimate transceivers often tend to ``escape" or ``hide" themselves from jammers. As an example, frequency-hopping spread spectrum (FHSS)~\cite{Yazicigil2018Ultra}-\cite{Cheng2018Mode} allows a wireless device to quickly switch its operating frequency to other frequency channels. As soon as jammer attacks the channel, the device will quickly change its operating frequency, thereby avoiding the jamming attack. In~\cite{Yazicigil2018Ultra}, the authors proposed an integrated bit-level FHSS for low-power wireless communication systems. The key idea of this approach is exploiting the frequency agility of bulk acoustic wave resonators. In~\cite{Mpitziopoulos2007Hybrid}, the authors proposed a hybrid approach using FHSS and direct sequence spread spectrum (DSSS) to cope with fast-following jammers. Using a stochastic game framework, the authors of \cite{Wang2011Anti} studied the strategic interaction between jammers and legitimate users. In particular, the jammer and the transmitter are considered as two players playing with each other to obtain the optimal attack and defense strategies, respectively. Through the minimax-Q learning algorithm, the transmitter can gradually obtain the optimal defense policy, i.e., how to switch between different channels. The simulation results demonstrated that the proposed framework can maximize the spectrum-efficient throughput. Similarly, a game theory based anti-jamming framework for frequency hopping wireless communications was considered in~\cite{Gao2018Game}. Nevertheless, these game models require complete information of the jammer, which may not be available in advance in practice. In~\cite{Kong2018Reinforcement} and~\cite{Liu2018Anti}, the authors adopt the Q-learning and deep Q-learning algorithms that allow the transmitter to choose frequencies to hop when the jammer attacks the channel. In~\cite{Cheng2018Mode}, the authors propose a mode-frequency hopping scheme which jointly uses the mode hopping and the traditional FH for anti-jamming in cognitive radio networks. However, the main limitation of the FHSS technique is that it requires extra spectrum resources (for hopping to evade the jammers). In addition, with powerful jammers which can attacks multiple channels simultaneously, FHSS may be less effective.

Besides the FHSS and DSSS techniques, the rate adaptation (RA) technique is also widely adopted, e.g., ~\cite{Pelechrinis2009RA}-\cite{Noubir2011RA}. The key idea of the RA technique is to proactively or adaptively account for jamming attacks by operating at a lower transmission rate. In~\cite{Pelechrinis2009RA}, the authors proposed an RA algorithm together with power control to mitigate jamming. Specifically, the algorithm consists of two modules: (i) a rate module for rate adaptation and (ii) a power control module for controlling the transmit power at legitimate transmitters. The experimental results demonstrated that the proposed algorithm can improve the network throughput by 150\% under jamming attacks.
However, in~\cite{Firouzbakht2012RA}, the authors revealed that the RA technique is not effective on a single channel. In~\cite{Noubir2011RA}, the authors investigated the performance of several state-of-the-art RA algorithms under different scenarios. The experimental results demonstrated that the existing RA algorithms are not effective to combat smart jamming attacks. Similar to the RA method, legitimate transmitters can also adapt (i.e., bump/push) their transmit power or beamforming vectors/matrices (to improve the effective SINR) to overcome or reject the effect of excessive interference. Nevertheless, this solution is either power-inefficient or not viable for low-power or hardware-constrained devices (e.g., in IoT applications).

In~\cite{Hanawal2016Joint}, the authors proposed a joint RA and FHSS technique to mitigate attacks from a reactive-sweep jammer. In particular, the jammer can sweep through a set of channels and sense the activities of legitimate transmitters to attack. To combat the jammer, the legitimate transmitters can either hop to a new channel and/or adapt their transmission rates. The authors modeled the system as a zero-sum Markov game and obtained the optimal policies for the transmitters by solving a constrained Nash equilibrium problem. Similar to~\cite{Wang2011Anti,Gao2018Game}, this work also assumed complete information of the jammer in deriving the defense strategy. Another widely adopted approach in the literature is to use the ultra-wideband communications to hide the legitimate signals \cite{Mpitziopoulos2009Survey} in the noise.

It is worth noting that, to allow the transmitters to effectively escape or hide from the jammer, most aforementioned solutions require additional resources (in spectrum bandwidth, or transmit power, or hardware capability). This fact limits the practical applications of these conventional methods, especially for low-power and/or low-cost communications systems (e.g., in IoT). In this paper, we present a novel anti-jamming framework for these low-power and/or wireless-power communications devices. Such a framework allows these wireless transceivers to not only survive jamming attacks without requiring additional resources but also leverage the jamming signal to improve their transmission rate. To that end, we first observe that most existing anti-jamming solutions are reactive ones that are constrained by the lack of timely knowledge of jamming attacks (especially from smart jammers). Bringing together the latest advances in neural network architectures and ambient backscattering communications, this work allows wireless nodes to effectively ``face" the jammer (instead of escaping) by first learning its jamming strategy, then adapting the rate and transmitting information right on the jamming signal (i.e., backscattering modulated information on the jamming signal).  In our design, transmitters are augmented with an ambient backscattering communication circuit~\cite{Liu2013Ambient,Huynh2018Survey} and an energy harvester. When a jammer attacks the communication channel, the transmitter can leverage the jamming signal to backscatter information to the gateway or harvest energy from the jamming signal.

Our key idea is inspired by the latest advances in ambient backscattering communications (ABC) and RF energy harvesting. An ABC-capable transceiver can modulate/backscatter the RF ambient signals (e.g., FM, AM radio signals) to transmit its own information. Interested readers of ABC are referred to~\cite{Liu2013Ambient},~\cite{Huynh2018Survey} and therein references. Note that ABC, by contrast to bistatic backscattering communications, does not require a dedicated RF signal source. Specifically, in bistatic backscattering communications, e.g., \cite{Kimionis2012Bistatic},~\cite{Kimionis2014Increased},~\cite{Alevizos2014Channel},~\cite{Daskalakis2016Soil}, backscatter devices transmit information by backscattering the RF signals generated by a dedicated RF source which is controllable. With the recent development of RF energy harvesting, the transmitter can harvest energy from RF signals with a high efficiency. In particular, in~\cite{Olgun2010Low},~\cite{Olgun2010Wireless}, and~\cite{Scorcioni2012Optimized}, the authors propose novel designs for the rectenna, i.e., rectifier and antenna, and RF-DC converter to improve the amount of harvested RF energy from RF energy sources. The experimental results demonstrate that with the proposed rectenna, a tag can harvest RF energy and convert the harvested energy to DC with 70\% efficiency under a wide range of input power. In~\cite{Zhang2010Low}, low power circuit designs for the voltage regulator and resistor to digital converter are also proposed. Differently, in~\cite{Hu2017Optimal} and~\cite{Shafie2015Cooperative}, the authors aim to maximize the amount of harvested energy by considering the joint information and energy cooperative problem with channel constraints.

To deal with the uncertainty (or unknowns) of jamming attacks and ambient RF signals, existing work often relies on reinforcement learning algorithms, e.g,. Q-learning under the framework of a Markov decision process (MDP). However, the Q-learning algorithm is notorious for its slow convergence to the optimal policy, especially when the system state and action spaces are large. This makes the Q-learning algorithm pragmatically inapplicable. To overcome this problem, we design a novel deep reinforcement learning algorithm using a new dueling neural network architecture. Our proposed algorithm allows the transmitter to effectively learn about the jammer and attain the optimal countermeasures (e.g., adapt the transmission rate or backscatter or harvest energy or stay idle) thousand times faster than that of the conventional Q-learning algorithm. The transmitters not only successfully defeat jamming attacks, but also leverage jamming signal to significantly improve the performance for the system. Specifically, extensive simulation results show that our design (using ambient backscattering and the deep dueling neural network architecture) can improve the average throughput (under smart and reactive jamming attacks) by up to 426\% and reduce the packet loss by 24\%. By augmenting the ambient backscattering capability on devices and using our algorithm, it is interesting to observe that the (successful) transmission rate increases with the jamming power.

\subsection{Main Contributions}
The major contributions of this paper can be summarized as follows.
\begin{itemize}
	\item We propose a novel smart anti-jamming design using ambient backscatter, energy harvesting, and rate adaption techniques to defeat smart and reactive jammers. Based on this method, when a jammer attacks the channel, the transmitter can leverage jamming signal to transmit information using the ambient backscattering communication technique or harvest energy from the jamming signal. Alternatively, the transmitter can also choose to adapt the transmission rate to \emph{actively}{\footnote{It refers to the conventional radio transmission that is different from the ambient backscattering transmission.}} transmit data. 
		
	\item To lay a theoretical foundation for our design, we develop a dynamic approach using the MDP framework and Q- and deep Q-learning algorithms to maximize the average long-term throughput for systems under the uncertainty of jamming attacks and ambient RF signals. Unlike existing rate adaption methods that require the explicit or implicit knowledge of the interference/jamming level, our reinforcement learning-based rate adaptation framework does not require such information. It is also worth emphasizing that such a rate adaptation method is not susceptible/subject to the imperfect estimation/observations of the jamming signal (e.g., misdetection or false alarms).      
		
		
	\item To provide a practical solution for the theoretical Q-learning based approach above, we design a deep dueling neural network architecture which allows the system to quickly approach to the optimal defend policy. The key idea of this approach is implementing two streams of fully-connected hidden layers to separately and concurrently estimate the values of states and advantages of actions. As a result, the proposed deep dueling algorithm converges thousands times faster than Q-learning based algorithms.
		
	\item Finally, we perform extensive simulations with the aims of not only demonstrating the efficiency of proposed solutions in comparison with other conventional methods, but also providing insightful analytical results for the implementation of our framework. 
\end{itemize}
	
The rest of paper is organized as follows. Section~\ref{Sec.System} and Section~\ref{Sec:prob} describe the system model and the problem formulation, respectively. Section~\ref{sec:QDeepQ} presents the Q-learning and deep Q-learning algorithms. Then, the deep dueling algorithm is presented in Section~\ref{Sec:deepdueling}. After that, the evaluation results are discussed in Section~\ref{sec:evaluation}. Finally, conclusions are drawn in Section~\ref{sec:conclusion}.

\section{System Model}
\label{Sec.System}
	
\begin{figure*}[!]
	\centering
	\includegraphics[scale=0.2]{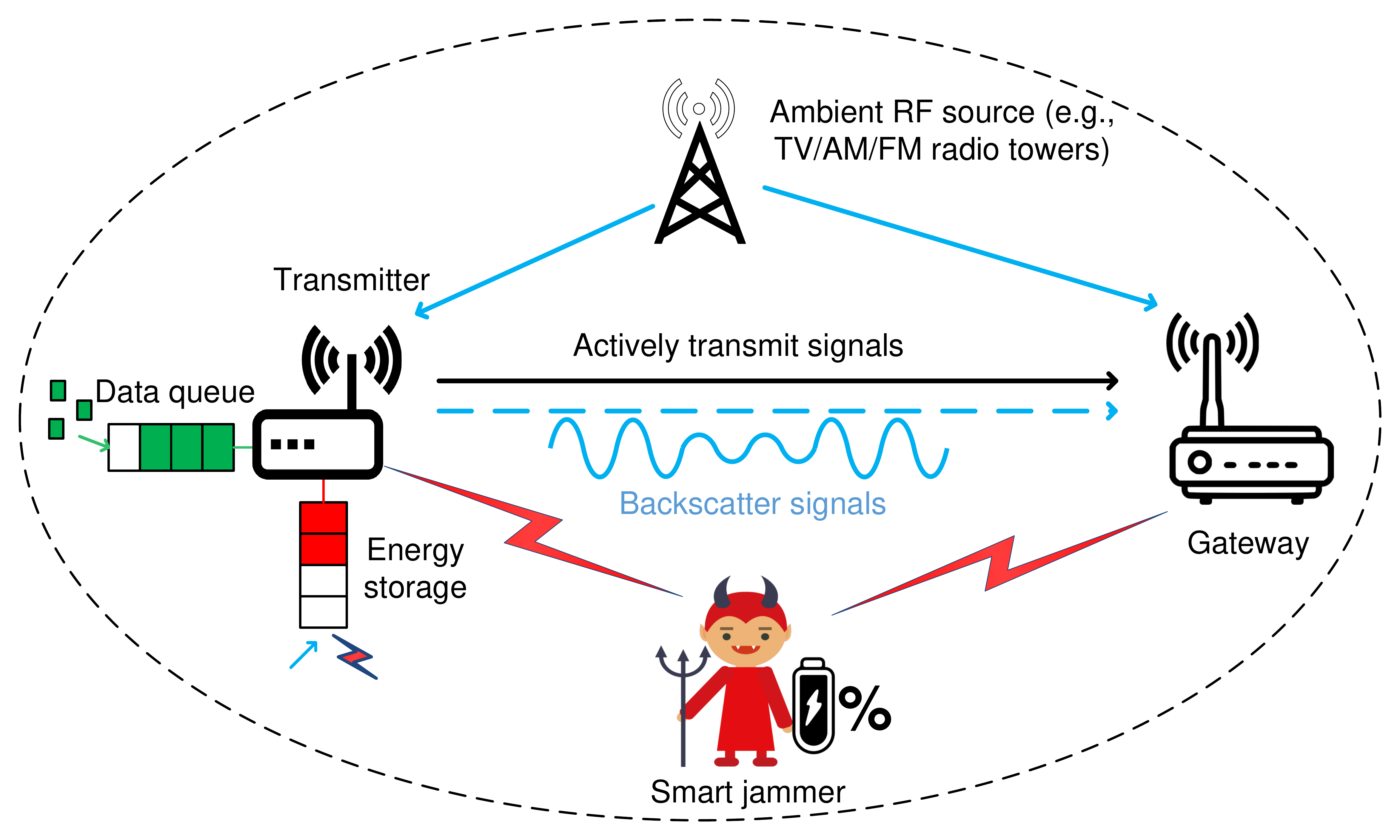}
	\caption{System model}
	\label{Fig.system_model}
\end{figure*}

We consider a wireless system considering of a gateway and a transmitter as illustrated in Fig.~\ref{Fig.system_model}. The transmitter is equipped with a data buffer to store data before transmitting to the gateway. In addition, we assume that the transmitter is equipped with an energy harvesting circuit and an energy storage. The energy harvesting circuit is used to harvest energy from surrounding signals, and then the harvested energy will be stored in the energy storage for future use. We consider an ambient RF source, e.g., an FM radio tower, that is located near the system, and thus the transmitter can harvest energy from the RF energy source when the source is active, i.e., broadcasts signals. Then, the transmitter can use the harvested energy to transmit data to the gateway when the RF energy source becomes idle. This transmission scheme is also known as the harvest-then-transmit protocol that is well known in the literature~\cite{Huynh2018Survey}.
	
\subsection{Smart and Reactive Jammer with Self-Interference Suppression Capability}
\label{subsec:jammer}
We consider a\footnote{Note that our system model can be extended straightforwardly to the case with multiple jammers who perform attacks to the channel cooperatively, i.e., only one jammer attacks the channel at a time.} smart and reactive jammer with self-interference suppression (SiS) capability. With the latest advances in SiS \cite{Wang2011Anti}, the jammer can ``listen" to the channel while jamming. That allows the jammer to instantaneously discern its jamming outcome and reactively optimize its jamming strategy to maximize the disruption of the victims. The SINR at the gateway is formally calculated as follows~\cite{Hanawal2016Joint},~\cite{Firouzbakht2012RA}: 
\begin{equation}
\label{eq:SIRN}
\theta = \frac{P^\mathrm{R}}{\phi P^\mathrm{J} + \rho^2},
\end{equation}
where $P^\mathrm{R}$ is the received power from the transmitter at the gateway, $P^\mathrm{J}$ is the jamming power transmitted by the jammer, $\rho^2$ is the variance of additive white Gaussian noise, and $\phi P^\mathrm{J}$ expresses the jamming power received at the gateway in which $0 \le \phi \le 1$ is an attenuation factor.

In practice, the jammer can adjust its pulse duty cycle factor to achieve the maximum degradation on the target channel while maintaining a time-average power constraint $P_{\mathrm{avg}}$. Note that the average power $P_{\mathrm{avg}}$ should be less than the peak jamming power $P_{\max}$, i.e., $P_{\mathrm{avg}} \leq P_{\max}$~\cite{Firouzbakht2012RA}. Specifically, let $\mathbf{P}_{\mathrm{J}} = \{P^{\mathrm{J}}_0, \ldots, P^{\mathrm{J}}_n, \ldots, P^{\mathrm{J}}_N\}$ denote the vector of discrete jamming power levels. In each time slot, the jammer can select any transmit power level $P^{\mathrm{J}}_n$ as long as its the average power constraint is satisfied. If we denote $\mathbf{x} \triangleq (x_0,\ldots, x_n, \ldots, x_N)$ as a probability vector, then the strategy space of the jammer, denoted by $\mathbf{J}_\mathrm{s}$, can be defined as follows:
\begin{equation}
\begin{aligned}
\mathbf{J}_s \triangleq \Big\{(x_0,\ldots, x_n, \ldots, x_N), \sum_{n=0}^{N} x_n = 1, x_n \in [0,1], \mbox{} \forall n \in \{0,\ldots,N\}, \mathbf{x} \mathbf{P}_{\mathrm{J}}^\top \leq P_{\mathrm{avg}}\Big\}.
\end{aligned}
\end{equation}
	
To find the defense policy in the worst case, as mentioned above, we consider a smart jammer that would know the information of the transmitter, e.g., how many packets the transmitter can transmit/backscatter and how many packets it can bring down if the jamming is successful (thanks to the SiS capability). In such a case, based on this information and its given average power constraint $P_{\mathrm{avg}}$, the jammer will find an optimal strategy to attack the channel in order to maximize the disruption. In particular, we assume that the jammer receives a reward $w^{\mathrm{J}}_n$ if it attacks the channel with power level $P^{\mathrm{J}}_n$. $w^{\mathrm{J}}_n$ can be referred as the number of packets that have been completely corrupted (i.e., not being successfully received/decoded, hence not ACKed by the receiver) if the jamming power is $P^{\mathrm{J}}_n$. Let $\mathbf{w}_{\mathrm{J}} = \{ w^{\mathrm{J}}_0, \ldots, w^{\mathrm{J}}_n, \ldots, w^{\mathrm{J}}_N \}$ denote the reward vector of the jammer. Thus, the objective function of the jammer can be defined as follows:
\begin{equation}
\begin{aligned}
& \max_{\mathbf{x}} \phantom{5}	\mathbf{x} \mathbf{w}_{\mathrm{J}}^\top	 , \\
& \mbox{s.t.}	\left\{	\begin{array}{ll}
\sum_{n=0}^{N} x_n = 1 , \\
x_n \in [0,1], \mbox{} \forall n \in \{0,\ldots,N\} , \\
\mathbf{x}\mathbf{P}_\mathrm{J}^\top \leq P_{\mathrm{avg}} .
\end{array}	\right. 		
\end{aligned}
\label{eq:obj_jammer}
\end{equation}

Note that the ambient RF signal is an external source that is out of control of the jammer and the transmitter and also not impacted by neither the jammer or the node. Additionally, in this paper, we consider the case in which the transmitter operates as a secondary user that only actively transmits data when the ambient RF source is inactive. As such, Eq.~(\ref{eq:SIRN}) does not need to account for the ambient RF signals. Note that our analysis is also applicable for general cases in which both the ambient RF source and the transmitter are licensed devices operating on different licensed frequencies.

\subsection{Ambient Backscattering-Augmented Communications}
	
\begin{figure}[!]
	\centering
	\includegraphics[scale=0.2]{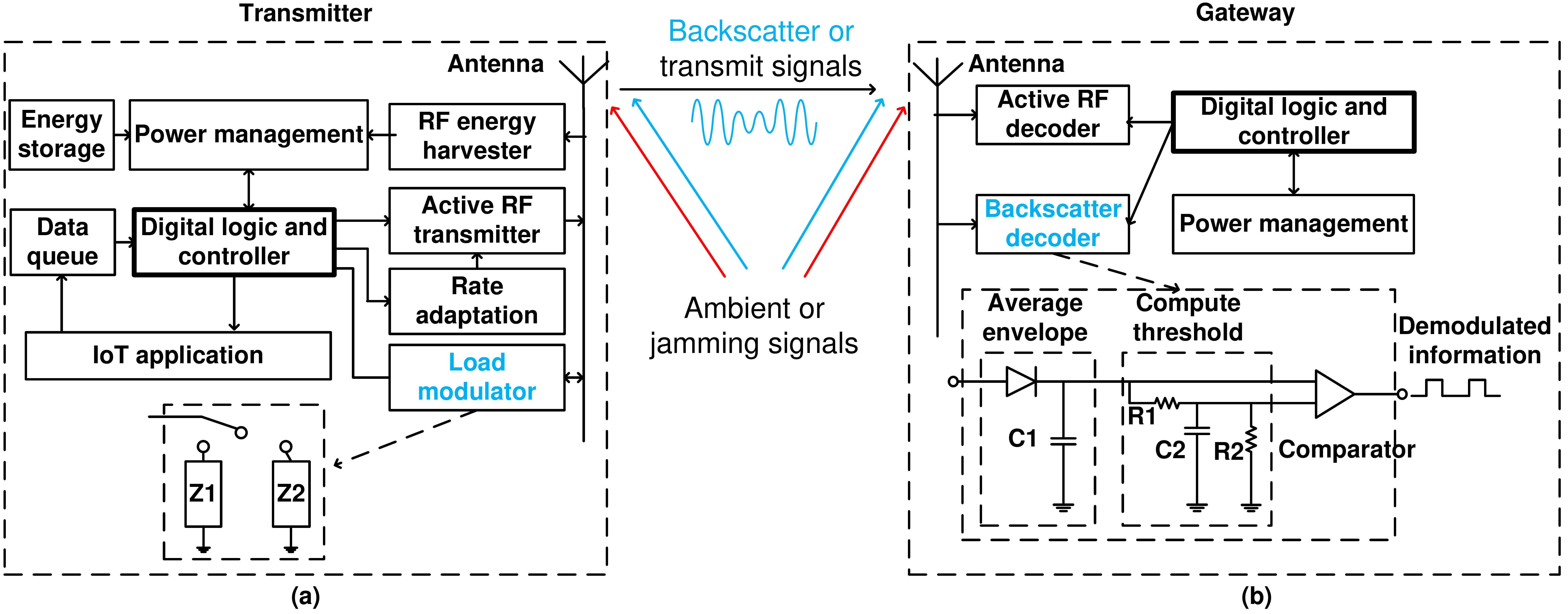}
	\caption{Function and circuit diagram of the proposed anti-jamming system}
	\label{Fig.circuit_diagram}
\end{figure}

To defeat the above smart and reactive jamming attack, we propose a novel communications scheme, namely ``ambient backscatter-augmented communications". Our high-level circuit architecture is shown in Fig.~\ref{Fig.circuit_diagram}. 
	
In Fig.~\ref{Fig.circuit_diagram}(a), we show the proposed circuit diagram that allows the transmitter to be able to harvest energy, perform active RF transmission, and backscatter information. The circuit design for the integration of energy harvesting and backscattering communication has also been considered in several research works in the literature such as~\cite{Liu2013Ambient} and~\cite{Boyer2014Backscatter}. The architecture of transmitter consists of an antenna, a controller, an energy harvester, an RF transmitter for active transmissions, an energy storage, a data buffer, and a load modulator together with a backscatter decoder for ambient backscatter communications. The controller takes responsibilities to make decisions, e.g., stay idle, transmit data, backscatter data, and harvest energy, for the transmitter. When the ambient RF source is active and/or the jammer attacks the channel, if the transmitter chooses to harvest energy, it will use the energy harvester to harvest energy from the ambient signal or the jamming signal. The harvested energy is then stored in the energy storage and used when the transmitter decides to actively transmit data to the gateway. In contrast, if the transmitter chooses to backscatter to immediately transmit data to the gateway, the transmitter will modulate and reflect the ambient RF signal or the jamming signal by using the load modulator~\cite{Liu2013Ambient}. In particular, the load modulator consists of an RF switch, e.g., ADG902, directly connected to the antenna. The input of the load modulator is a stream of one and zero bits which is generated by the controller depending on the application. When the input bit is zero, the load modulator switches to load $Z_1$, and thus the transmitter is in the non-reflecting state. Otherwise, when the input bit is one, the load modulator turns to load $Z_2$, and thus the transmitter is in the reflecting state. By doing so, the transmitter can backscatter its data to the gateway. It is worth noting that while operating in the backscatter mode, the transmitter still can harvest energy (in the non-reflecting state), but the amount of the harvested energy is relatively small and only suitable for operations in the backscatter mode~\cite{Liu2013Ambient},~\cite{Huynh2018Survey}.
	
To allow the gateway to decode backscattered signals, the transmitter backscatters information at a lower rate than the ambient signal, i.e., jamming signal and ambient signal. For completeness, we formally describe the principle as follows. Assuming that we have a digital receiver that samples the received signal at the Nyquist-information rate, e.g., using ADC~\cite{Liu2013Ambient}. The received signal at the gateway is sampled to $y[n]$ as:
\begin{equation}
y[n] = x[n] + \zeta B[n]x[n] + l[n],
\end{equation}
where $x[n]$s are the samples of the ambient signal (i.e., the jamming signal, the ambient RF signal, or both of them) received at the gateway, $l[n]$ is the noise, $\zeta$ is the complex attenuation of the backscattered signal, and $B[n]$s are the bits transmitted by the transmitter (through the load modulator). If the transmitter sends information at a fraction of the rate, say $\frac{1}{N}$, then $B[Ni+j]$ are all equal for $j=1$ to $N$~\cite{Liu2013Ambient}. Then, the gateway averages powers of $N$ received samples as follows:
\begin{equation}
\label{receive_power}
\frac{1}{N}\sum_{i=1}^{N} {|y[n]|}^{2} = \frac{1}{N}\sum_{i=1}^{N} {|x[n]+\zeta Bx[n] + w[n]|}^2,
\end{equation} 
where $B$ takes a value of `0' or `1' depending on the non-reflecting and reflecting states, respectively. As $x[n]$ is uncorrelated with the noise $w[n]$, (\ref{receive_power}) can be expressed as follows:
\begin{equation} \label{receive_data}
\frac{1}{N}\sum_{i=1}^{N} {|y[n]|}^{2} = \frac{{|1+\zeta B|}^2}{N} \sum_{i=1}^{N} {|x[n]|}^2 + \frac{1}{N} \sum_{i=1}^{N} {w[n]}^2.
\end{equation}
Denote $P=\frac{1}{N}\sum_{i=1}^{N}{|x[n]|}^2$ as the average power of the received jamming signal (or ambient signal). Ignoring the noise, the average power at the receiver is ${|1+\zeta|}^2 P$ and $P$ when the backscatter transmitter is at the reflecting ($B=1$) state and the non-reflecting ($B=0$) state, respectively. Based on the differences between ${|1+\zeta|}^2 P$ and $P$, the backscatter receiver can decode the data from the backscattered signal with a conventional digital receiver~\cite{Liu2013Ambient}.
	
However, the ADC consumes a significantly amount of power to sample the received signal. For a low-power design, in Fig.~\ref{Fig.circuit_diagram}(b), we describe a circuit diagram using only analog components to decode the backscattered signal. In particular, the gateway is equipped with an antenna to receive information (either active transmission or backscatter information) from the transmitter. Based on the received signal, the gateway will decide to use a suitable mode to decode information from the transmitter. In particular, the active RF decoder is used to decode active transmission signal from the transmitter. If the transmitter transmits data by using the backscatter mode, the gateway will use the backscatter decoder to extract the information from the backscattered signal. Specifically, at the backscatter decoder, the backscattered signal is first smoothed by the envelope-averaging circuit. After that, the compute-threshold circuit produces an output voltage between low and high levels of the smoothed signal. Then, the comparator compares the signal with a predefined threshold to derive output bits zero and one properly. The more detailed information about hardware designs as well as decoding algorithms at the receiver can be found in~\cite{Liu2013Ambient}.

Note that although we consider a single transmitter in this paper, the proposed model and analysis can be extended to the case with multiple transmitters. In such a case, one can adopt popular scheduling mechanisms, e.g., TDMA, to avoid the collision between transmitters. Another approach is that transmitters backscatter data at different rates~\cite{Liu2013Ambient},~\cite{Huynh2018Survey}. In this way, the gateway can decode the information in the backscattered signal by leveraging the difference in communication rates.

\subsection{System Operation}
\label{Subsec.systemop}
	
We denote the probability of the ambient RF source being idle in each time slot by $\eta$. Due to the constraints on average power $P_{\mathrm{avg}}$ and maximum transmit power $P_{\max}$, the jammer may attack the channel with different power levels at different time. When the jammer attacks the channel and the ambient RF source is idle, the transmitter can choose one of the following actions (i) go to sleep mode, i.e., stay idle, (ii) harvest energy from the jamming signal, (iii) backscatter information based on the jamming signal, or (iv) adapt its transmission rate by using rate adaption (RA) techniques~\cite{Hanawal2016Joint,Firouzbakht2012RA}. Depending on the transmit power level $P^{\mathrm{J}}_n$ of the jammer, the transmitter can harvest $e^{\mathrm{J}}_n$ units of energy and backscatter maximum $\widehat{d}^{\mathrm{J}}_n$ packets through the jamming signal. In practice, the more power the jammer uses to attack the channel, the more energy the transmitter can successfully harvest from the jamming signal~\footnote{Based on the Friis equation~\cite{Balanis2012Antenna}, we also can obverse the proportional relationship between the amount of harvested energy and the transmission power of energy source, i.e., the jammer.}. In addition, through many real experiments and analysis on backscatter communication systems in the literature~\cite{Liu2013Ambient,Boyer2014Backscatter,Kimionis2012Bistatic}, it can be observed that the more power the jammer uses to attack the channel, the more energy per information bit the transmitter can backscatter to the gateway, and thus the less Bit Error Rate (BER) of backscatter communication is. This also implies that the more packets the transmitter can successfully transmit to the gateway by backscattering the jamming signal when the jammer uses higher power levels to attack. We denote $\mathbf{e}=\{e^{\mathrm{J}}_0,\ldots, e^{\mathrm{J}}_n, \ldots, e^{\mathrm{J}}_N\}$ as the amount of energy that the transmitter can successfully harvest from the jamming signal when the jammer attacks the channel with power level $\mathbf{P}_{\mathrm{J}} = \{P^{\mathrm{J}}_0, \ldots, P^{\mathrm{J}}_n, \ldots, P^{\mathrm{J}}_N\}$, respectively. Similarly we denote $\widehat{\mathbf{d}}=\{\widehat{d}^{\mathrm{J}}_0,\ldots,\widehat{d}^{\mathrm{J}}_n, \ldots,\widehat{d}^{\mathrm{J}}_N\}$ as the number of packets that the transmitter can successfully transmit to the gateway when the jammer attacks the channel with power level $\mathbf{P}_{\mathrm{J}} = \{P^{\mathrm{J}}_0, \ldots, P^{\mathrm{J}}_n, \ldots, P^{\mathrm{J}}_N\}$, respectively.

In practice, when the jammer attacks the channel and the ambient RF source does not transmit data, the transmitter still can transmit its data by reducing its data rate. Specifically, based on jamming power $P^{\mathrm{J}}_\mathrm{n}$, the transmitter can actively transmit data at maximum rate $r_m$. We then denote $\mathbf{r}=\{r_1, \ldots, r_m, \ldots,  r_M\}$ as the set of available transmission rates that the transmitter can choose to transmit data when the jammer attacks the channel. At each rate $r_m$, the transmitter can transmit maximum $\widehat{d}^{\mathrm{r}}_m$ packets. Note that, for $m=1, \ldots, M$, when $\gamma_{m-1} \leq \theta < \gamma_m $ with $\gamma_m$ is the value of SINR, the gateway only can decode packets sent at rates $r_0, r_1, \ldots, r_{m-1}$, and the packets sent at rate $r_m$ or higher will be completely lost~\cite{Hanawal2016Joint}. To detect the states of the ambient RF source and the jammer, several detection techniques can be adopted, e.g., energy detection~\cite{Digham2007On},~\cite{Zhang2009Optimization}. Note that there are miss detection and false alarm probabilities when detecting the states of channels. However, our proposed algorithm can learn these probabilities and dynamically adjust its optimal policy.

In this work, we define the packet delivery ratio (PDR) as the ratio of packets that are successfully delivered to the gateway over the total number of packets arriving at the system. The arrival data process follows the Poisson distribution with mean rate $\lambda$. The maximum data queue size and energy storage capacity are denoted by $D$ and $E$, respectively. If a packet arrives at the system when the data queue is full, it will be dropped. To consider a low-latency system, if a packet stays in the queue longer than a latency threshold, i.e., $t_{\mathrm{th}}$, it will be discarded. 
	
If at least one of the sources (i.e., either the ambient RF source, or the jammer, or both of them) is active, the transmitter can choose to backscatter data or harvest energy. The transmitter then observes the results of the taken action, i.e., the total number of packet backscattered or the total amount of harvested energy, and update the learning function. Based on the states of the ambient RF source and the jammer, the operations of our system can be expressed as follows:
\begin{itemize}
	\item \emph{When the ambient RF source is idle and the jammer does not attack the channel}: the transmitter can (i) transmit maximum $\widehat{d}_{\mathrm{t}}$ packets if it has enough energy (each packet requires $e_\mathrm{t}$ units of energy to be successfully transmitted) or (ii) stay idle.	
	\item \emph{When the ambient RF source is idle and the jammer attacks the channel with power level $P^{\mathrm{J}}_n$}: the transmitter can (i) use the RA technique to transmit maximum $\widehat{d}^{\mathrm{r}}_m$ packets if it has enough energy, (ii) backscatter maximum $\widehat{d}^{\mathrm{J}}_n$ packets, (iii) harvest $e^{\mathrm{J}}_n$ units of energy, or (iv) stay idle.	
	\item \emph{When the ambient RF source is active and the jammer does not attack the channel}: the transmitter can choose to (i) backscatter maximum $\widehat{d}_\mathrm{b}$ packets, (ii) harvest $e_\mathrm{h}$ units of energy, or (iii) stay idle.
	\item \emph{When the ambient RF source is active and the jammer attacks the channel with the power level $P^{\mathrm{J}}_n$}: the transmitter can choose to (i) backscatter $d_\mathrm{sum}$ packets with $ d_{\mathrm{min}} \leq  d_\mathrm{sum} \leq d_\mathrm{max}$ where $d_\mathrm{min} = \min(\widehat{d}_\mathrm{b}, \widehat{d}^{\mathrm{J}}_n)$ and $d_\mathrm{max} = \widehat{d}_\mathrm{b} + \widehat{d}^{\mathrm{J}}_n${\footnote{The backscatter rate when both sources are active, $d_\mathrm{sum}$, should be in between the minimum of the backscatter rates of individual sources and the summation of them. In general, we assume that $d_\mathrm{sum}$ is unknown and captured by a random variable with a given distribution in the above range.}}, (ii) harvest $e_\mathrm{sum}$ units of energy with $e_\mathrm{min} \leq e_\mathrm{sum} \leq e_\mathrm{max}$ where $e_\mathrm{min}=\max(e_\mathrm{h}, e^{\mathrm{J}}_n)$ and $e_\mathrm{max} = e_\mathrm{h} + e^{\mathrm{J}}_n$~\cite{Yang2017Riding}{\footnote{The harvested energy when both sources are active, $e_\mathrm{sum}$, should be in between the maximum of the energy harvested from each individual source and the summation of them. In general, we assume that $e_\mathrm{sum}$ is unknown and captured by a random variable with a given distribution in the above range.}}, or (iii) stay idle. 
\end{itemize}

In this paper, time is slotted. In each time slot, given a particular channel condition and states of the ambient RF source and the jammer, the amount of harvested energy and the number of backscattered/transmitted packets, i.e., $\widehat{d}_{\mathrm{t}}$,  $\widehat{d}^{\mathrm{r}}_m$, $\widehat{d}^{\mathrm{J}}_n$, $e^{\mathrm{J}}_n$, $\widehat{d}_\mathrm{b}$, $e_\mathrm{h}$, $d_\mathrm{sum}$, $e_\mathrm{sum}$, can be observed after interacting with the environment. Our proposed deep dueling algorithm does not require this explicit information in advance. Instead, the algorithm learn these values and converge to the optimal policy for the transmitter.

\section{Problem Formulation}
\label{Sec:prob}
	
To deal with the uncertainty of jamming attacks and ambient RF signal, we adopt the Markov decision process (MDP) framework to formulate the optimization problem of the system. This framework allows the transmitter to dynamically make optimal actions based on its observations to maximize its average long-term reward. The MDP is defined by a tuple $<\mathcal{S}, \mathcal{A}, r>$ where $\mathcal{S}$ is the state space, $\mathcal{A}$ is the action space, and $r$ is the immediate reward of the system. 
	
\subsection{State Space}
We define the state space of the system as follows:
\begin{equation}
\begin{aligned}
\mathcal{S} \triangleq \Big\{ (c,j,d,e): c \in \{0,1\}; j \in \{0,1\}; d \in \{0,\ldots, D\}; e \in \{0,\ldots, E \} \Big\},
\end{aligned}
\end{equation}
where $c$ represents the state of the ambient RF channel, i.e., $c = 1$ when the ambient RF channel is busy and $c = 0$ otherwise. $j$ represents the state of the jammer, i.e., $j=1$ when the jammer is active and $j=0$ otherwise. $d$ and $e$ represent the number of packets in the data queue and the energy units in the energy storage of the  transmitter, respectively. $D$ and $E$ are the maximum data queue size and energy storage capacity, respectively. The system state is then defined as a composite variable $s = (c,j,d,e) \in \mathcal{S}$.

\subsection{Action Space}
	
The transmitter can perform one of the $(M+4)$ actions, i.e., stay idle, actively transmit data, harvest energy from the ambient signals, backscatter data from the ambient signals, harvest energy from the jamming signals, backscatter data from the jamming signals, or actively transmit data when then channel is attacked with one of $M$ transmission rates by using the RA technique. Then, the action space of the transmitter can be defined by $\mathcal{A} \triangleq \{a:a \in \{1,\ldots, M+4\} \}$, where
	\begin{equation}
	a 	=	\left\{	\begin{array}{ll}
	1,	&	\mbox{the transmitter stays idle},	\\
	2,	&	\mbox{the transmitter transmits data},	\\
	3,	&	\mbox{the transmitter harvests energy},\\
	4,	&	\mbox{the transmitter backscatters data},\\
	4+m, &  \mbox{the transmitter adapts its transmission to rate $r_m$ with $m \in \{1, \ldots, M\}$.}
	\end{array}	\right.
	\end{equation}

\subsection{Immediate Reward}
	
We define the reward for the system as the number of packets that are successfully transmitted to the gateway. Thus, the immediate reward of the system after the transmitter makes an action $a$ at state $s$ can be defined as follows:
\begin{equation}
r(s,a)	=	\left\{	\begin{array}{ll}
d_\mathrm{t},	& \text{if} \phantom{5} c=0, j=0, d>0, e \geq e_\mathrm{t}, \text{and} \phantom{5} a=2,	\\
d_\mathrm{b},	& \text{if} \phantom{5} c=1, j = 0, d>0, \text{and} \phantom{5} a=4,	\\
d^{\mathrm{J}}_n,	& \text{if} \phantom{5} j=1, c = 0,  d>0, \text{and} \phantom{5} a=4,	\\
d_\mathrm{sum}, & \text{if} \phantom{5} j = 1, c = 1, d >0, \text{and} \phantom{5} a=4\\
d^{\mathrm{r}}_m, & \text{if} \phantom{5} c=0, j=1, d>0, e>0, \text{and} \phantom{5} a=4+m, \\
0	,						&	\mbox{otherwise}	. \end{array}	\right.
\end{equation}
	
In the above, when the ambient RF source is idle, the jammer does not attack the channel, and the number of data and energy units are sufficient for active transmission, the transmitter can actively transmit $0 < d_\mathrm{t} \leq \widehat{d}_\mathrm{t}$ packets to the gateway (i.e, $a=2$). When the ambient RF source is active, the jammer is idle, and the transmitter has data to transmit, it can choose to backscatter $0 < d_\mathrm{b} \leq \widehat{d}_\mathrm{b}$ packets(i.e., $a=4$). Similarly, when the jammer attacks the channel, the RF source is idle, and the transmitter has data to transmit, if it choose to backscatter, it can transmit maximum $0 < d^{\mathrm{J}}_n \leq \widehat{d}^{\mathrm{J}}_n$ packets(i.e., $a=4$). If the ambient RF source is idle, the jammer attacks the channel, and the transmitter has enough energy and data in the queues, it can choose to adapt its rate (i.e., $a=4+m; m \in \{1, \ldots, M\}$) and actively transmit $0<d^{\mathrm{r}}_m \leq \widehat{d}^{\mathrm{r}}_m$ packets to the gateway. If both the jammer and the RF source are active, and the transmitter has data to transmit, if it choose to backscatter data, it can transmit $ d_{\mathrm{min}}\leq  d_\mathrm{sum} \leq d_\mathrm{max}$ to the gate way~\cite{Yang2017Riding}. Finally, the immediate reward is equal to $0$ if the transmitter cannot successfully transmit any packet to the gateway.

Note that after performing an action, the transmitter will observe the results from the environment including reward, i.e., number of packets that are successfully transmitted based on ACK messages sent from the gateway. In other words, $d_\mathrm{t}$, $d_\mathrm{b}$, $d^{\mathrm{J}}_n$, $d_\mathrm{sum}$, and $d^{\mathrm{r}}_m$ are the actually received packet at the gateway, i.e., successfully-ACKed packets. For that, the reward function captures the overall path between the source and the tag-receiver, e.g., fading, end-to-end SNR, BER, or the packet error rate.

\subsection{Optimization Formulation}

We formulate an optimization problem to obtain the optimal policy, denoted by $\pi^*$, that maximizes the average long-term reward for the system. Specifically, the optimal policy is a mapping from a state to an action taken by the transmitter. In other words, given the current system state, i.e., data queue, energy level, and channel states, the policy determines an optimal action to maximize the average long-term reward for the system. The optimization problem is then expressed as follows:
\begin{eqnarray} 
	\label{eq:average_reward}
	\max_\pi	& &	{\mathcal{R}}(\pi)	=	\lim_{T \rightarrow \infty} \frac{1}{T} \sum_{k=1}^{T} {\mathbb{E}} \left( r_k (s_k, \pi(s_k)) \right),	\label{eq:cmdp_obj}
\end{eqnarray}
where ${\mathcal{R}}(\pi)$ is the average reward of the transmitter under the policy $\pi$ and $r_k (s_k, \pi(s_k))$ is the immediate reward under policy $\pi$ at time step $k$. Clearly, the state space $\mathcal{S}$ contains only one communicating class, i.e., from a given state the process can go to any other states after $k$ steps. In other words, the MDP with states in $\mathcal{S}$ is irreducible. Thus, for every $\pi$, the average throughput $\mathcal{R}(\pi)$ is well defined and does not depend on the initial state~\cite{CompetitiveBook}.
%
%
%

\section{Learning about Jammer and Its Strategies with Reinforcement Learning Approaches}
\label{sec:QDeepQ}
\subsection{Q-Learning Approach}
\begin{figure}[!]
	\centering
	\includegraphics[scale=0.3]{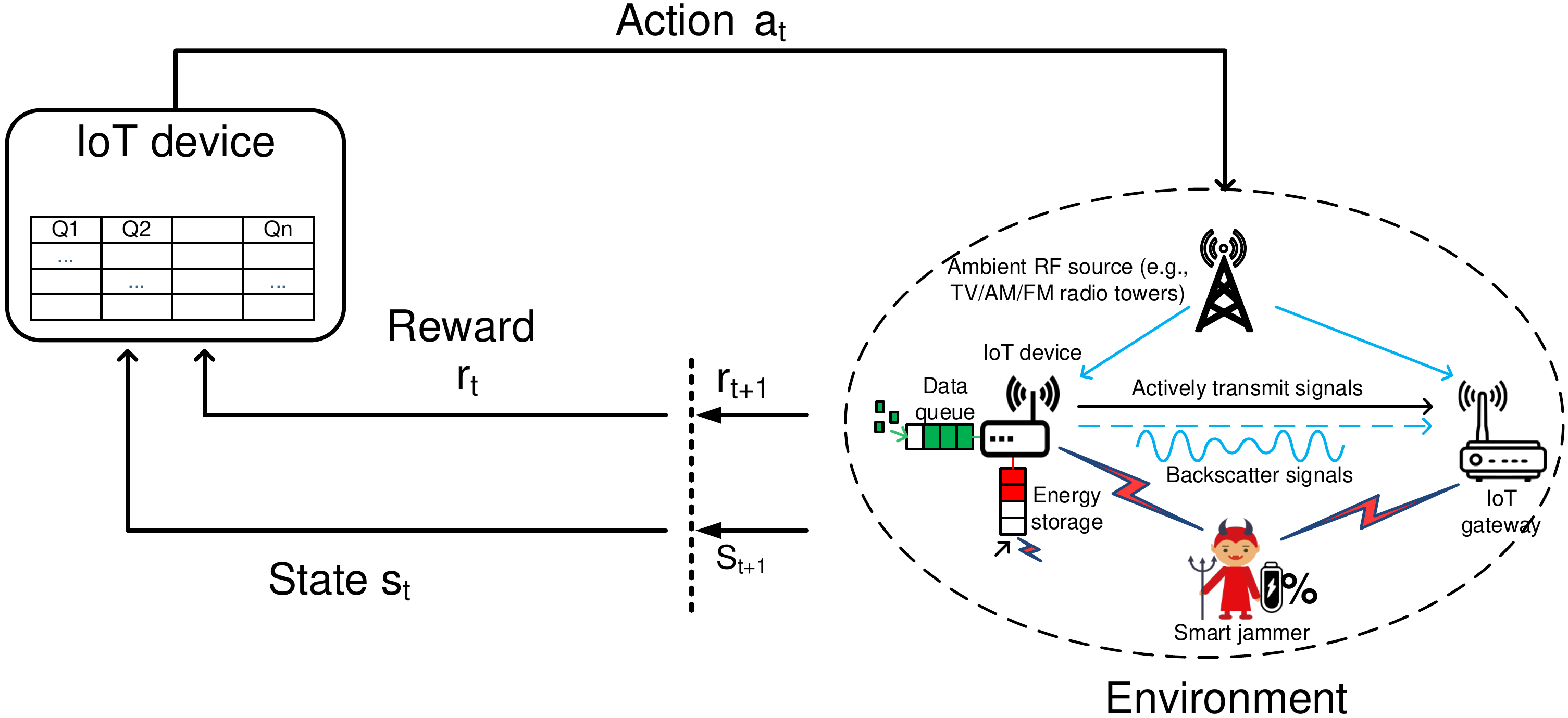}
	\caption{Q-learning based model.}
	\label{Fig.qlearning}
\end{figure}

In the system under consideration, the transmitter cannot obtain the information about the jammer, e.g., jamming capacity, as well as the ambient RF signal, e.g., channel operation frequency, in advance to find the optimal policy. Thus, this section introduces Q-learning~\cite{Watkins1992QLearning}, a reinforcement learning algorithm, which can help the transmitter find the optimal policy without requiring prior information about the jammer as well as the channel. In particular, as illustrated in Fig.~\ref{Fig.qlearning}, the Q-learning algorithm implements a Q-table to store state-action pair values. Given a current state, the algorithm will select an action based on its current strategy. After performing the selected action, the Q-learning algorithm observes the immediate reward and next state, and updates the Q-values based on the Q-value function. In this way, the Q-learning algorithm can learn from its decisions, and it was proved that the Q-learning algorithm will converge to the optimal policy after a finite number of iterations~\cite{Watkins1992QLearning}.
	
In this paper, we aim to find the optimal policy $\pi^*:\mathcal{S} \rightarrow \mathcal{A}$, i.e., a mapping from states to their corresponding actions, for the transmitter to optimize its performance, i.e., maximize its long-term average throughput and minimize its packet loss, under the jamming attacks. Let's denote $\mathcal{V}^\pi(s): \mathcal{S} \rightarrow \mathbb{R}$ as the expected value function obtained by policy $\pi$ from a state $s \in \mathcal{S}$, that can be defined as follows:
\begin{equation}
\begin{aligned}
\mathcal{V}^\pi(s) = \mathbb{E}_\pi \Big [ \sum_{t=0}^{\infty} \gamma r_t(s_t, a_t)|s_0=s\Big ] =\mathbb{E}_\pi\Big [ r_t(s_t, a_t) + \gamma\mathcal{V}^\pi(s_{t+1})|s_0=s\Big ],
\end{aligned}
\end{equation}
where $0\leq \gamma < 1$ is the discount factor which represents the importance of long-term reward~\cite{Watkins1992QLearning}. Specifically, if $\gamma$ is close to 0, the algorithm is likely to select actions to maximize its short-term reward. In contrast, when $\gamma$ is close to 1, the algorithm will make actions such that its long-term reward is maximized. $r_t(s_t, a_t)$ is the immediate reward achieved after performing action $a_t$ at state $s_t$. 
To find the optimal policy $\pi^*$, at each state, an optimal action has to be found through the following optimal value function.
\begin{equation}
\mathcal{V}^*(s) = \max_{a} \Big \{ \mathbb{E}_\pi[r_t(s_t, a_t)+ \gamma\mathcal{V}^\pi(s_{t+1})] \Big\} ,\quad \forall s \in \mathcal{S}.
\end{equation}
For all state-action pairs, the optimal Q-functions are denoted by:
\begin{equation}
\mathcal{Q}^*(s,a) \triangleq r_t(s_t, a_t) + \gamma\mathbb{E}_\pi[\mathcal{V}^\pi(s_{t+1})] , \quad \forall s \in \mathcal{S}.
\end{equation}
	
Then, the optimal value function $\mathcal{V}^*(s)$ can be written as $\mathcal{V}^*(s) = \max_{a} \{ \mathcal{Q}^*(s,a)\}.$
By making samples iteratively, the problem is reduced to determining the optimal value of Q-function, i.e., $\mathcal{Q}^*(s,a)$, for all state-action pairs. In particular, the Q-function is updated according to (\ref{Eq:qfunction}).
\begin{equation}
\label{Eq:qfunction}
\begin{aligned}
\mathcal{Q}_{t+1}(s_t,a_t) = \mathcal{Q}_t(s_t,a_t) + \tau_t \Big [ r_t(s_t, a_t) + \gamma\max_{a_{t+1}} \mathcal{Q}_t(s_{t+1}, a_{t+1})- \mathcal{Q}_t(s_t,a_t)\Big ].
\end{aligned}
\end{equation}
In particular, (\ref{Eq:qfunction}) is used to find the temporal difference between the predicted Q-value, i.e., $r_t(s_t, a_t) + \gamma \max_{a_{t+1}} \mathcal{Q}_t(s_{t+1}, a_{t+1})$ and its current value, i.e., $\mathcal{Q}_t(s_t,a_t)$. The learning rate $\tau_t$ determines the impact of new information to the existing value. During the learning process, the learning rate can be adjusted dynamically, or it can be chosen to be a constant. However, to guarantee the convergence for the Q-learning algorithm, the learning rate $\tau_t$ is deterministic, nonnegative, and satisfies the following conditions~\cite{Watkins1992QLearning}:
\begin{equation}
\label{Eq:rules}
\tau_t \in [0,1), \sum_{t=1}^{\infty}\tau_t = \infty, \mbox{ and } \sum_{t=1}^{\infty} ( \tau_t  )^{2} < \infty.
\end{equation}

\begin{algorithm}
		\caption{Optimal Defense Strategy with Q-learning Algorithm}
		\label{algorithm0}
		\begin{algorithmic}[1]
			\State \textbf{Inputs:} For each state-action pair $(s, a)$, initialize the table entry $\mathcal{Q}(s, a)$ arbitrarily, e.g., to zero. Observe the current state $s$, initialize a value for the learning rate $\tau$ and the discount factor $\gamma$.
			\For{\textit{t=1 to T}}
			\State From the current state-action pair $(s_t, a_t)$, execute action $a_t$ and obtain the immediate reward $r_t$ 
			\State and new state $s_{t+1}$. Select an action $a_{t+1}$ based on the state $s_{t+1}$ and then update the table entry 
			\State for $\mathcal{Q}(s_t, a_t)$ as follows:
			\begin{equation}
			\begin{aligned}
			\mathcal{Q}_{t+1}(s_t,a_t) = \mathcal{Q}_t(s_t,a_t) + \tau_t\Big [ r_t(s_t, a_t) + \gamma\max_{a_{t+1}} \mathcal{Q}_t(s_{t+1}, a_{t+1})- \mathcal{Q}_t(s_t,a_t)\Big ].
			\end{aligned}
			\end{equation}
			\State Replace $s_t \leftarrow s_{t+1}$.
			\EndFor
			\State {\textbf{Outputs:}} $\pi^*(s) = \arg\max_{a} \mathcal{Q}^*(s,a)$.
		\end{algorithmic}
	\end{algorithm}
	
The details of the Q-learning algorithm is provided in Algorithm~\ref{algorithm0}. Specifically, from the current state $s_t$, the algorithm will choose an action $a_t$ and observe results after performing this action. In practice, to select action $a_t$, $\epsilon$-greedy algorithm~\cite{Sutton1998Reinforcement} is often adopted. In particular, this technique chooses a random action with probability $\epsilon$, and selects an action that maximizes the $\mathcal{Q}(s,a_{s})$ with probability $1-\epsilon$. After performing the chosen action, the Q-learning algorithm observes the next state and reward, and then updates the table entry for $\mathcal{Q}(s_t, a_t)$ based on Eq.~(\ref{Eq:qfunction}). When all $Q$-values converge or a certain number of iterations is reached, the learning process will be terminated. The Q-learning algorithm yields the optimal policy indicating an action to be taken at each state such that $\mathcal{Q}^*(s,a)$ is maximized for all states in the state space, i.e., $\pi^*(s) = \arg\max_{a} \mathcal{Q}^*(s,a)$. Under the conditions of $\tau_t$ stated in Eq.~(\ref{Eq:rules}), in Theorem~\ref{theo:convergeQ}, we show that the Q-learning algorithm will converge to the optimum action-values with probability one.
\begin{theorem}
	\label{theo:convergeQ}
	Under the conditions of $\tau_t$ in Eq.~(\ref{Eq:rules}), the Q-learning algorithm converges to the optimum action-values with probability one.
\end{theorem}
The proof of Theorem~\ref{theo:convergeQ} is provided in Appendix~\ref{appendix:convergeQ}. It is worth noting that the Q-learning algorithm can converge to the optimal policy in a reasonable time when the state space and the action space are small. Nonetheless, for a complicated system with thousands of state-action pairs, the convergence rate of the Q-learning algorithm is usually slow. That makes the Q-learning algorithm practically inapplicable~\cite{LillicrapContinuous}, especially for our considered system model which needs to learn activities from both the jammer and ambient RF source. Thus, in the following, we introduce deep Q-learning and deep dueling algorithms to quickly obtain the optimal defend policy for the transmitter, thereby effectively defeating the jamming attacks and optimizing performance for the system.

\subsection{Deep Q-Learning based Adaptation}
\begin{figure*}[!]
	\centering
	\includegraphics[scale=0.28]{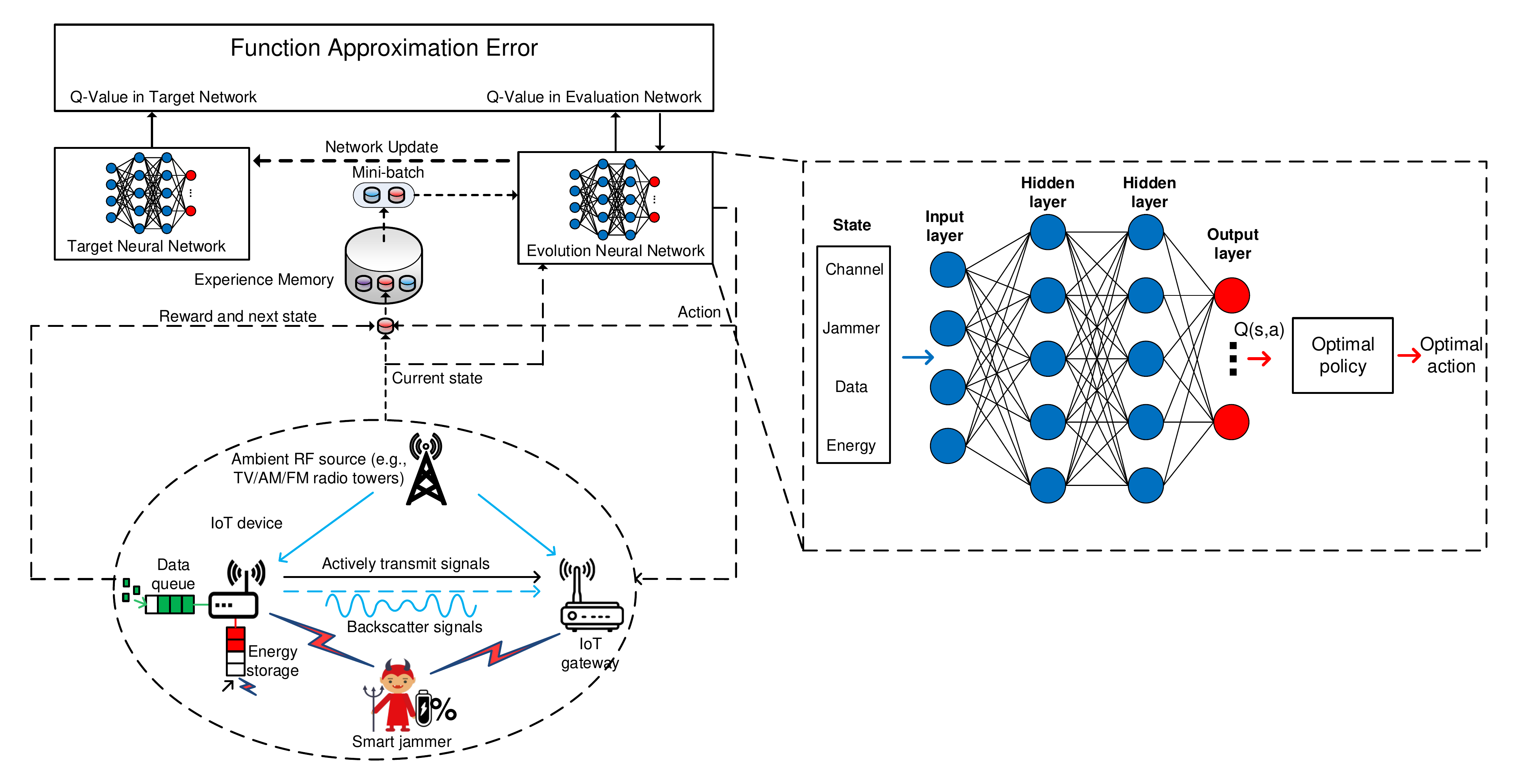}
	\caption{Deep Q-learning based model}
	\label{Fig.deepqlearning}
\end{figure*}

In this section, we propose the deep Q-learning algorithm~\cite{Mnih2015Human} to cope with the low-convergence problem of the Q-learning algorithm introduced in Section~\ref{sec:QDeepQ}. Intuitively, the deep Q-learning algorithm was introduced by Google DeepMind in 2015~\cite{Mnih2015Human} to teach machines to play games without human intervention. The deep Q-learning algorithm implements a deep neural network instead of the Q-table to find the approximated values of $\mathcal{Q}^*(s,a)$ as illustrated in Fig.~\ref{Fig.deepqlearning}.
	
According to~\cite{Mnih2015Human}, the performance of reinforcement learning approaches might not be stable or even diverges when using a nonlinear function approximator. The reason is that with a small change of Q-values, the data distribution and correlations between the Q-values and the target values, i.e., $r+\gamma \max_{a} \mathcal{Q}(s,a)$, are varied, and thus the policy is greatly affected. To address this issue, we use three mechanisms, i.e., experience replay, target Q-network, and feature set.
	
\begin{itemize}
	\item \textit{Experience replay mechanism:} The algorithm implements a replay memory $\mathbf{D}$, i.e., memory pool, to store transitions $(s_t, a_t, r_t, s_{t+1})$ instead of running on state-action pairs as they occur during experience. Random samples from the memory pool are then fed to the deep neural network for training. In this way, the algorithm can efficiently learn from previous experiences many times and remove the correlations between observations~\cite{Mnih2015Human}.
		
	\item \textit{Target Q-network:} Obviously, the Q-values will be changed during the training process. As a result, the value estimations can be out of control if a constantly shifting set of values is used to update the Q-network resulting in the destabilization of the algorithm. To overcome this issue, the deep Q-learning algorithm implements a target Q-network to frequently but slowly update to the primary Q-network. As such, the correlations between the target and estimated Q-values are significantly eliminated, thereby stabilizing the algorithm.
		
	\item \textit{Feature set: } For each state, we determine four features including the activities of the jammer and the ambient RF source, i.e., active or idle, as well as the status of data and energy queues of the transmitter. These features are then fed to the deep neural network to approximate Q-values for each state-action pair. Doing so, all aspects of each state are trained resulting in a high convergence rate.
\end{itemize}
	
\begin{algorithm}
		\caption{Deep Q-learning Based Anti-jamming Algorithm}
		\label{deepqlearning}
		\begin{algorithmic}[1]
			\State Initialize replay memory $\mathbf{D}$ to capacity $\mathcal{D}$.
			\State Initialize the Q-network $\mathcal{Q}$ with random weights $\theta$.
			\State Initialize the target Q-network $\hat{\mathcal{Q}}$ with weight $\theta^-=\theta$.
			\For{\textit{episode=1 to I}}
			\State With probability $\epsilon$ select a random action $a_t$, otherwise select $a_t=\arg \max \mathcal{Q}^*(s_t, a_t; \theta)$
			\State Perform action $a_t$ and observe reward $r_t$ and next state $s_{t+1}$
			\State Store transition $(s_t, a_t, r_t, s_{t+1})$ in the replay memory $\mathbf{D}$
			\State Sample random mini-batch of transitions $(s_j, a_j, r_j, s_{j+1})$ from $\mathbf{D}$
			\State $y_j=r_j+\gamma\max_{a_{j+1}}\hat{\mathcal{Q}}(s_{j+1},a_{j+1};\theta^-)$
			\State Perform a gradient descent step on $(y_j-\mathcal{Q}(s_j, a_j; \theta))^2$ with respect to the network parameter $\theta$.
			\State Every $C$ steps reset $\hat{\mathcal{Q}} = \mathcal{Q}$
			\EndFor
		\end{algorithmic}
\end{algorithm}

Algorithm~\ref{deepqlearning} provides the details of the deep Q-learning algorithm. In particular, as shown in Fig.~\ref{Fig.flowchart}, the training phase consists of multiple episodes. In each episode, given the current state, the algorithm chooses an action based on the epsilon greedy algorithm. The algorithm will start with a fairly randomized policy and later slowly move to a deterministic policy. In other words, at the first episode, $\epsilon$ is set at a large value, e.g., 0.9, and gradually decayed to a small value, e.g., 0.1. After that, the algorithm performs the selected action and observes results from taking this action, i.e., next state and reward. This transition is then stored in the replay memory for training process at later episodes.
	
In the learning process, random samples of transitions from the replay memory will be fed into the neural network. The algorithm then updates the neural network by minimizing the following lost function.
\begin{equation}
\label{lossfunction}
\begin{aligned}
L_i(\theta_i)=\mathbb{E}_{(s,a,r,s')\sim U(\mathbf{D})}\bigg[ \bigg( r + \gamma\max_{a'}\mathcal{Q}(s',a';\theta_i^-) -\mathcal{Q}(s,a;\theta_i)\bigg)^2\bigg],
\end{aligned}
\end{equation}
where $\gamma$ is the discount factor, $\theta_i$ are the parameters of the Q-networks at episode $i$ and $\theta_i^-$ are the parameters of the target network, i.e., $\hat{\mathcal{Q}}$. Differentiating the loss function in~(\ref{lossfunction}) with respect to the parameters of the neural networks, we have the following gradient:
\begin{equation}
\label{gradient_loss}
\begin{aligned}
\nabla_{\theta_i}L(\theta_i)=\mathbb{E}_{(s,a,r,s')}\bigg[\bigg(r+\gamma \max_{a'}\mathcal{Q}(s',a';\theta_i^-)-\mathcal{Q}(s,a;\theta_i)\nabla_{\theta_i}\mathcal{Q}(s,a;\theta_i)\bigg)\bigg].
\end{aligned}
\end{equation}
	
To minimize the loss function in~(\ref{lossfunction}), one can use the \textit{Stochastic Gradient Descent} algorithm~\cite{Goodfellow2016Deep}, which is very important algorithm to power nearly all of deep learning algorithms, to calculate the gradient in~(\ref{gradient_loss}). In general, the cost function used by a machine learning algorithm is decayed by a sum over training examples of some per-example loss function. For instance, the negative conditional log-likelihood of the training data can be expressed as:
\begin{figure}[!]
	\centering
	\includegraphics[scale=0.13]{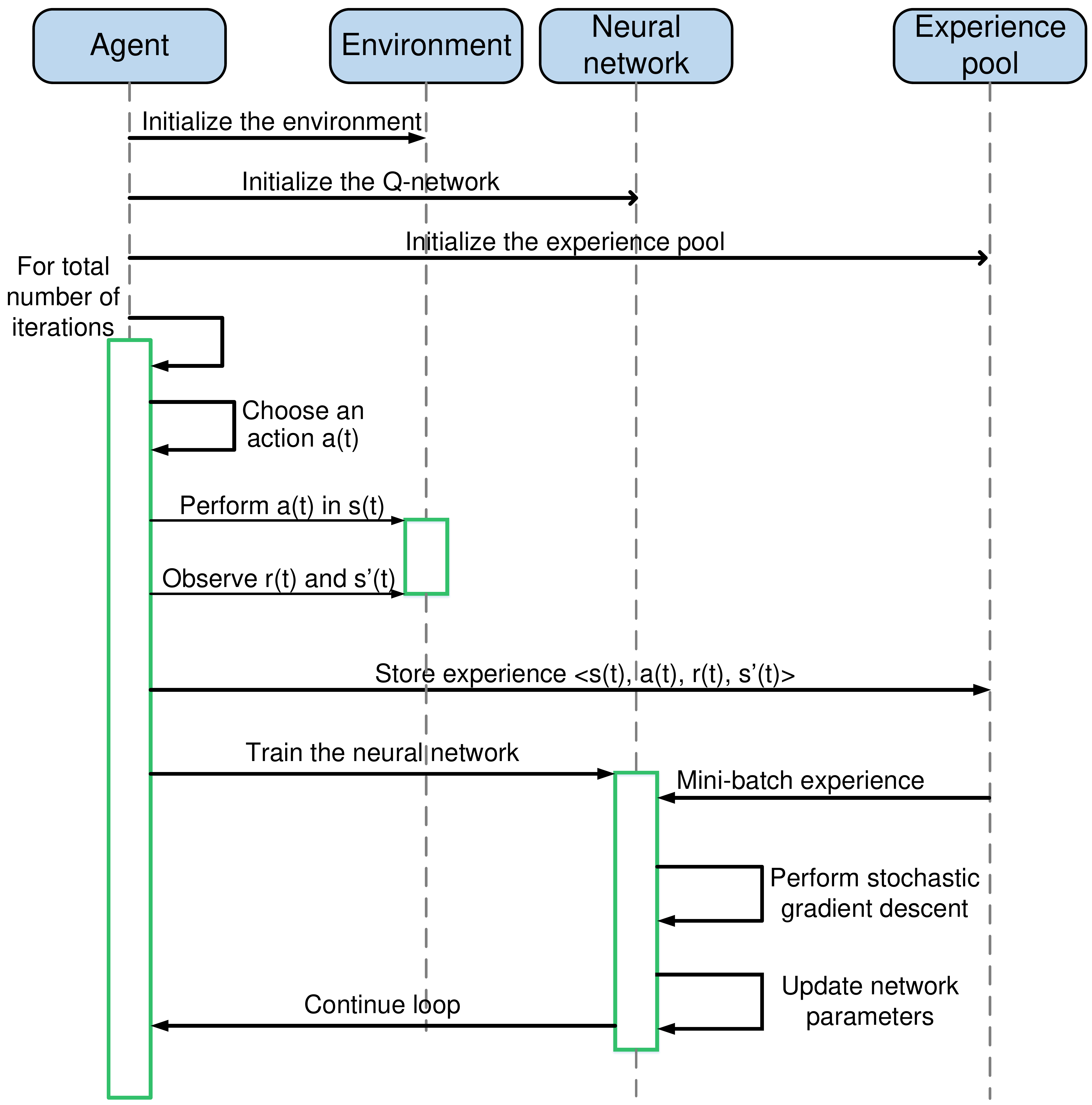}
	\caption{Flow chart of the deep Q-learning algorithm}
	\label{Fig.flowchart}
\end{figure}
\begin{equation}
\begin{aligned}
J(\theta) = \mathbb{E}_{(s,a,r,s') \sim U(\mathbf{D})}L\big((s,a,r,s'),\theta\big) =\frac{1}{\mathcal{D}}\sum_{i=1}^{\mathcal{D}}L\big((s,a,r,s')^{(i)}, \theta\big),
\end{aligned}
\end{equation}
where $\mathcal{D}$ is the size of the memory pool. For these additive cost function, gradient descent requires computing as follows:
\begin{equation}
\label{eq:cost}
\nabla_\theta J(\theta) = \frac{1}{\mathcal{D}}\sum_{i=1}^{\mathcal{D}} \nabla_\theta L\big((s,a,r,s')^{(i)}, \theta\big).
\end{equation}
The computational cost for the operation in Eq.~(\ref{eq:cost}) is $O(\mathcal{D})$. Thus, as the size $\mathcal{D}$ of the replay memory is increased, the time to take a single gradient step becomes prohibitively long. As a result, in this paper, we adopt the stochastic gradient descent technique. The key idea of using stochastic gradient descent is that the gradient is an expectation. Clearly, the expectation can be approximately estimated by using a small set of samples. In particular, we can uniformly sample a mini-batch of experiences from the replay memory $\mathbf{D}$ at each step of the algorithm. In general, the mini-batch size can be set to be relative small number of experiences, e.g., from one to a few hundred. As such, the training time is significantly fast. The estimate of the gradient under the stochastic gradient descent is then formulated as follows:
\begin{equation}
g=\frac{1}{N} \nabla_\theta\sum_{i=1}^{N}L\big((s,a,r,s')^{(i)}, \theta \big),
\end{equation}
where $N$ is the mini-batch size. The stochastic gradient descent algorithm then follows the estimated gradient downhill as in Eq.~(\ref{eq:downhill}).
\begin{equation}
\label{eq:downhill}
\theta \leftarrow \theta- \nu g,
\end{equation}
where $\nu$ is the learning rate of the algorithm.
	
After every $C$ steps, the algorithm updates the target network parameters $\theta_i^-$ with the Q-network parameters $\theta_i$. The target network parameters remain unchanged between individual updates. Fig.~\ref{Fig.flowchart} shows the flowchart of the deep Q-learning algorithm.
	
\section{Fighting Jammer with Deep Dueling Neural Network Architecture}
	\label{Sec:deepdueling}
\subsection{Deep Dueling Neural Network Architecture}
	
\begin{figure}[!]
	\centering
	\includegraphics[scale=0.15]{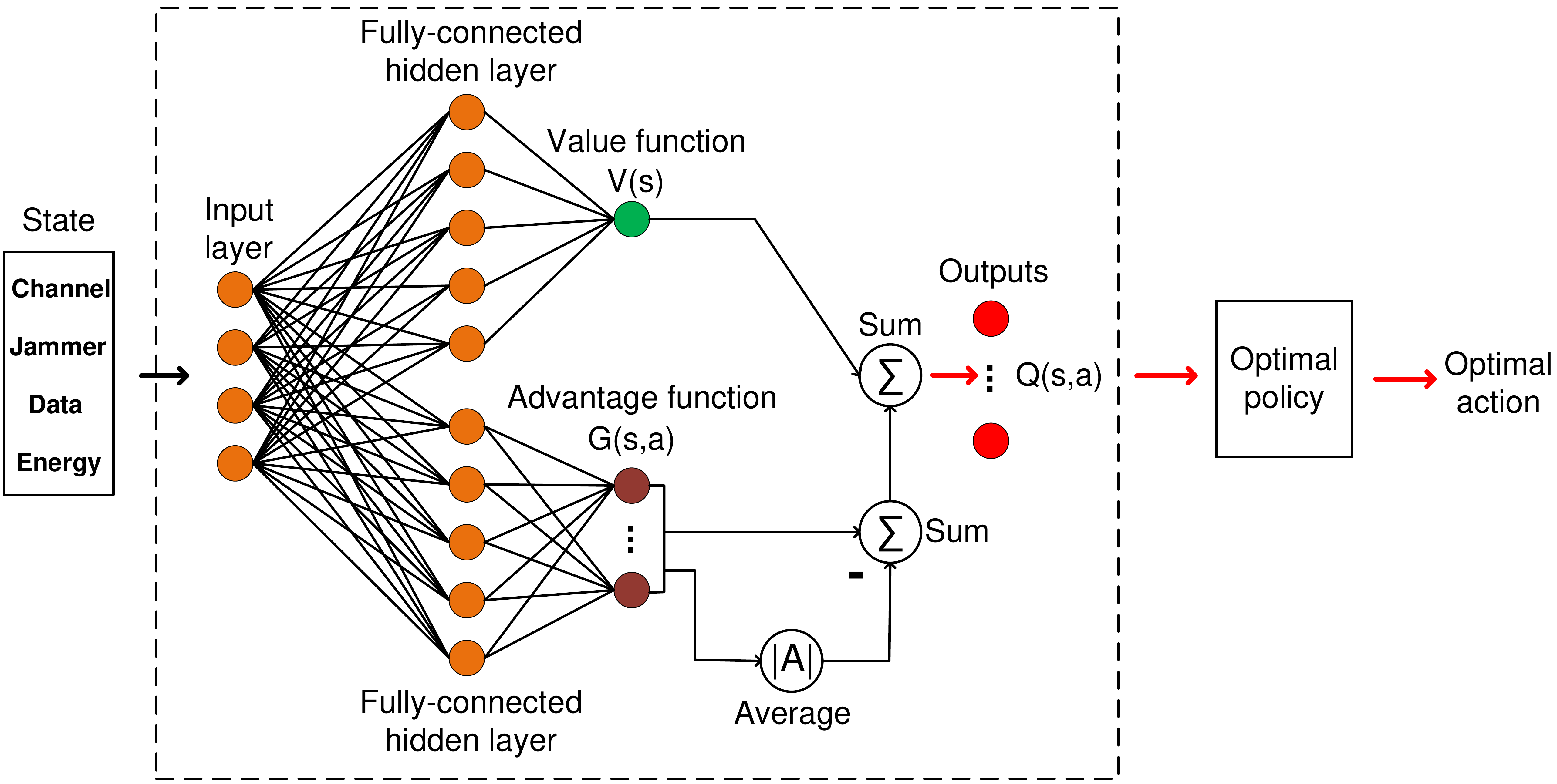}
	\caption{Deep dueling neural network architecture based solution}
	\label{Fig.deepduelingqlearning}
\end{figure}
	
According to~\cite{Wang2015Dueling}, the convergence rate of the deep Q-learning algorithm is still limited due to the overestimation of optimizers, especially in systems with large action and state spaces as considered in this work. Therefore, we propose deep dueling algorithm~\cite{Wang2015Dueling}, which was also originally developed by Google DeepMind in 2016, to further improve the system's convergence speed. The key idea making the deep dueling superior to conventional approaches is its novel neural network architecture. Clearly, in many states, it is unnecessary to estimate the value of corresponding actions as the choice of these actions has no repercussion on what happens~\cite{Wang2015Dueling}. For example, the rate adaptation actions only matter when the jammer attacks the channels with low power levels. Hence, instead of estimating the action-value function, i.e., Q-function, the algorithm divides the deep neural network into two sequences, i.e., streams, of fully connected layers to separately estimate the values of states and advantages of actions{\footnote{The value function represents how good it is for the system to be in a given state. The advantage function is used to measure the importance of a certain action compared with others \cite{Wang2015Dueling}.}}. The values and advantages are then combined at the output layer as shown in Fig.~\ref{Fig.deepduelingqlearning}. In this way, the deep dueling algorithm can achieve more robust estimates of state value, and thus significantly improving its convergence rate as well as stability. It is worth noting that the flowchart of the deep dueling algorithm is the same as in the deep Q-learning. The main difference between the deep dueling algorithm and other conventional deep reinforcement learning algorithms is the deep dueling neural network. In the following, we present details of separating the Q-value into the value and the advantage functions.
	
Recall that given a stochastic policy $\pi$, the values of state-action pair $(s,a)$ and state $s$ are as follows:
\begin{equation}
\begin{aligned}
\mathcal{Q}^{\pi}(s,a) &= \mathbb{E} \big[r_t|s_t=s, a_{t}=a,\pi\big], \\
\mathcal{V}^{\pi}(s) &= \mathbb{E}_{a \sim \pi(s)}\big[\mathcal{Q}^{\pi}(s,a)\big].
\end{aligned}
\end{equation}
The advantage function of actions can be expressed as:
\begin{equation}
\mathcal{G}^{\pi}(s,a) =  \mathcal{Q}^{\pi}(s,a) - \mathcal{V}^{\pi}(s).
\end{equation}
Specifically, the value function $\mathcal{V}$ corresponds to how \emph{good it is to be in a particular state $s$}~\cite{Wang2015Dueling}. The state-action pair, i.e., Q-function, calculate the value of performing action $a$ in state $s$. The advantage function decouples the state value from the Q-function to measure the importance of each action.
	
To estimate values of $\mathcal{V}$ and $\mathcal{G}$ functions, we use a dueling neural network in which one stream of fully-connected layers outputs a scalar $\mathcal{V}(s;\beta)$ and the other stream estimates an $|\mathcal{A}|$-dimensional vector $\mathcal{G}(s, a;\alpha)$, where $\alpha$ and $\beta$ are the parameters of fully-connected layers. These two sequences are then combined at the output layer to obtain the Q-function by Eq.~(\ref{combined}).
\begin{equation}
\label{combined}
\mathcal{Q}(s, a;\alpha, \beta) = \mathcal{V}(s;\beta) + \mathcal{G}(s, a;\alpha).
\end{equation}
Note that Eq.~(\ref{combined}) applies to all $(s, a)$ instances. Thus, to express equation (\ref{combined}) in matrix form, one needs to replicate the scalar, $\mathcal{V}(s;\beta)$, $|\mathcal{A}|$ times. Importantly, $\mathcal{Q}(s, a;\alpha, \beta)$ is a parameterized estimate of the true Q-function, and given $\mathcal{Q}$, we cannot obtain $\mathcal{V}$ and $\mathcal{G}$ uniquely. In other words, adding a constant to $\mathcal{V}(s;\beta)$ and subtracting the same constant from $\mathcal{G}(s, a;\alpha)$ result in the same Q-value. Therefore, Eq.~(\ref{combined}) is unidentifiable resulting in poor performance. To address this problem, the combining module of the network is implemented the following mapping:
\begin{equation}
\label{ouput_max}
\mathcal{Q}(s,a;\alpha,\beta) = \mathcal{V}(s;\beta) + \big(\mathcal{G}(s,a;\alpha)-\max_{a \in \mathcal{A}}\mathcal{G}(s,a;\alpha)\big).
\end{equation}
In this way, the advantage function estimator has zero advantage when choosing action. Intuitively, given $a^*=\arg\max_{a \in \mathcal{A}} \mathcal{Q}(s,a;\alpha,\beta)=\arg\max_{a \in \mathcal{A}}\mathcal{G}(s,a;\alpha)$, we have $\mathcal{Q}(s, a^*;\alpha, \beta)=\mathcal{V}(s;\beta)$. Therefore, we can convert (\ref{ouput_max}) into a simple form by replacing the max operator with an average as follows:
\begin{equation}
\label{output_average}
\mathcal{Q}(s,a;\alpha,\beta) = \mathcal{V}(s;\beta) + \big(\mathcal{G}(s,a;\alpha)- \frac{1}{|\mathcal{A}|}\sum_{a}^{}\mathcal{G}(s, a;\alpha)\big).
\end{equation}
Note that subtracting the mean in Eq.~(\ref{output_average}) solves the unidentifiable problem. However, it does not change the relative rank of the advantage function values, and hence the $\mathcal{Q}$ values for actions at each state.
\begin{algorithm}[!]
		\caption{Deep Dueling Neural Network Based Anti-jamming Algorithm}
		\label{deepduelingqlearning}
		\begin{algorithmic}[1]
			\State Initialize replay memory $\mathbf{D}$ to capacity $\mathcal{D}$.
			\State Initialize the primary network $\mathcal{Q}$ including two fully-connected layers with random weights $\alpha$ and $\beta$.
			\State Initialize the target network $\hat{\mathcal{Q}}$ as a copy of the primary Q-network with weights $\alpha^- = \alpha$ and $\beta^- = \beta$.
			\For{\textit{episode=1 to I}}
			\State Base on the $\epsilon$-greedy algorithm, with probability $\epsilon$ select a random action $a_{t}$ at state $s_t$. Otherwise, 
			\State select $a_{t}=\arg \max \mathcal{Q}^*(s_t, a_t; \alpha, \beta)$
			\State Perform action $a_t$ and observe reward $r_t$ and next state $s_{t+1}$
			\State Store transition $(s_t, a_t, r_t, s_{t+1})$ in the replay memory
			\State Sample random mini-batch of transitions $(s_j, a_{j}, r_j, s_{j+1})$ from the replay memory
			\State Combine the value function and advantage functions as follows:
			\begin{equation}
			\begin{aligned}
			\mathcal{Q}(s_j,a_{j};\alpha,\beta) = \mathcal{V}(s_j;\beta) + \big(\mathcal{G}(s_j,a_{j};\alpha) - \frac{1}{|\mathcal{A}|}\sum_{a_{j}}^{}\mathcal{G}(s_j, a_{j};\alpha)\big).
			\end{aligned}
			\end{equation}
			\State $y_j=r_j+\gamma\max_{a_{j+1}}\hat{\mathcal{Q}}(s_{j+1},a_{j+1}; \alpha^-, \beta^-)$
			\State Perform a gradient descent step on $(y_j-\mathcal{Q}(s_j, a_{j}; \alpha, \beta))^2$
			\State Every $C$ steps reset $\hat{\mathcal{Q}} = \mathcal{Q}$
			\EndFor
		\end{algorithmic}
\end{algorithm}
Based on Eq.~(\ref{output_average}) and the advantages of the deep reinforcement learning, we propose the deep dueling algorithm as shown in Algorithm~\ref{deepduelingqlearning}. It is worth noting that Eq.~(\ref{output_average}) is viewed and implemented as a part of the network and not as a separated algorithmic step~\cite{Wang2015Dueling}. In addition, $\mathcal{V}(s;\beta)$ and $\mathcal{G}(s,a;\alpha)$ are estimated automatically without any extra supervision or modifications in the algorithm.
	
\subsection{Complexity Analysis and Implementation}
Deep reinforcement learning (and deep learning in general) is well known for solving complicated and intractable problems with much higher performance compared to conventional methods. However, it typically requires many CPU and GPU resources, especially for large-scale and complex problems which require more hidden layers. To implement deep learning in resource-constrained IoT devices, several solutions have been proposed in the literature~\cite{Complexity}. In~\cite{Han2015Learning}, the authors proposed a novel method, called network compression, to convert a densely connected neural network into a sparsely connected network. In this way, the storage and computation load can be reduced by a factor of 10. The authors show that the network compression can allow deep learning algorithm to be implemented on commonly used IoT platforms from Qualcomm, Intel, and NVidia. Other approaches to reducing computational load such as approximate computing and deploy specialized accelerator hardware also make deep learning suitable to run on the transmitter~\cite{Complexity}. Note that our proposed deep dueling algorithm only uses a simple deep neural network architecture with only one hidden layer for each stream. Together with recent advances in network compression, approximate computing, and accelerators, our proposed solution can be successfully implemented in common IoT devices.

Nevertheless, for ultra low-power transmitters, deep reinforcement learning might not be able to be implemented. To solve this problem, we use a central controller running on the gateway which has sufficient computing power to run the algorithm. Note that although the number of iterations is significant, this is the iterations for running the deep-Q learning algorithm, not the physical/real interactions between the transmitter and the gateway. As such, the computation and communication overhead are practically manageable. Specifically, at each state, given the current optimal policy, the transmitter will take an action and observe the results from the environment. It then stores this experience, i.e., $(s_j ; a_j ; r_j ; s_{j+1})$, in a memory pool. After a certain period, e.g., a day, the transmitter will send experiences in the memory pool to the gateway. The gateway will use all these experiences to train the deep neural network, i.e., estimate the Q-value for each state-action pair, to obtain the optimal policy. Then, the optimal policy is sent back to the transmitter. This optimal policy is then stored at the transmitter to guide it in making real-time decisions. The proposed solution is thus also applicable to low-latency applications.

	
\section{Performance Evaluation}
\label{sec:evaluation}
\subsection{Parameter Setting}
In our system, the data queue of transmitter can store up to 20 packets with the packet size set at 300 bits~\cite{Blasco2013Learning}. The energy storage capacity is set to be 20 units. The fundamental energy unit is 60 $\mu J$~\cite{Papotto201490nm}. When the ambient RF source/channel is active, the transmitter can either harvest two units of energy or backscatter one packet to the gateway. When the channel is idle and the jammer does not attack the channel, if the transmitter performs active transmission, it can successfully transmit $4$ packets. Each transmitted packet requires one unit of energy. The jammer has four transmit power levels, i.e., $\mathbf{P}_{\mathrm{J}}$ = \{$0$W, $7$W, $15$W, $21$W\}, with $P_{\max}$= $21$W~\cite{21WJammer}. As explained in the Section~\ref{Sec.System}, as the jamming power increases, the transmitter can successfully harvest more energy or transmit more packets by backscattering jamming signal, and thus we set $\mathbf{e}=\{0,2,3,4\}$ and $\widehat{\mathbf{d}}=\{0,1,2,3\}$. In addition, when the jammer attacks the channel and the rate adaption technique is implemented, the transmitter can transmit $d^{\mathrm{r}}_m=$\{2, 1, 0\} packets when the jammer attacks under power levels $P^{\mathrm{J}}_n$=\{$7$W, $15$W, $21$W\}, respectively. In the case both the jammer and the ambient RF source are active, the total number of backscattered packets $d_\mathrm{sum}$ and the total amount of harvested energy $e_\mathrm{sum}$ follow the Poisson distribution with the means of $(d_\mathrm{min}+d_\mathrm{max})/2$ and $(e_\mathrm{min}+e_\mathrm{max})/2$, respectively. The latency threshold $t_\mathrm{th}$ is set at $3$ time units. Unless otherwise stated, the idle channel probability is $0.5$ and $P_\mathrm{avg}$ = $7$W. Note that the reinforcement learning algorithms, i.e., Q-learning, deep Q-learning, and deep dueling, do not require the information about jammer, e.g., jamming strategy, and the channel activity, i.e., the idle channel probability, in advance. These information can be learned through the real-time learning process.

The architecture of the deep neural network significantly affects the performance of the deep reinforcement learning algorithms, and thus it requires a thoughtful design. In particular, the complexity of the algorithms increases when the number of hidden layers increases. Nevertheless, if the number of hidden layers is small, the algorithms requires a long time to converge to the optimal policy. Similarly, if the size of hidden layers, i.e., numbers of neutrons, and the mini-batch size $N$ are large, the algorithms will need more time to estimate the Q-function. For deep Q-learning algorithms, we adopt parameters based on the common settings for designing neural networks~\cite{Mnih2015Human,Wang2015Dueling}. Specifically, for the deep Q-learning algorithm, two fully-connected hidden layers are implemented together with input and output layers (as illustrated in Fig.~\ref{Fig.deepqlearning}). For the deep dueling algorithm, the neural network is divided into two streams. Each stream consists of a hidden layer connected to the input and output layers as illustrated in Fig.~\ref{Fig.deepduelingqlearning}). The size of the hidden layers is $16$. The mini-batch size is set at $16$. The maximum size of the experience replay buffer is 10,000, and the target Q-network is updated every 1,000 iterations~\cite{Mnih2015Human,Goodfellow2016Deep}. All learning algorithms use the $\epsilon$-greedy scheme with the initial value of $\epsilon$ set at $1$ and its final value set at $0.1$~\cite{Watkins1992QLearning}.
	
To evaluate the proposed solutions, we compare their performance with two other schemes, i.e., HTT and WTJ. For the HTT scheme, the transmitter only implements harvest-then-transmit protocol without considering ambient backscatter communication technology. This scheme is to evaluate the impact of ambient backscatter communications to the system performance. For the WTJ, the transmitter can implement both harvest-then-transmit protocol and ambient backscatter communication technology only for the ambient RF signal. This scheme evaluates the system performance without leveraging the jamming signal. It is important to note that the optimal policies of both HTT and WTJ are also obtained by the deep dueling algorithm, i.e., Algorithm~\ref{deepduelingqlearning}, presented in Section~\ref{Sec:deepdueling}.

\subsection{Simulation Results}
\label{sec:evaluationB}
\begin{figure}[!]
	\centering
	\includegraphics[scale=0.3]{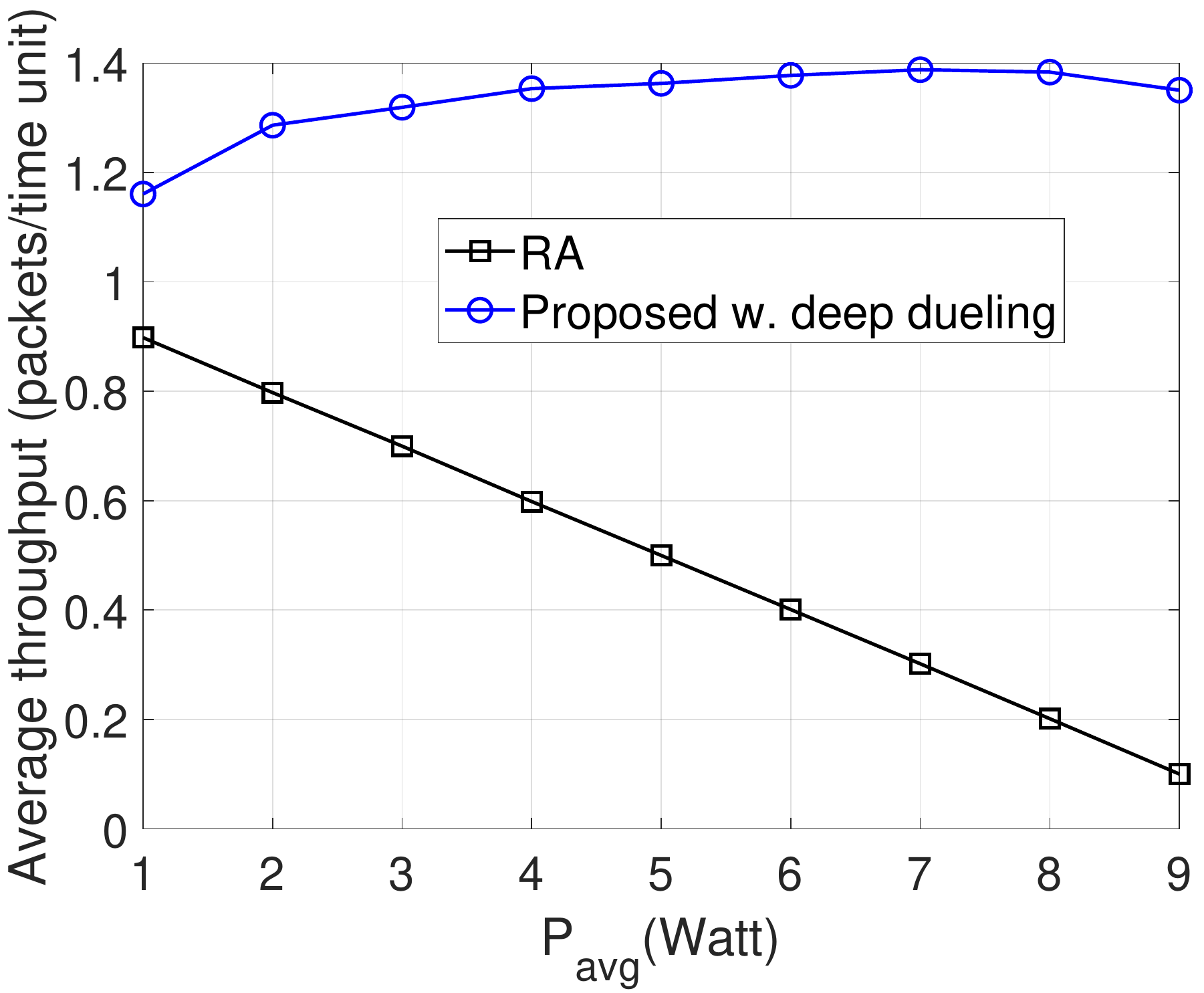}
	\caption{Average throughputs of the proposed solution and the RA technique vs. $P_\mathrm{avg}$.}
	\label{Fig.compare}
\end{figure}

\paragraph{Compare with Non-machine Learning Rate Adaptation Technique}
First, we compare our proposed solution with a non-machine learning technique, i.e., rate adaptation (RA). With RA technique, the transmitter has a fixed defend policy as follows: (i) when the jammer attacks the channel, the transmitter adapts its rate based on the power level of the jammer and (ii) otherwise, the transmitter harvests energy from the ambient RF signal. As shown in Fig.~\ref{Fig.compare}, the average throughput achieved by our proposed solution is much higher than that of the RA technique. The reason is that with RA technique, the transmitter transmits data at a low rate when the jammer attacks the channel with high power levels. Moreover, as stated in~\cite{Firouzbakht2012RA}, with the RA technique, the jammer can force the transmitter to always operate at the lowest rate by merely randomizing its power levels, provided that the average jamming power is above a given threshold. When $P_{avg}$ increases, i.e., the jammer has more opportunities to attack the channel with high power levels, the throughput achieved by the RA technique is significantly decreased. Our proposed solution, in contrast, can allow the transmitter to much more effectively adapt its defense strategy based on the environment condition and the attack strategy of the jammer. This is possible by implicitly learning the strategy of the jammer as well as unknown parameters that are often assumed to be available in the literature (e.g., the power constraint of the jammer, the jamming strategy...). Additionally, with the ambient backscatter capability, the transmitter can always transmit its data to the gateway by leveraging the strong jamming signal.

\paragraph{Performance Evaluation}
Next, we perform simulations to evaluate and compare the performance of proposed solutions with those of the HTT and WTJ schemes in terms of average throughput, packet loss, delay, and PDR. For the HTT and WTJ schemes, we adopt the Deep Dueling algorithm (with $4 \times 10^4$ iterations) to obtain the optimal policy for the transmitter. For the proposed solutions, we recruit both the Deep Dueling (with $4 \times 10^4$ iterations) and Q-learning algorithms (with $10^6$ iterations). 
	
\begin{figure*}[!]
		\centering
		\begin{subfigure}[b]{0.3\textwidth}
			\centering
			\includegraphics[scale=0.3]{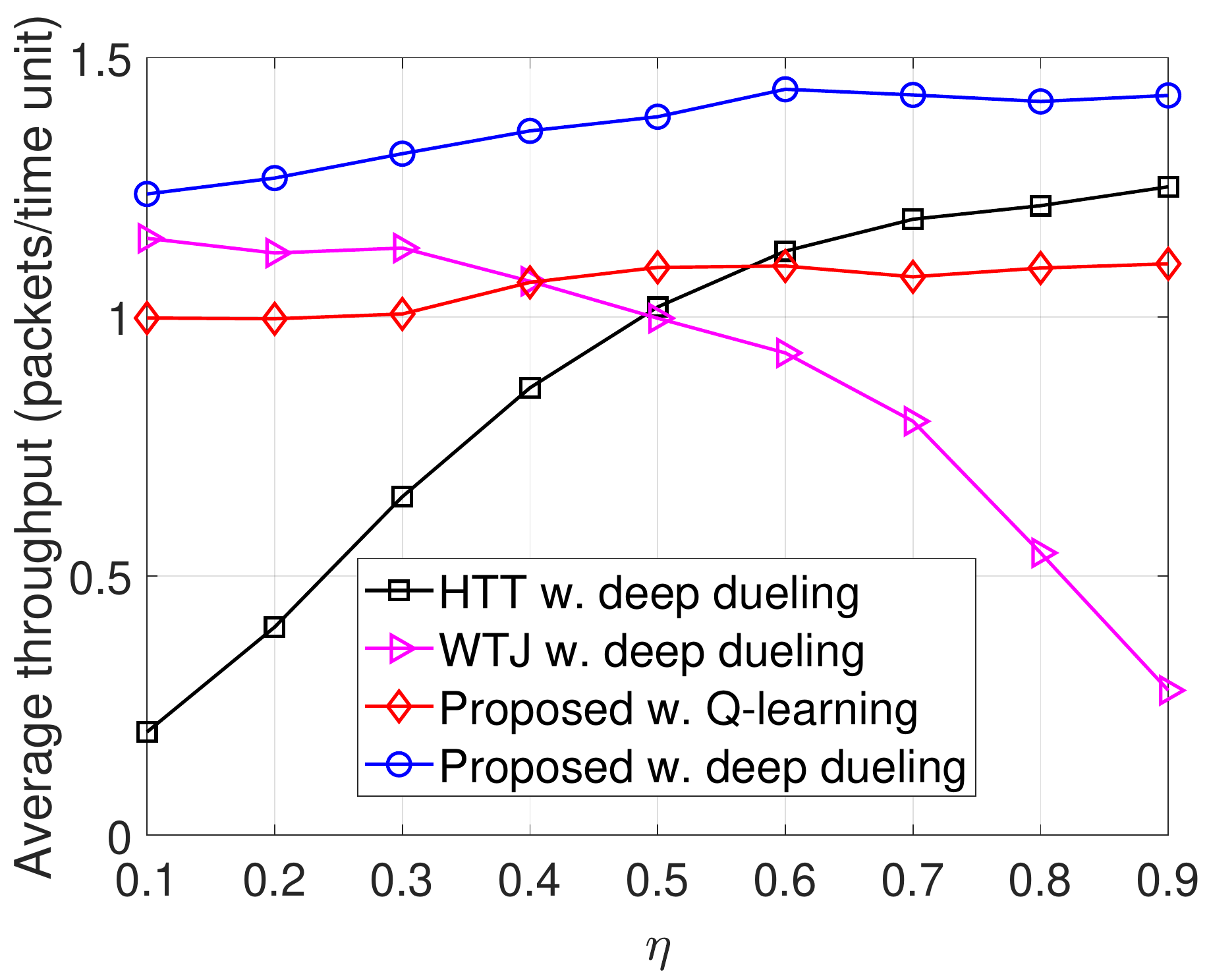}
			\caption{}
		\end{subfigure}%
		~ 
		\begin{subfigure}[b]{0.3\textwidth}
			\centering
			\includegraphics[scale=0.3]{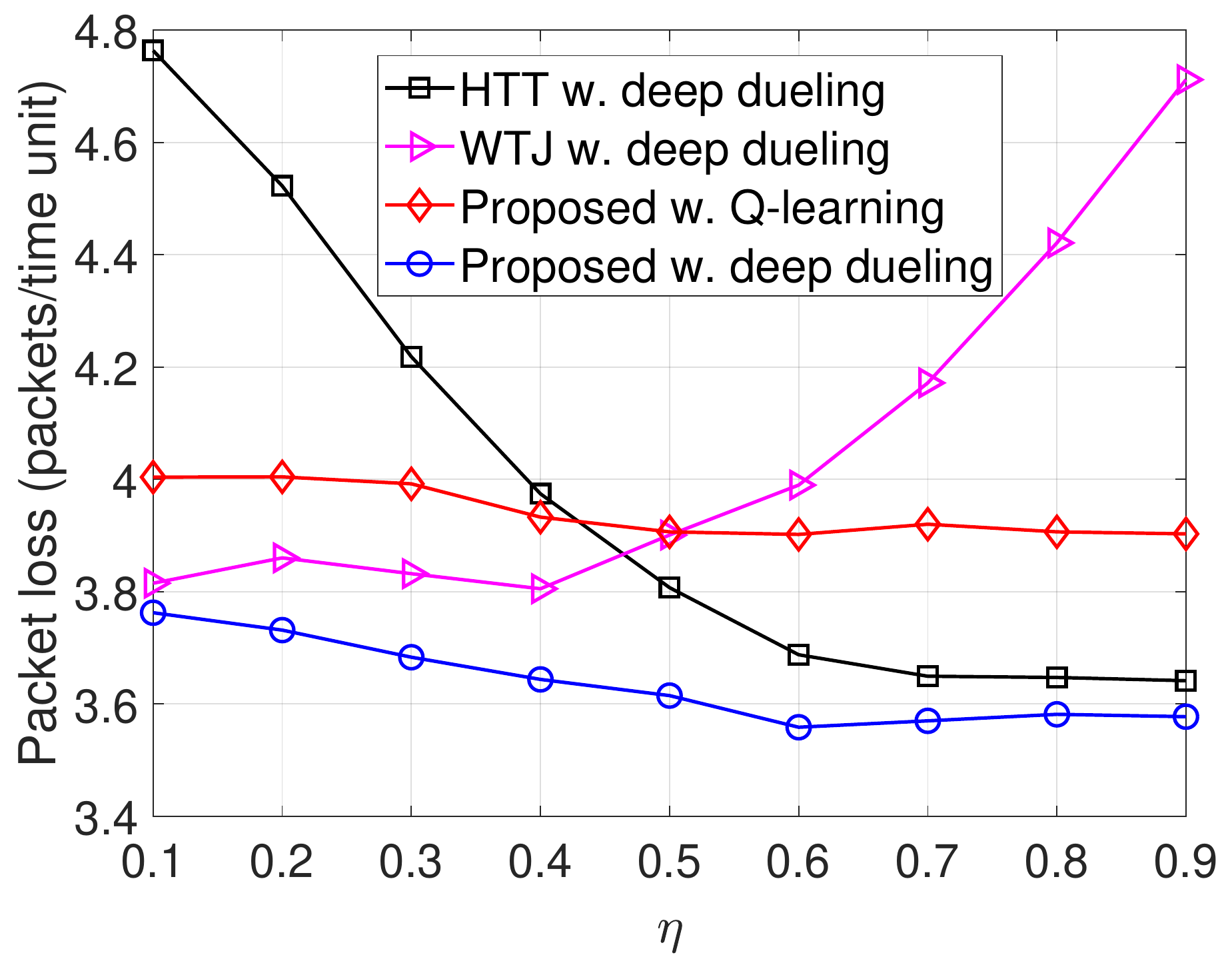}
			\caption{}
		\end{subfigure}%
		~
		\begin{subfigure}[b]{0.3\textwidth}
			\centering
			\includegraphics[scale=0.3]{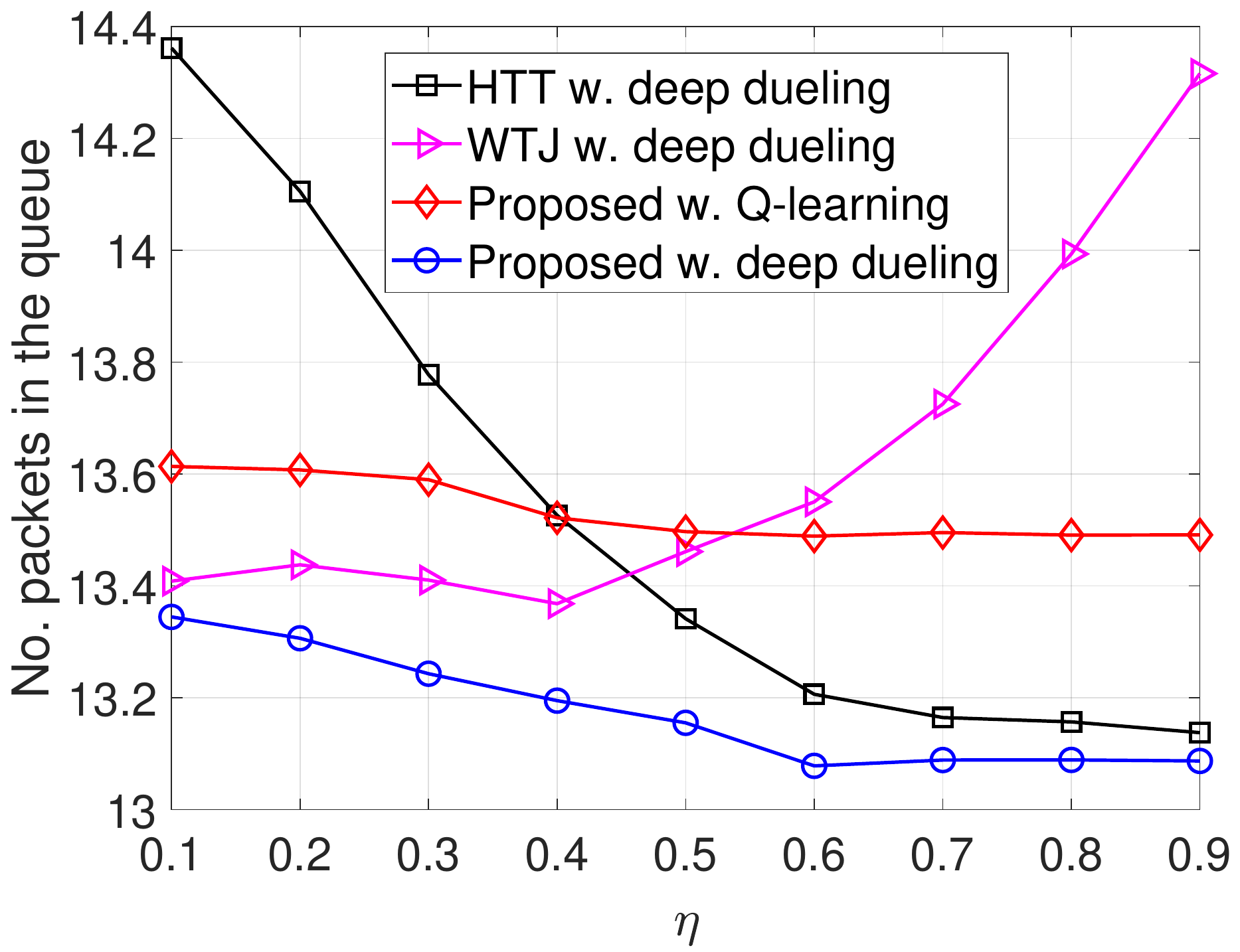}
			\caption{}
		\end{subfigure}%
		
		\begin{subfigure}[b]{0.3\textwidth}
			\centering
			\includegraphics[scale=0.3]{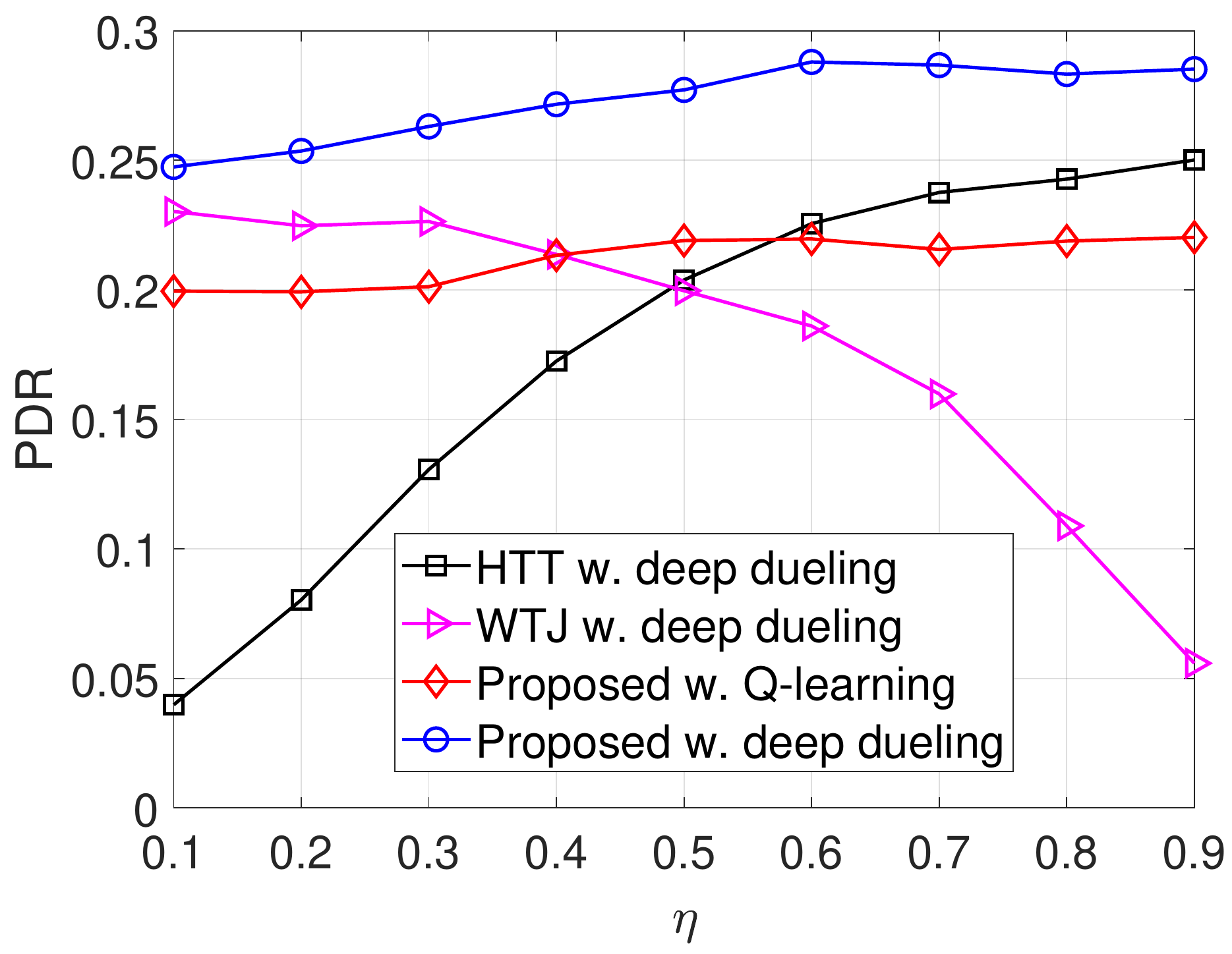}
			\caption{}
		\end{subfigure}%
		~
		\begin{subfigure}[b]{0.3\textwidth}
			\centering
			\includegraphics[scale=0.3]{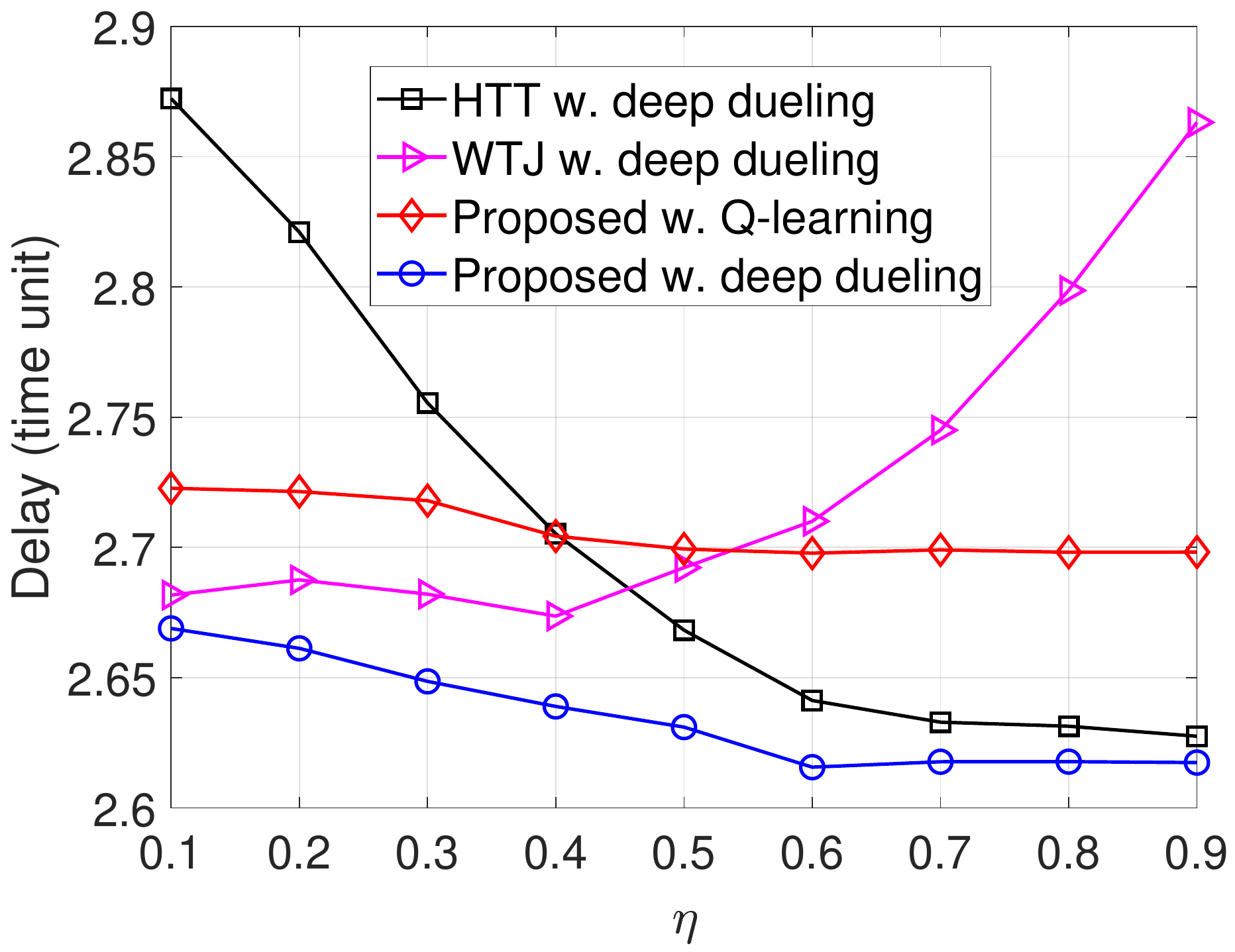}
			\caption{}
		\end{subfigure}%
		\caption{(a) Average throughput (packets/time unit), (b) Packet loss (packets/time unit), (c) Average number of packets in the data queue, (d) PDR, (e) Delay (time/units) vs. $\eta$.} 
		\label{fig:varyEta}
\end{figure*}
In Fig.~\ref{fig:varyEta}, we vary the idle channel probability of the ambient RF source $\eta$ and observe the performance of the system. Clearly, the throughput of the WTJ policy decreases when $\eta$ increases as shown in Fig.~\ref{fig:varyEta}(a). This is stemmed from the fact that when the ambient RF source is likely to be idle, the WTJ has less opportunities to harvest energy and backscatter data from the ambient signal. This also leads to the increase of packet loss and number of packets in the data queue as shown in Fig.~\ref{fig:varyEta}(b) and Fig.~\ref{fig:varyEta}(c), respectively. In contrast, as the idle channel probability increases, the average throughputs obtained by the HTT policy and the proposed solution increase, and their packet loss and number of packets in the data queue will be reduced. The reason is that the transmitter has more opportunities to harvest energy from the jamming signal and use the harvested energy to actively transmit data when the channel is idle. Additionally, the proposed solution can also backscatter data through both the jamming and ambient signals, thereby its throughput is considerably higher than that of the HTT scheme. In Fig.~\ref{fig:varyEta}(d), we observe the PDR of the system. Clearly, the proposed solution achieves the best PDR compared to the other schemes. It is worth noting that, the Q-learning algorithm cannot obtain the optimal policy in the first $10^6$ iterations, thereby the performance derived by the Q-learning algorithm is much lower than that of the Deep Dueling algorithm.
	
\begin{figure*}[!]
		\centering
		\begin{subfigure}[b]{0.3\textwidth}
			\centering
			\includegraphics[scale=0.3]{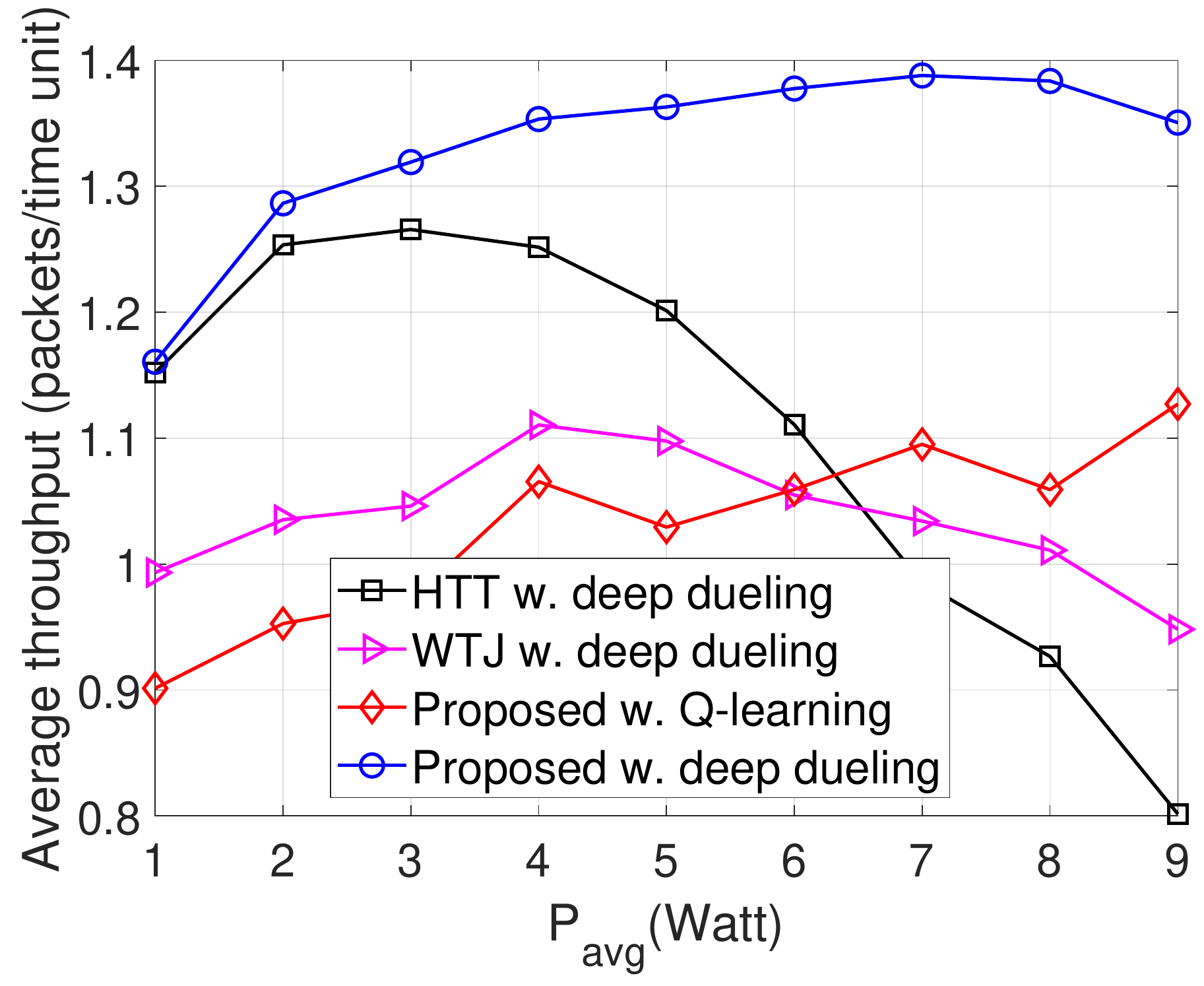}
			\caption{}
		\end{subfigure}%
		~ 
		\begin{subfigure}[b]{0.3\textwidth}
			\centering
			\includegraphics[scale=0.3]{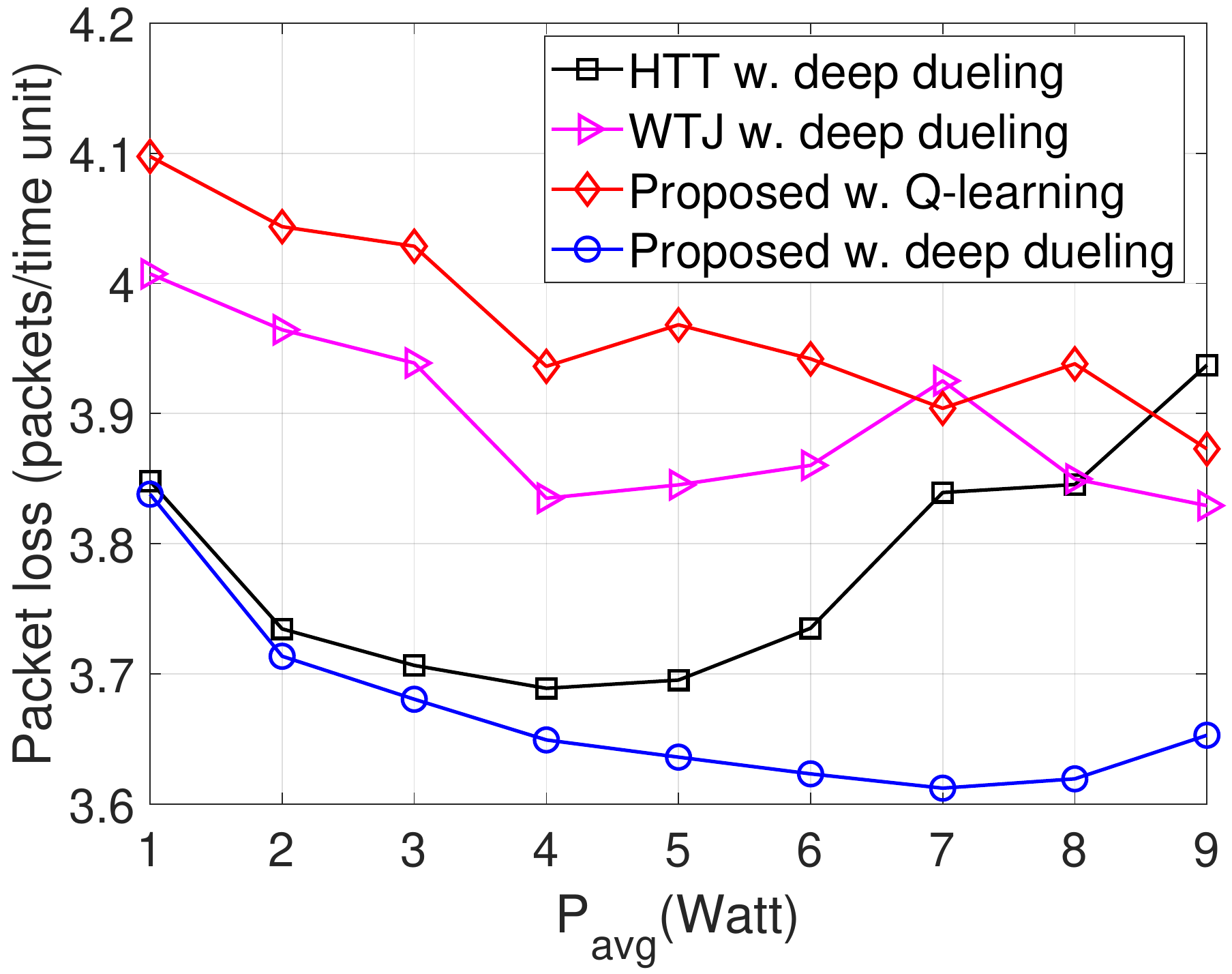}
			\caption{}
		\end{subfigure}%
		~
		\begin{subfigure}[b]{0.3\textwidth}
			\centering
			\includegraphics[scale=0.3]{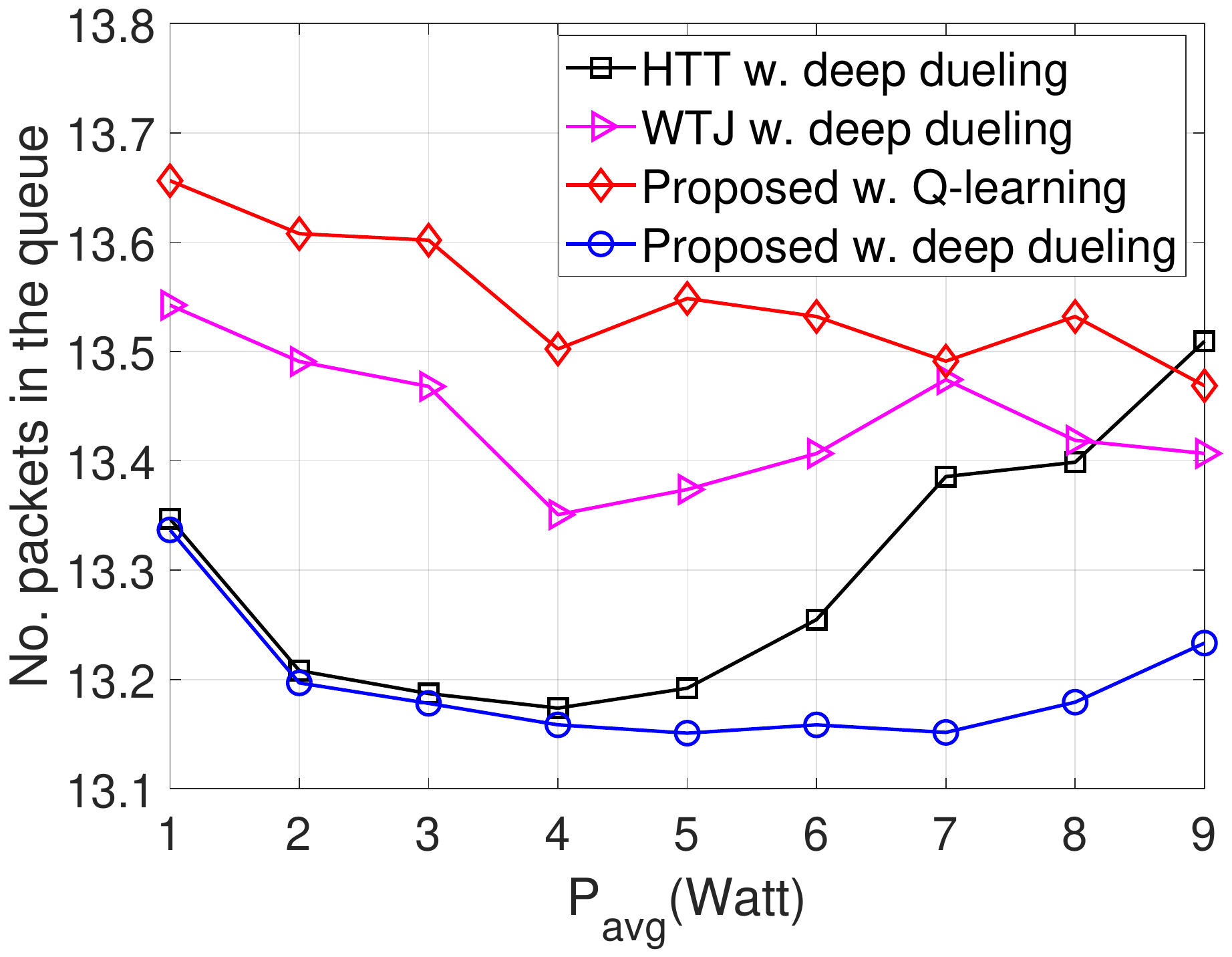}
			\caption{}
		\end{subfigure}%
		
		\begin{subfigure}[b]{0.3\textwidth}
			\centering
			\includegraphics[scale=0.3]{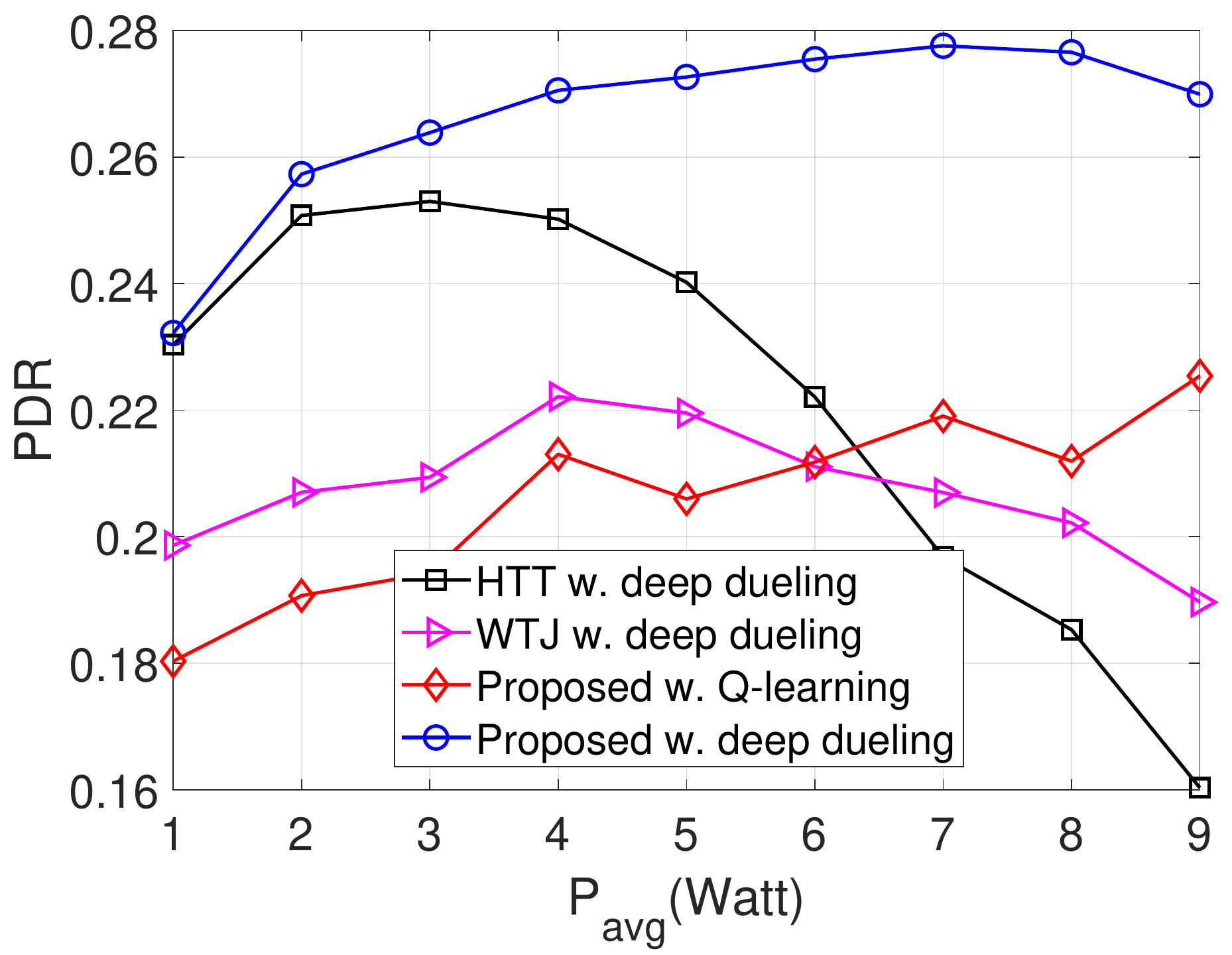}
			\caption{}
		\end{subfigure}%
		~
		\begin{subfigure}[b]{0.3\textwidth}
			\centering
			\includegraphics[scale=0.3]{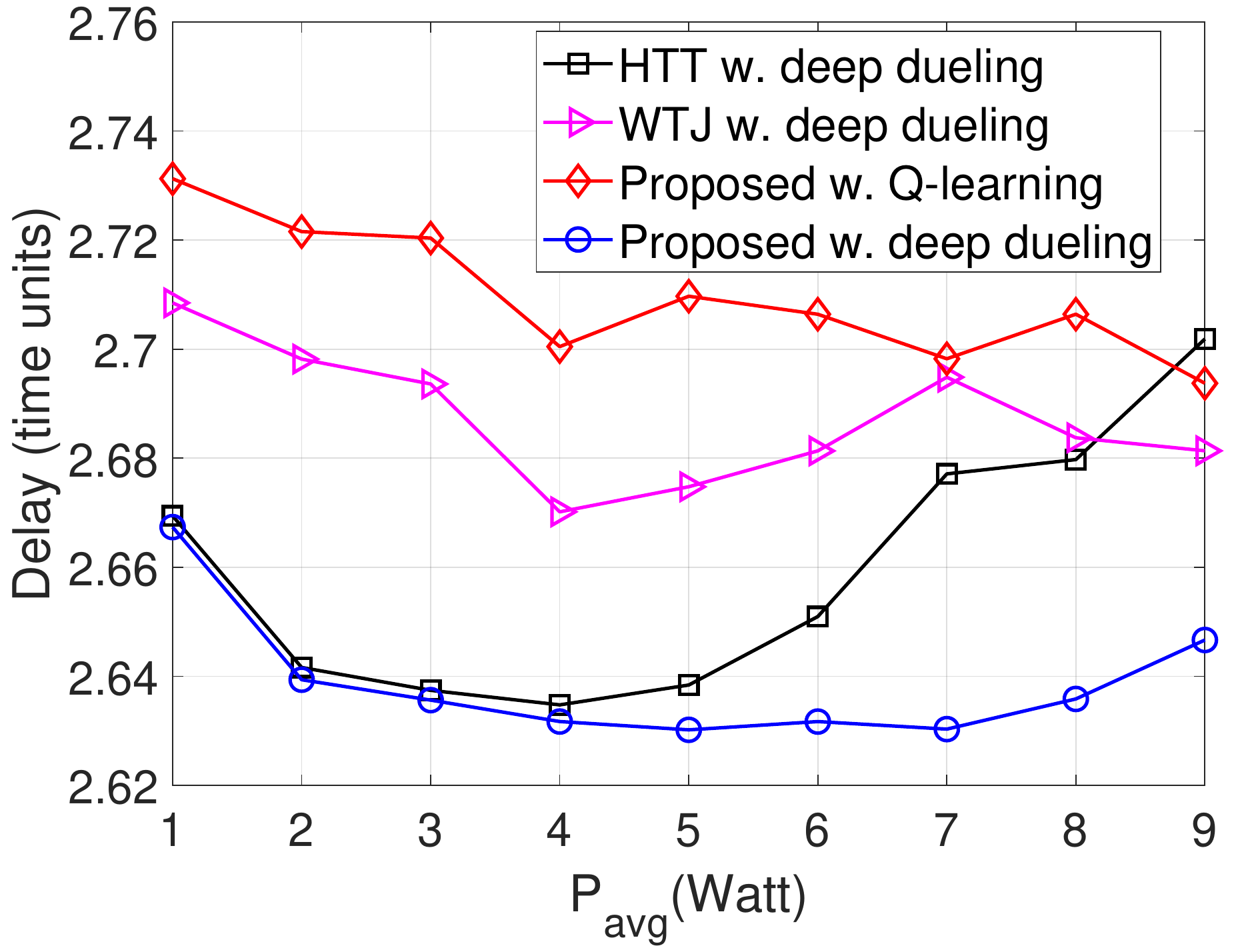}
			\caption{}
		\end{subfigure}%
		\caption{(a) Average throughput (packets/time unit), (b) Packet loss (packets/time unit), (c) Average number of packets in the data queue, (d) PDR, (e) Delay (time/units) vs. $P_\mathrm{avg}$.} 
		\label{fig:varyPavg}
\end{figure*}

\begin{figure*}[h]
	\centering
	\begin{subfigure}[b]{0.3\textwidth}
		\centering
		\includegraphics[scale=0.3]{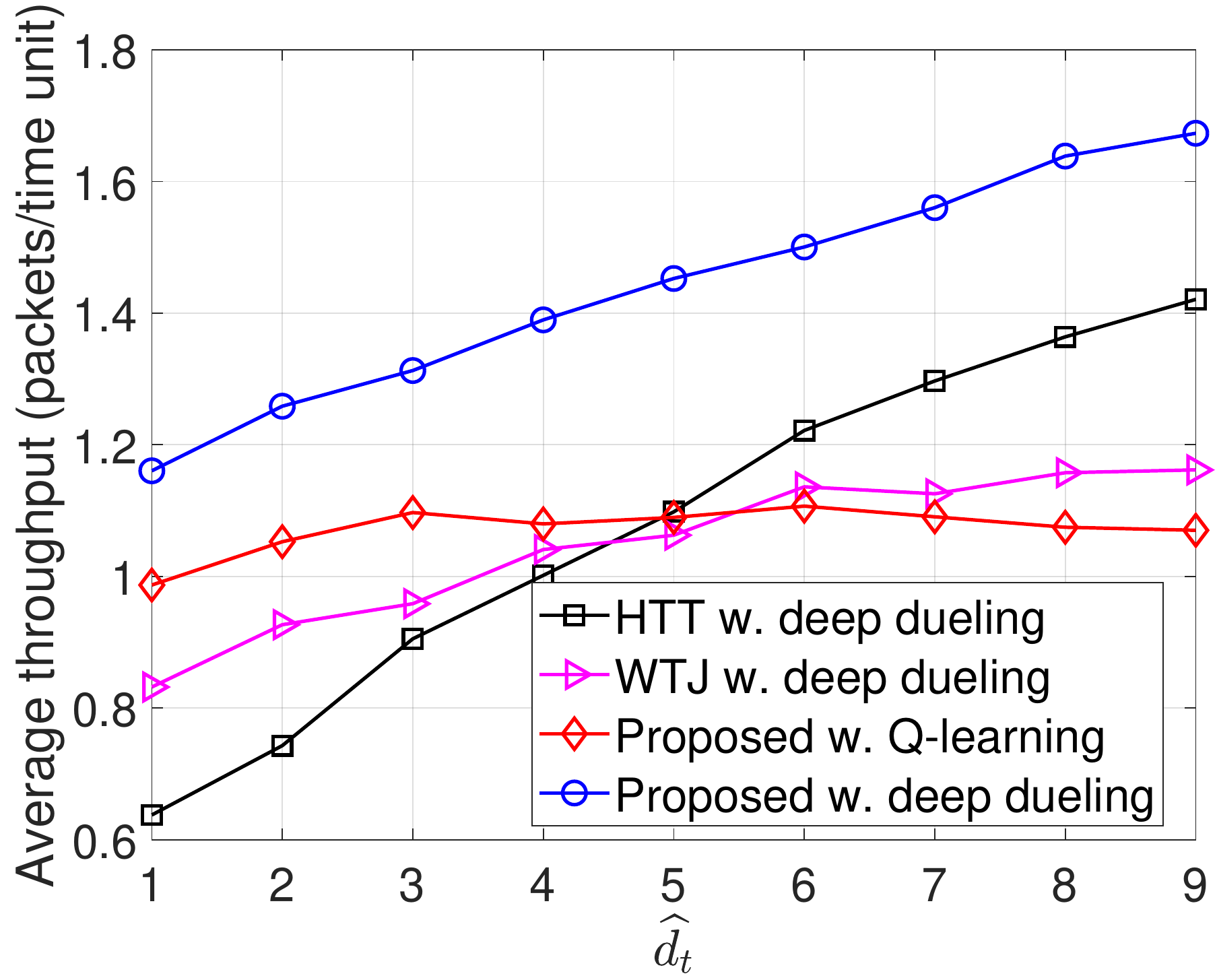}
		\caption{}
	\end{subfigure}%
	~ 
	\begin{subfigure}[b]{0.3\textwidth}
		\centering
		\includegraphics[scale=0.3]{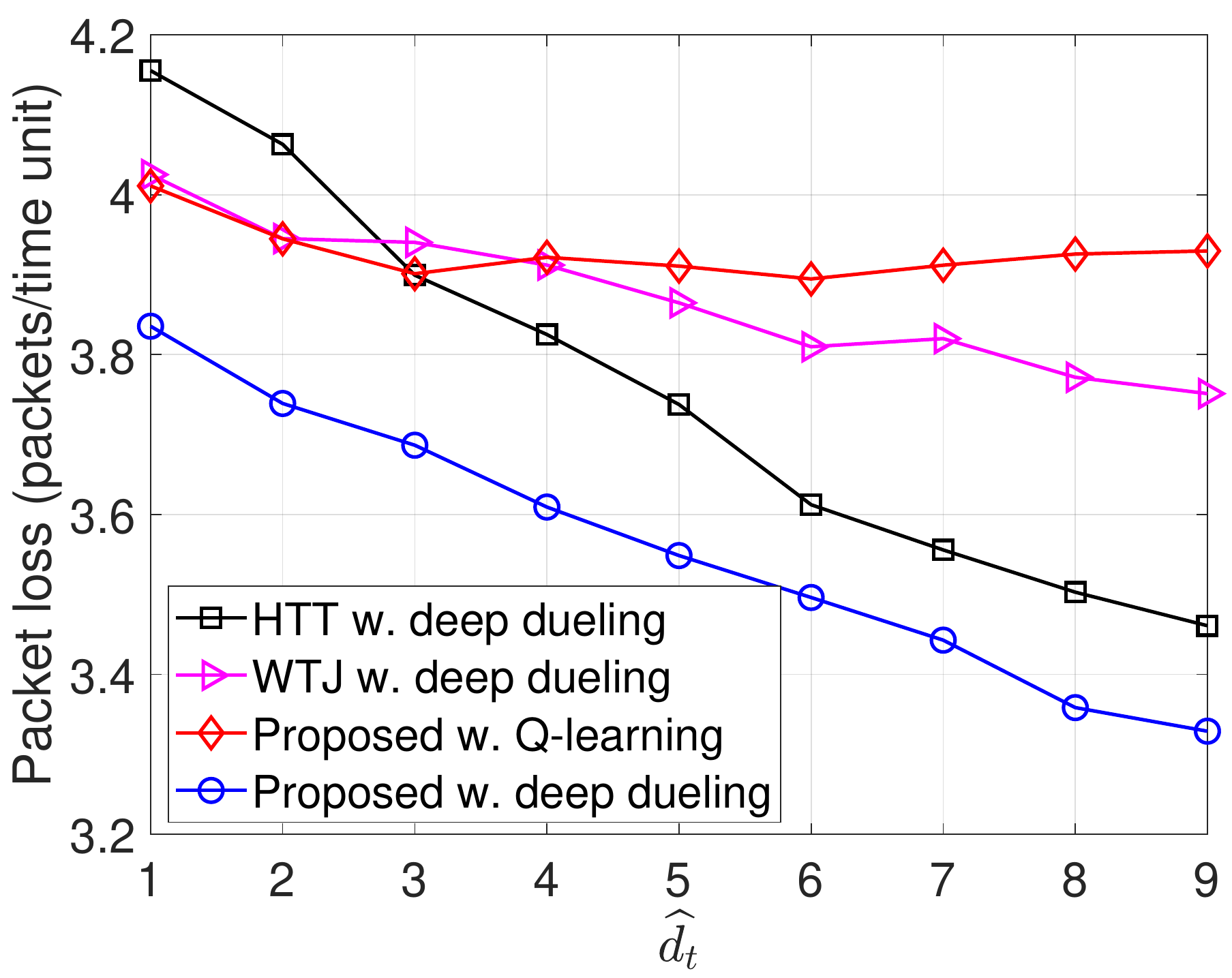}
		\caption{}
	\end{subfigure}%
	~
	\begin{subfigure}[b]{0.3\textwidth}
		\centering
		\includegraphics[scale=0.3]{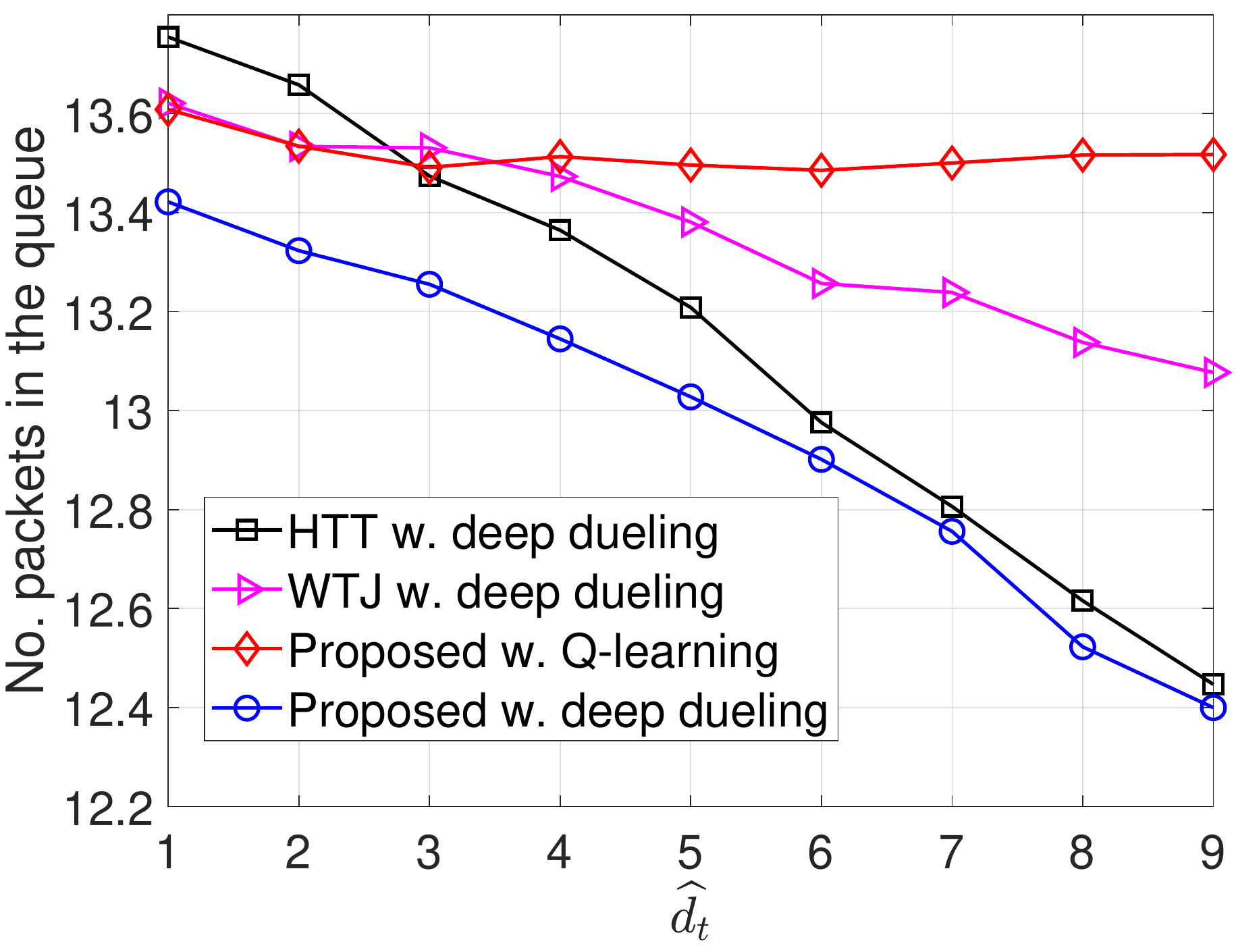}
		\caption{}
	\end{subfigure}%
	
	\begin{subfigure}[b]{0.3\textwidth}
		\centering
		\includegraphics[scale=0.3]{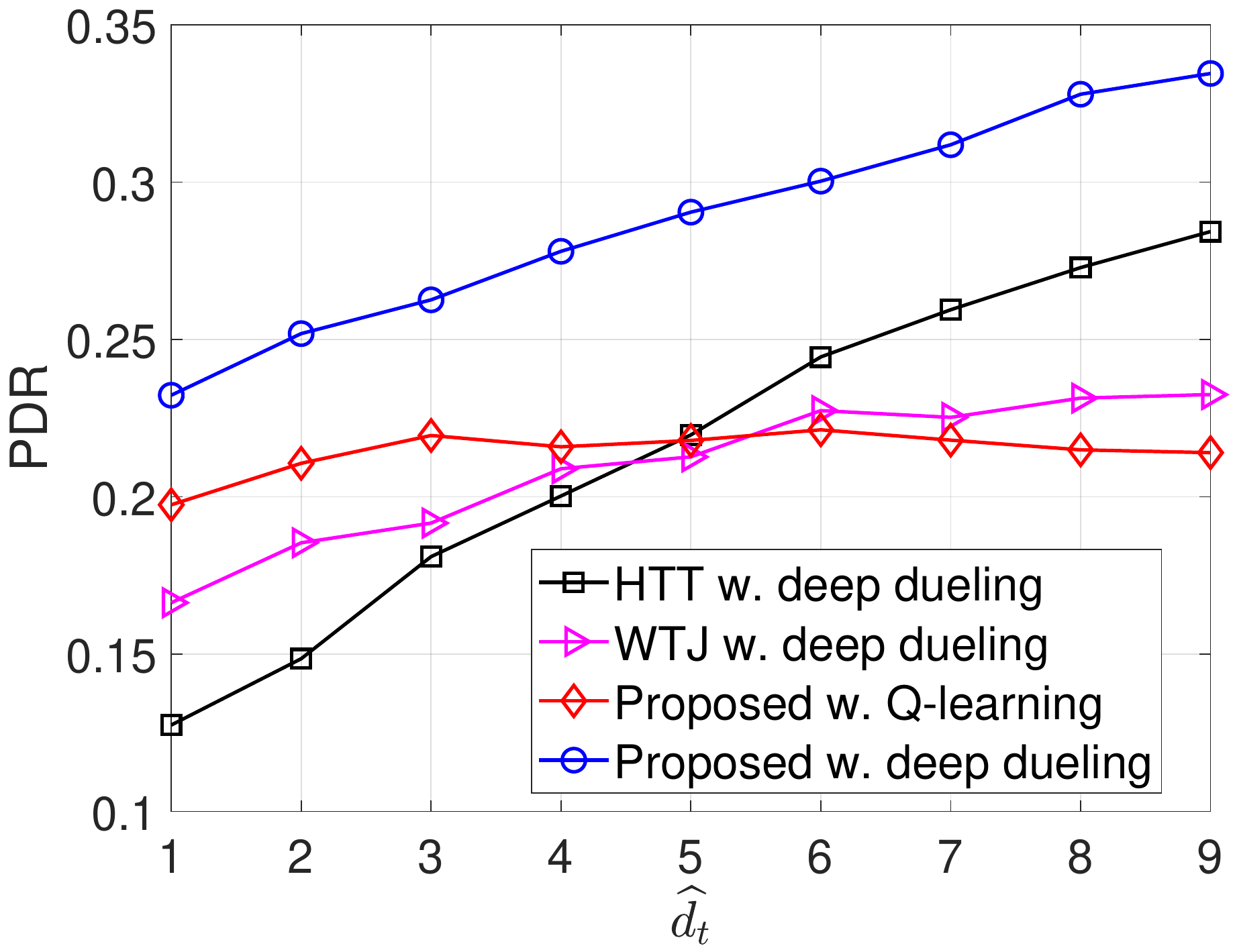}
		\caption{}
	\end{subfigure}%
	~
	\begin{subfigure}[b]{0.3\textwidth}
		\centering
		\includegraphics[scale=0.3]{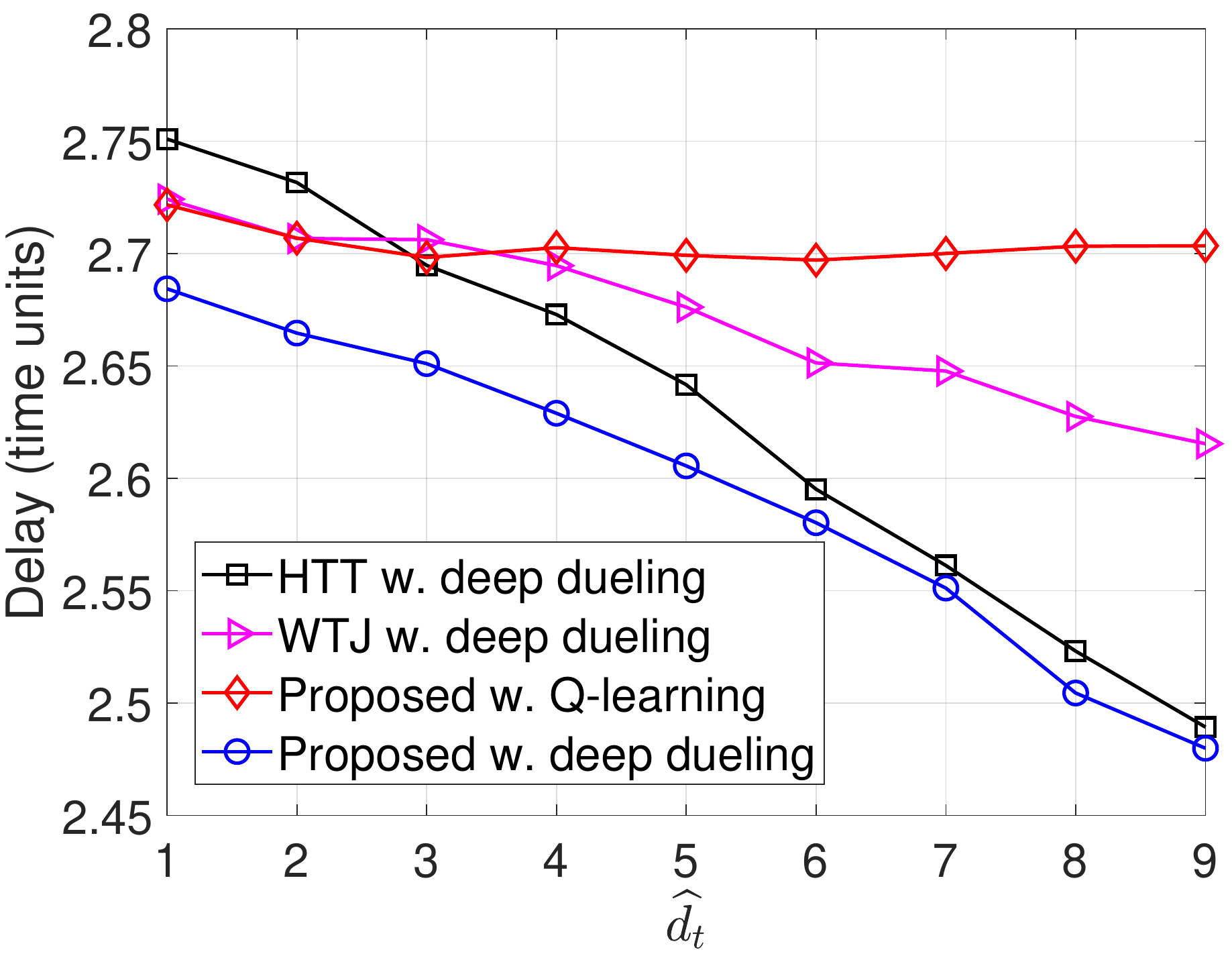}
		\caption{}
	\end{subfigure}%
	
	\caption{(a) Average throughput (packets/time unit), (b) Packet loss (packets/time unit), (c) Average number of packets in the data queue, (d) PDR, (e) Delay (time/units) vs. $\widehat{d}_\mathrm{t}$.} 
	\label{fig:varyDt}
\end{figure*}
	
In Fig.~\ref{fig:varyPavg}, we vary $P_\mathrm{avg}$ to evaluate the average throughput, packet loss, number of packets in the data queue, and PDR of the system. Obviously, when $P_\mathrm{avg}$ increases from 1W to 3W, the throughput of the HTT and WTJ policies increases. The reason is that the transmitter has more chances to harvest energy from the strong jamming signal and uses the harvested energy to transmit data when the jammer and the ambient RF source are idle. However, when $P_\mathrm{avg}$ is large (e.g., higher than $3$W), i.e., the jammer is more likely to attack the channel, the throughput of these policies decreases as the transmitter has less chance to actively transmit data to the gateway. In contrast, the throughput achieved by the proposed solution increases. The reason is that the proposed solution allows the transmitter to switch to the backscatter mode when the jammer is likely to attack the channel. Consequently, the proposed solutions achieve the best performance in terms of packet loss, number of packets in the queue, PDR, and delay as shown in Fig.~\ref{fig:varyPavg}(b), Fig.~\ref{fig:varyPavg}(c), Fig.~\ref{fig:varyPavg}(d), and Fig.~\ref{fig:varyPavg}(e), respectively. Again, the performance of the Q-learning algorithm is not as good as the Deep Dueling algorithm due to the slow-convergence problem.
\begin{figure*}[h]
	\centering
	\begin{subfigure}[b]{0.3\textwidth}
		\centering
		\includegraphics[scale=0.3]{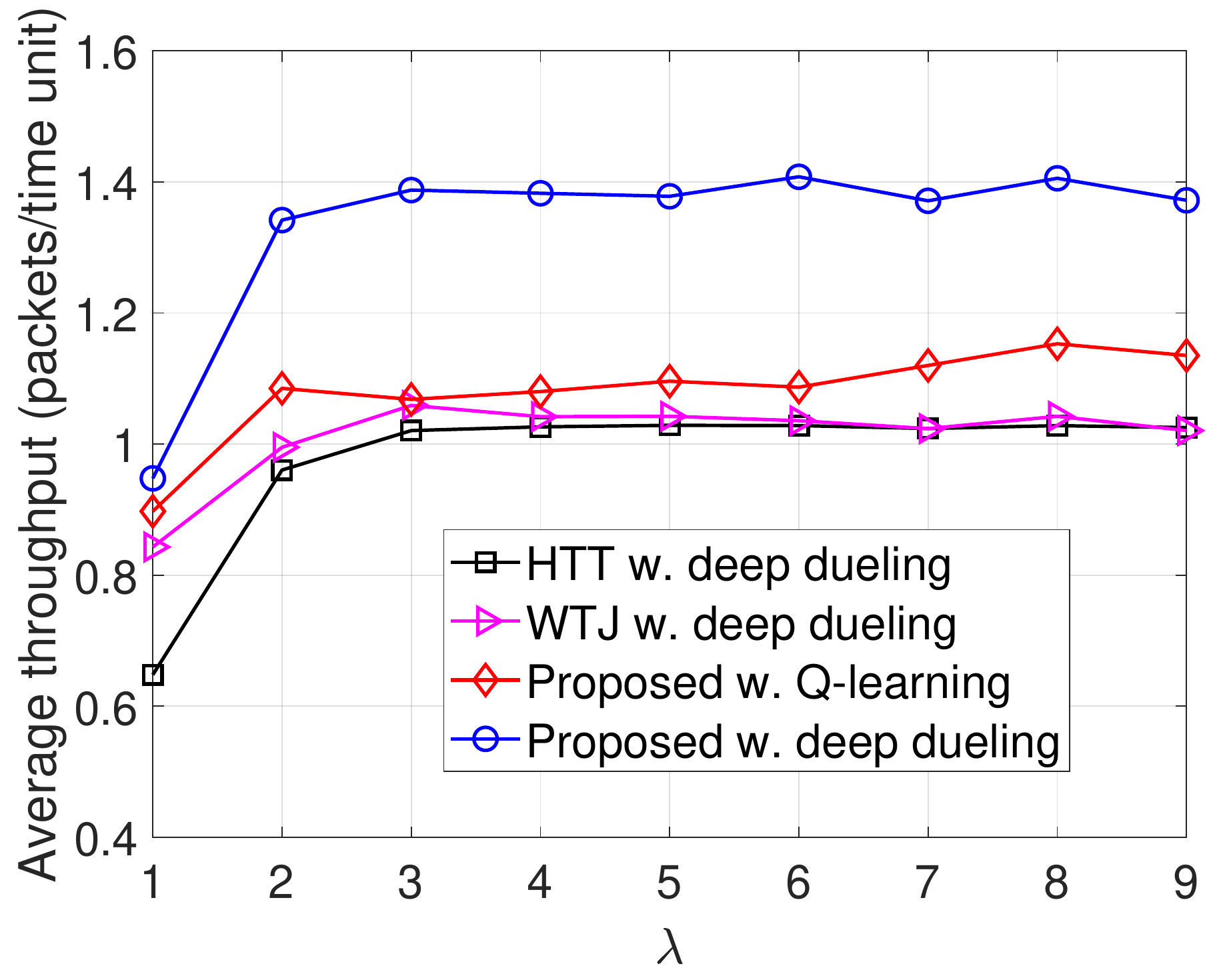}
		\caption{}
	\end{subfigure}%
	~ 
	\begin{subfigure}[b]{0.3\textwidth}
		\centering
		\includegraphics[scale=0.3]{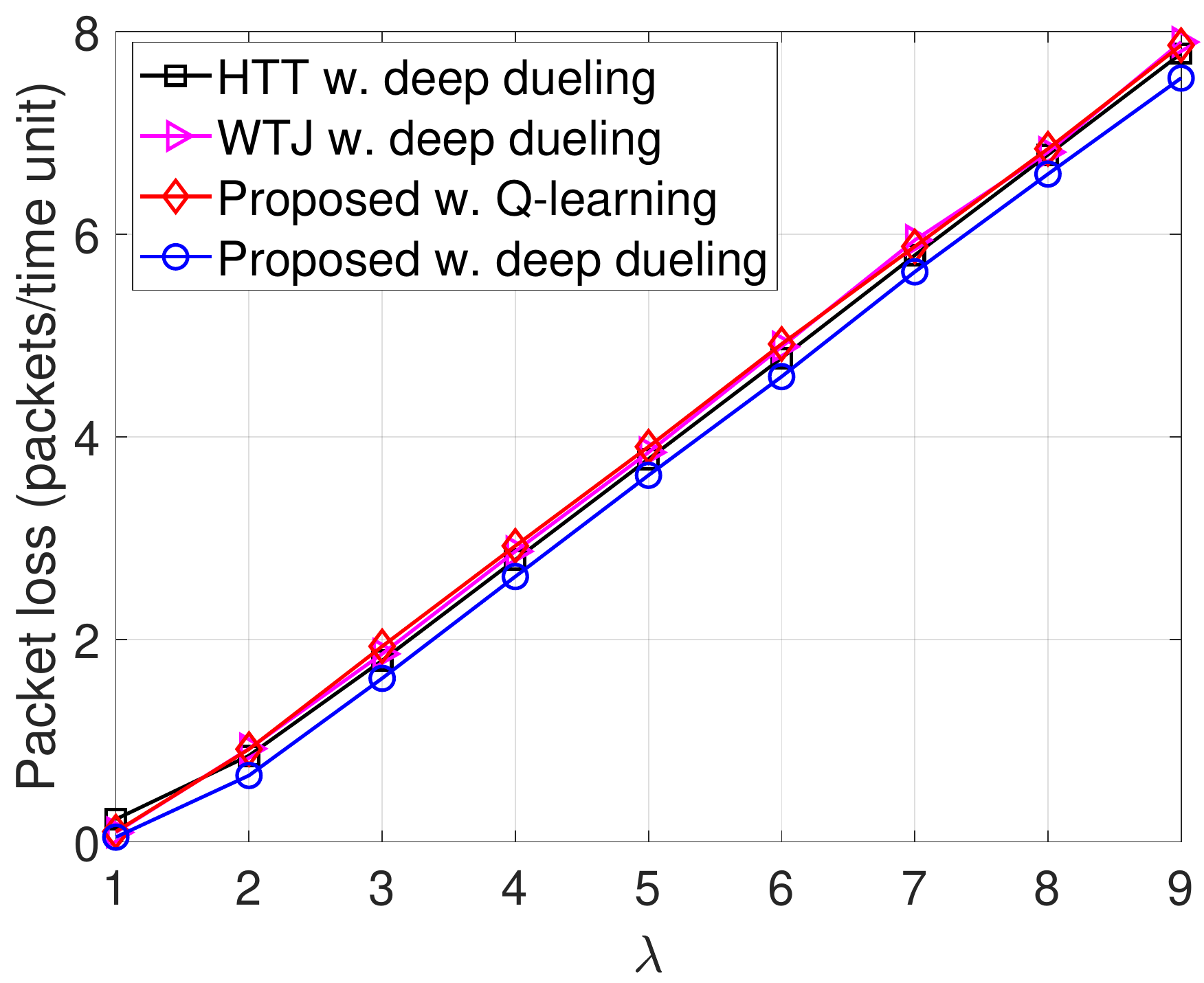}
		\caption{}
	\end{subfigure}%
	~
	\begin{subfigure}[b]{0.3\textwidth}
		\centering
		\includegraphics[scale=0.3]{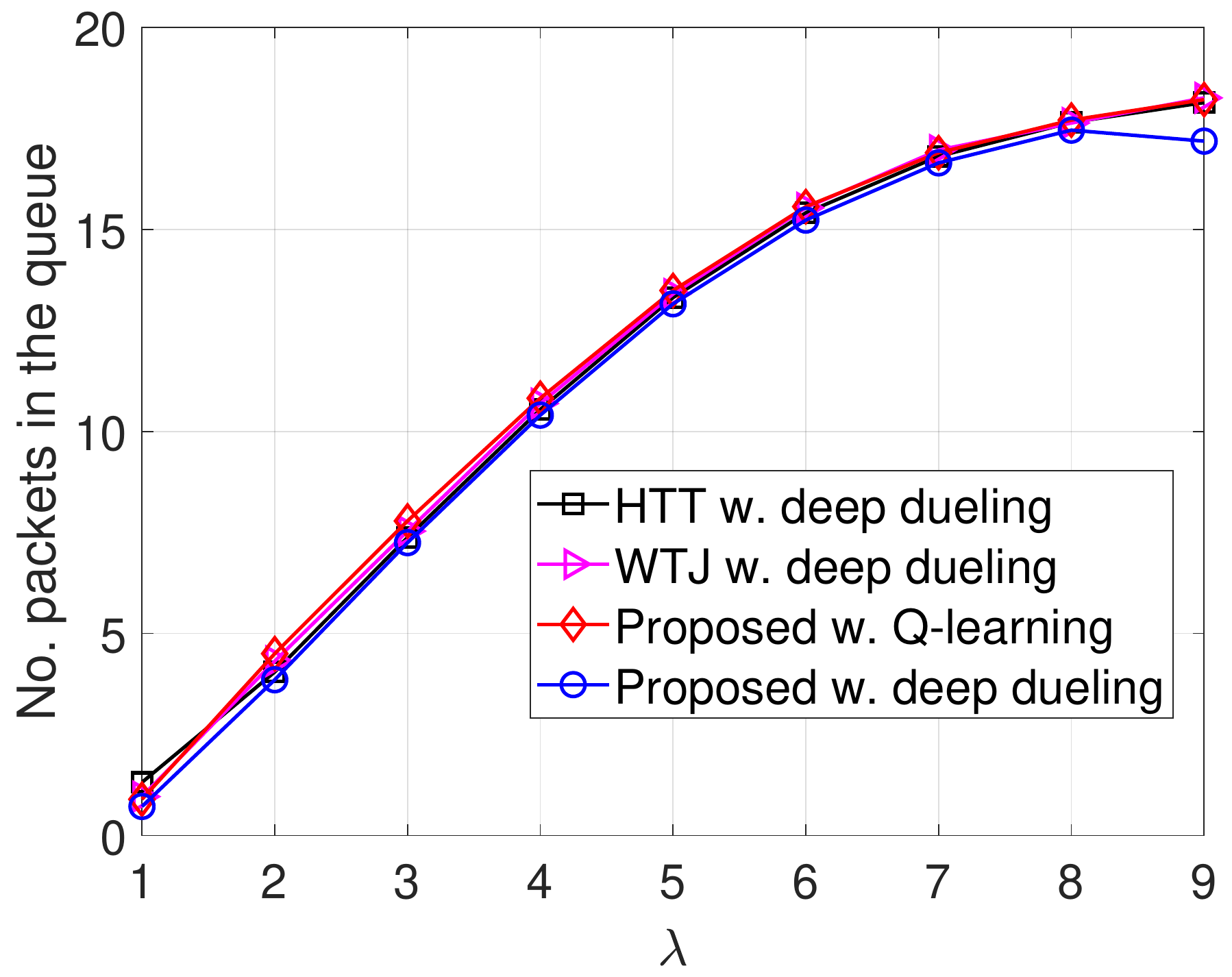}
		\caption{}
	\end{subfigure}%
	
	\begin{subfigure}[b]{0.3\textwidth}
		\centering
		\includegraphics[scale=0.3]{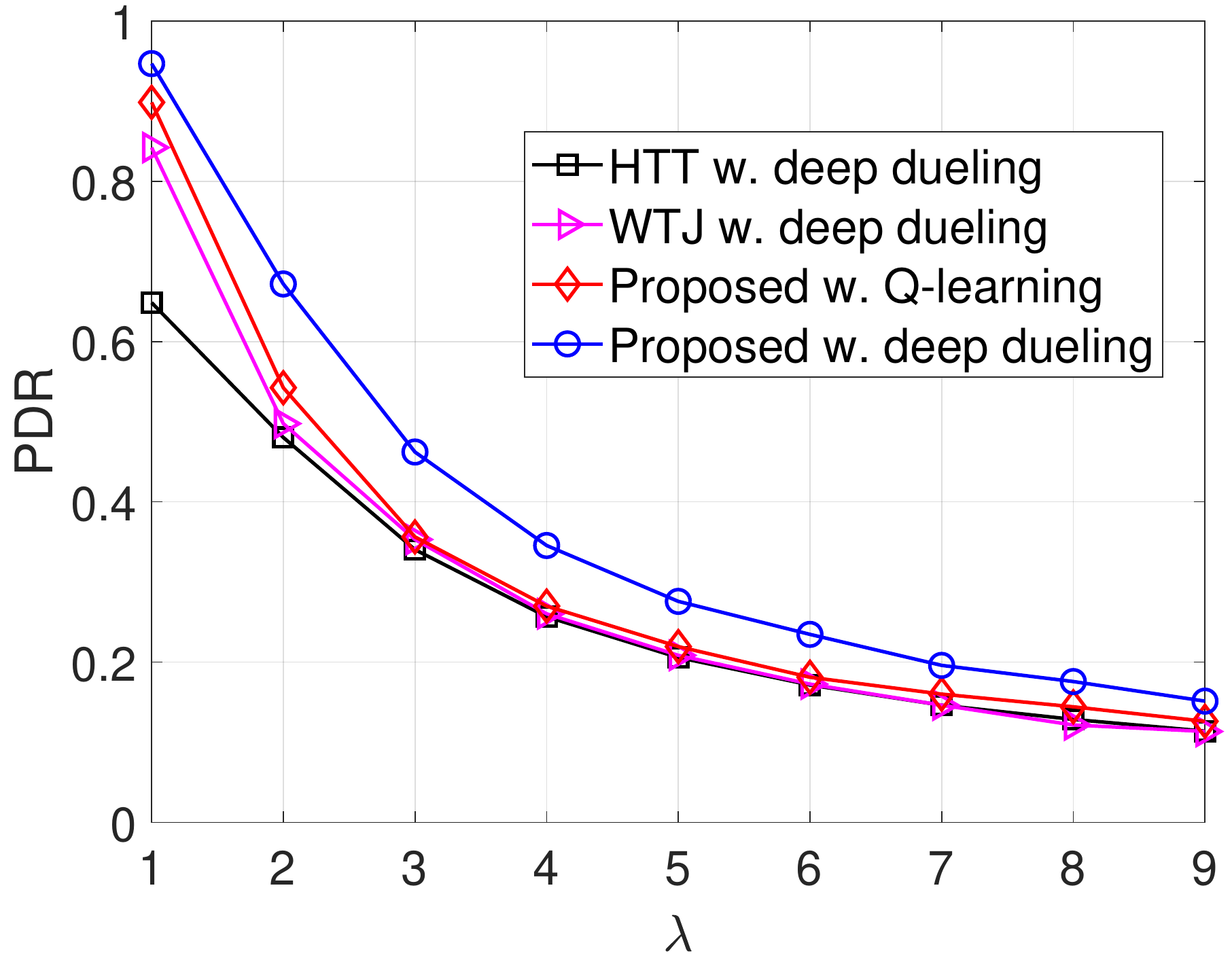}
		\caption{}
	\end{subfigure}%
	~
	\begin{subfigure}[b]{0.3\textwidth}
		\centering
		\includegraphics[scale=0.3]{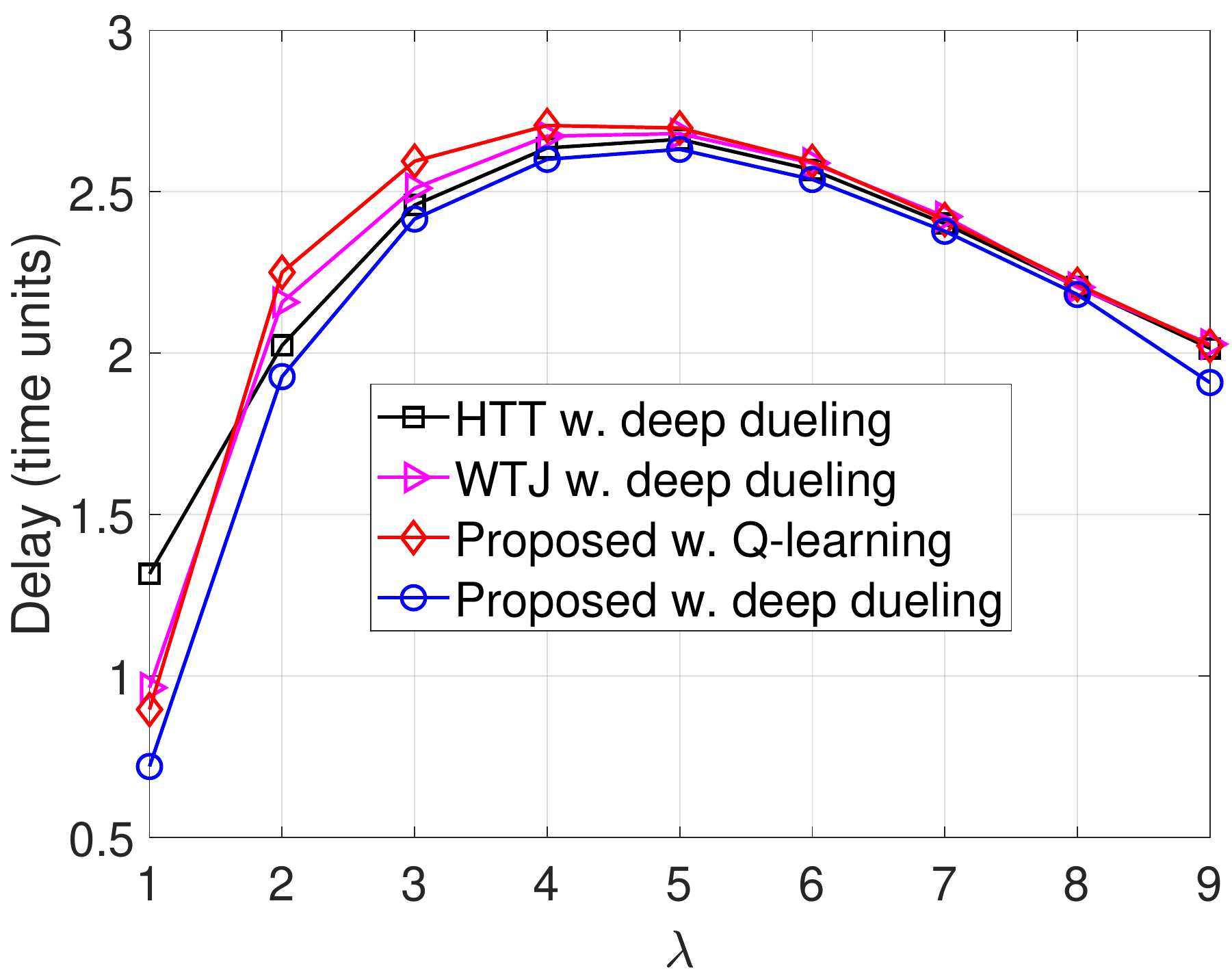}
		\caption{}
	\end{subfigure}%
	\caption{(a) Average throughput (packets/time unit), (b) Packet loss (packets/time unit), (c) Average number of packets in the data queue, (d) PDR, (e) Delay (time/units) vs. $\lambda$.} 
	\label{fig:varyLambda}
\end{figure*}
	
Next, we vary the maximum number of packets $\widehat{d}_\mathrm{t}$ that the transmitter can actively transmit to the gateway and evaluate the performance of the proposed solution as shown in Fig.~\ref{fig:varyDt}. As shown in Fig.~\ref{fig:varyDt}(a), as $\widehat{d}_\mathrm{t}$ increases, the throughput of WTJ scheme also increases and remains the same when $\widehat{d}_\mathrm{t} \geq 6$. This is due to the fact that the transmitter does not leverage the strong jamming signal (except when both the source are active), and thus the amount of harvested energy is limited. In contrary, the HTT policy allows the transmitter to harvest energy from both the ambient and jamming signal. As a result, its throughput increases and is higher than that of the WTJ policy. Importantly, by balancing the time for backscattering data and harvesting energy, the throughput achieved by the proposed solution is significantly higher than that of the HTT and WTJ schemes. This also leads to the reductions of the packet loss and number of packets waiting in the data queue as shown in Fig.~\ref{fig:varyDt}(b) and Fig.~\ref{fig:varyDt}(c), respectively. In Fig.~\ref{fig:varyDt}, we observe the PDR of obtained by the three schemes. Clearly, the proposed solution continues to achieve the best PDR compared to other schemes.
	
In Fig.~\ref{fig:varyLambda}, we vary the packet arrival rate $\lambda$ to evaluate the performance of the proposed solution. Clearly, when $\lambda$ increases to 2 packets/time slot, the throughputs of all three schemes are increased as the transmitter can transmit more packets. However, when $\lambda > 2$ packets/time slot, the throughputs remain the same as the transmitter obtains the optimal policy. Note that with energy harvesting and backscattering capabilities, the proposed solution can achieve the highest throughput among three schemes. As the transmitter cannot transmit all the arrival packets, the packet loss and the number of packets waiting in the data queue increase when $\lambda$ increases as shown in Fig.~\ref{fig:varyLambda}(b) and Fig.~\ref{fig:varyLambda}(c), respectively. With the total number of arrival packets increases, the PDRs of all three schemes are reduced as shown in Fig.~\ref{fig:varyLambda}(d). It is worth noting that in all cases the performance of the Q-learning algorithm is not as high as the Deep Dueling algorithm as it can not converge to the optimal policy within $10^6$ iterations. 

\begin{figure*}[!]
		\centering
		\begin{subfigure}[b]{0.3\textwidth}
			\centering
			\includegraphics[scale=0.3]{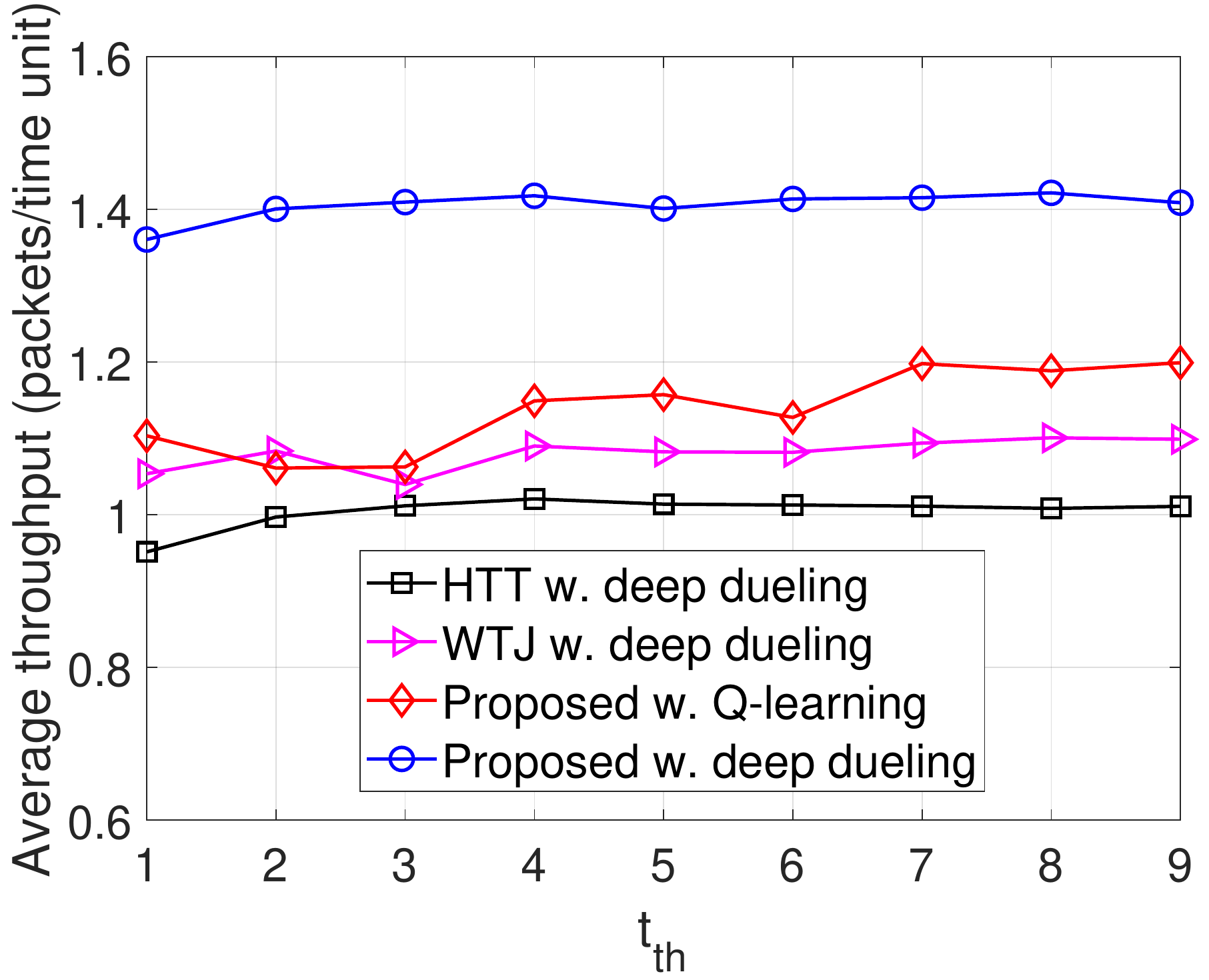}
			\caption{}
		\end{subfigure}%
		~ 
		\begin{subfigure}[b]{0.3\textwidth}
			\centering
			\includegraphics[scale=0.3]{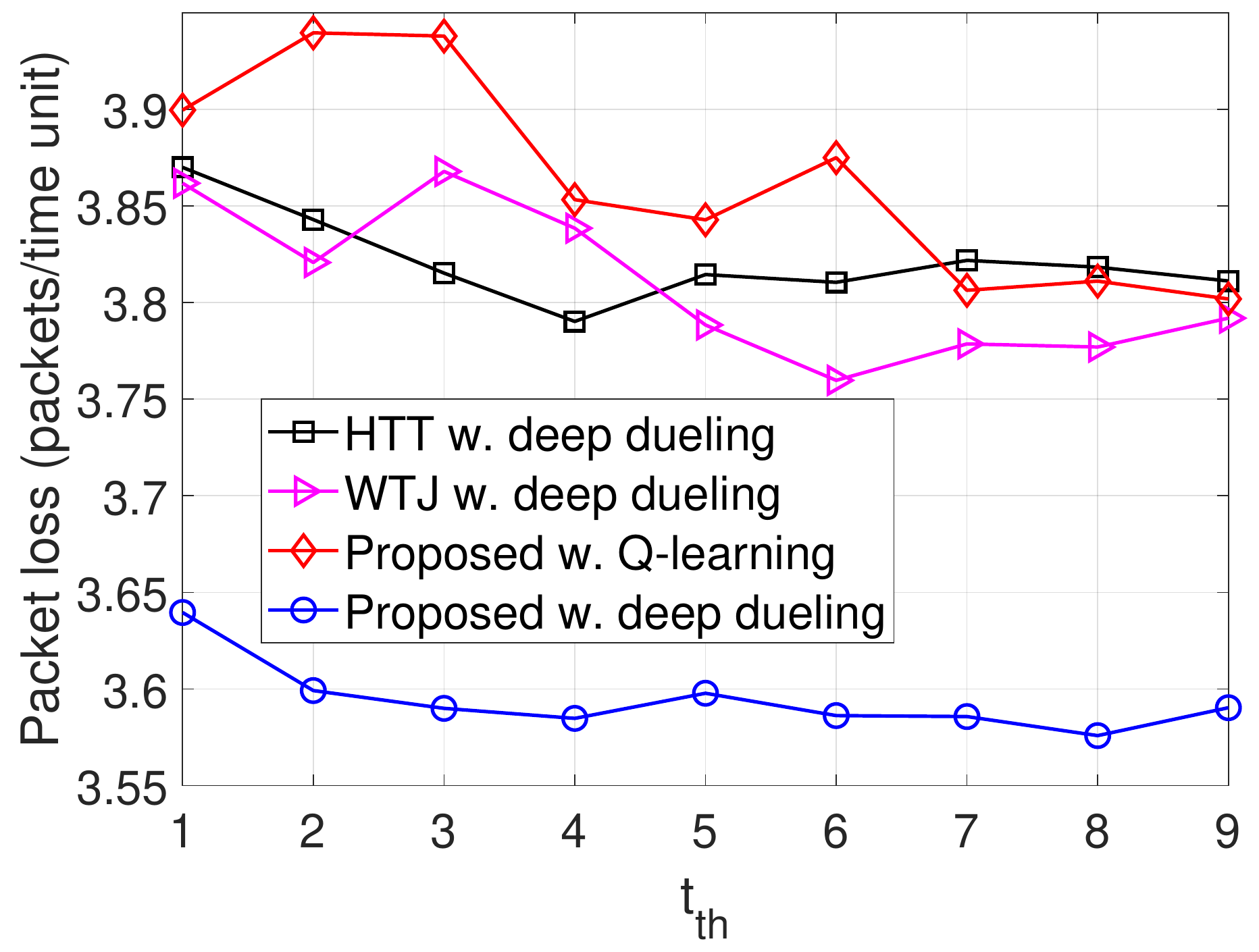}
			\caption{}
		\end{subfigure}%
		~
		\begin{subfigure}[b]{0.3\textwidth}
			\centering
			\includegraphics[scale=0.3]{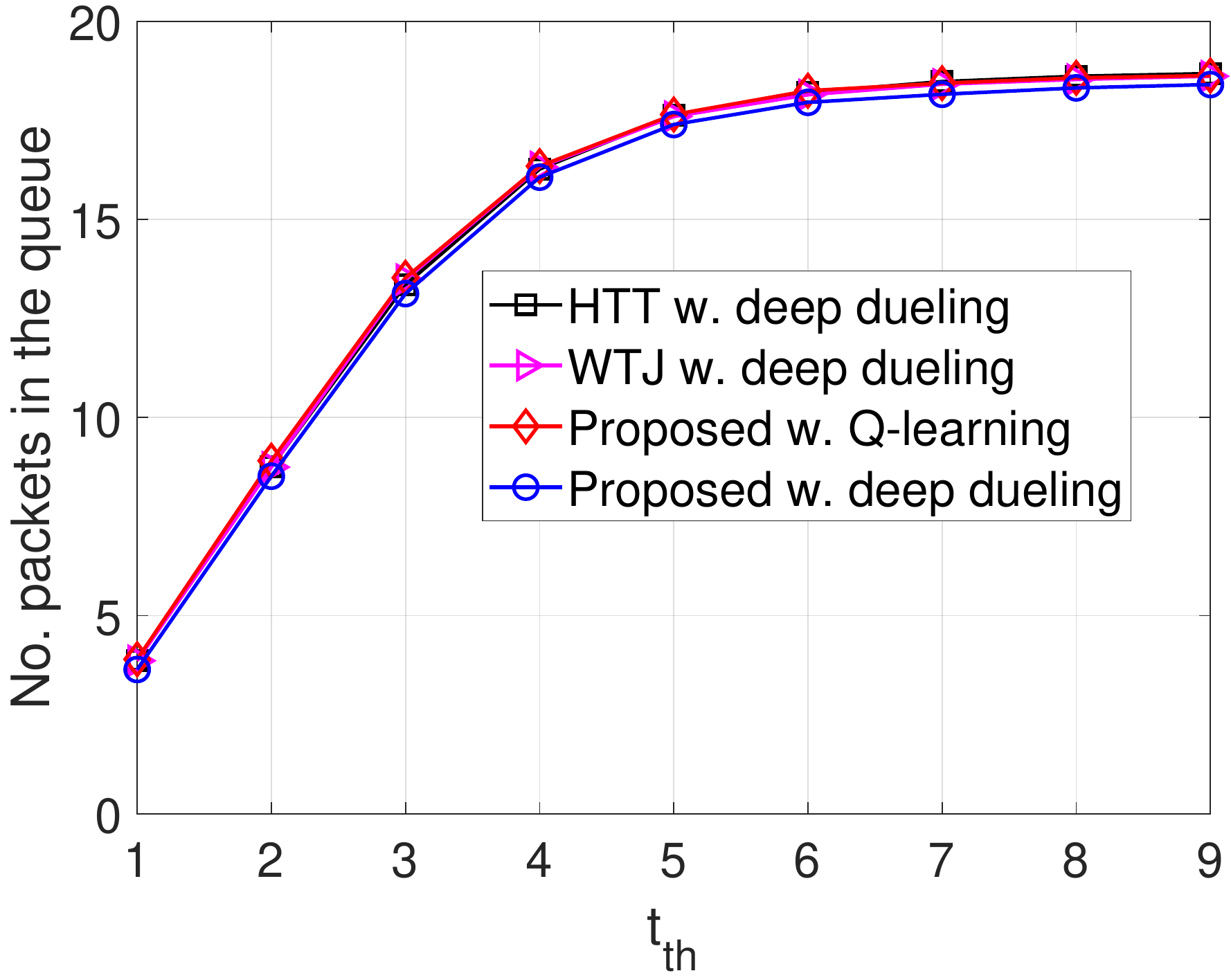}
			\caption{}
		\end{subfigure}%
		
		\begin{subfigure}[b]{0.3\textwidth}
			\centering
			\includegraphics[scale=0.3]{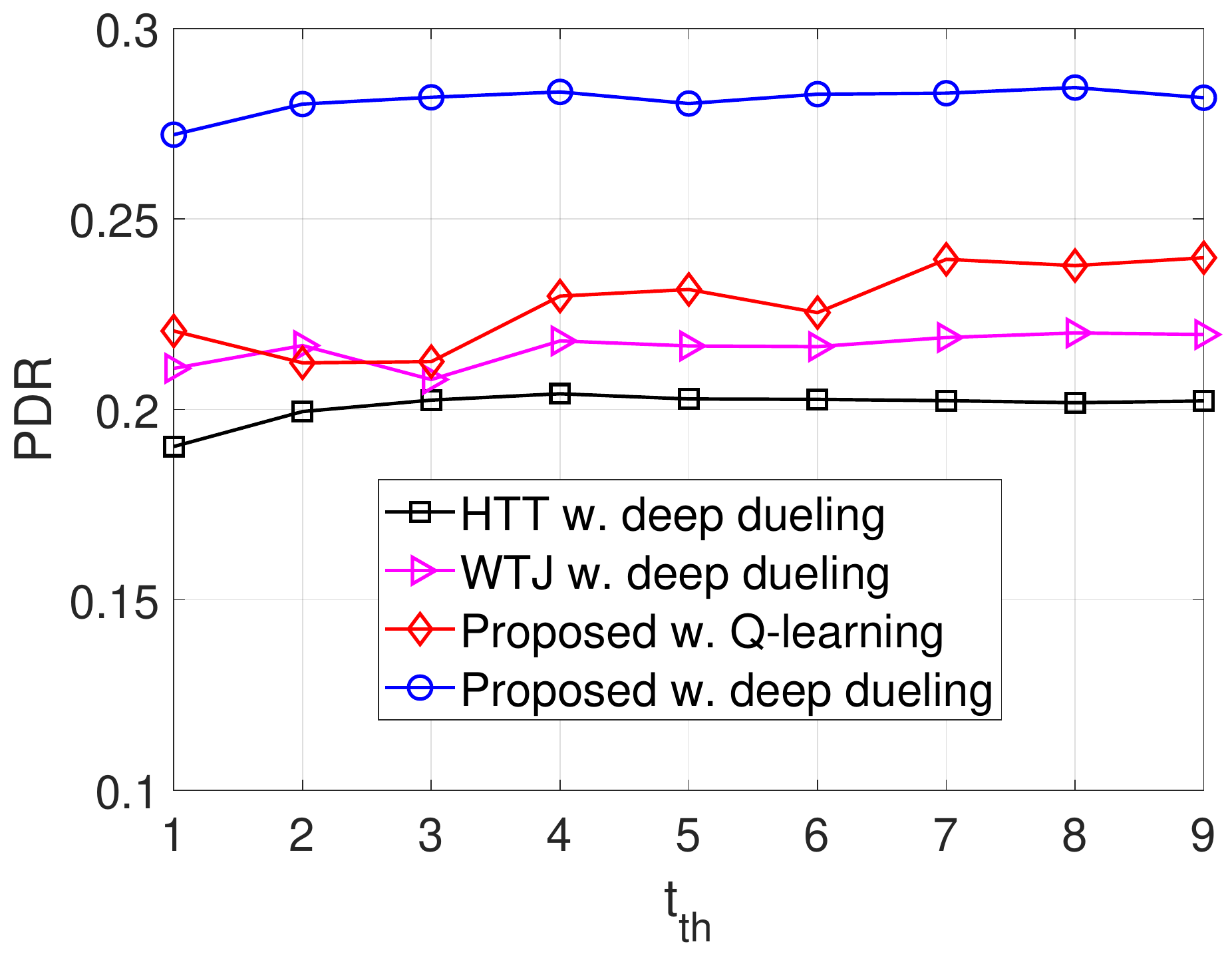}
			\caption{}
		\end{subfigure}%
		~
		\begin{subfigure}[b]{0.3\textwidth}
			\centering
			\includegraphics[scale=0.3]{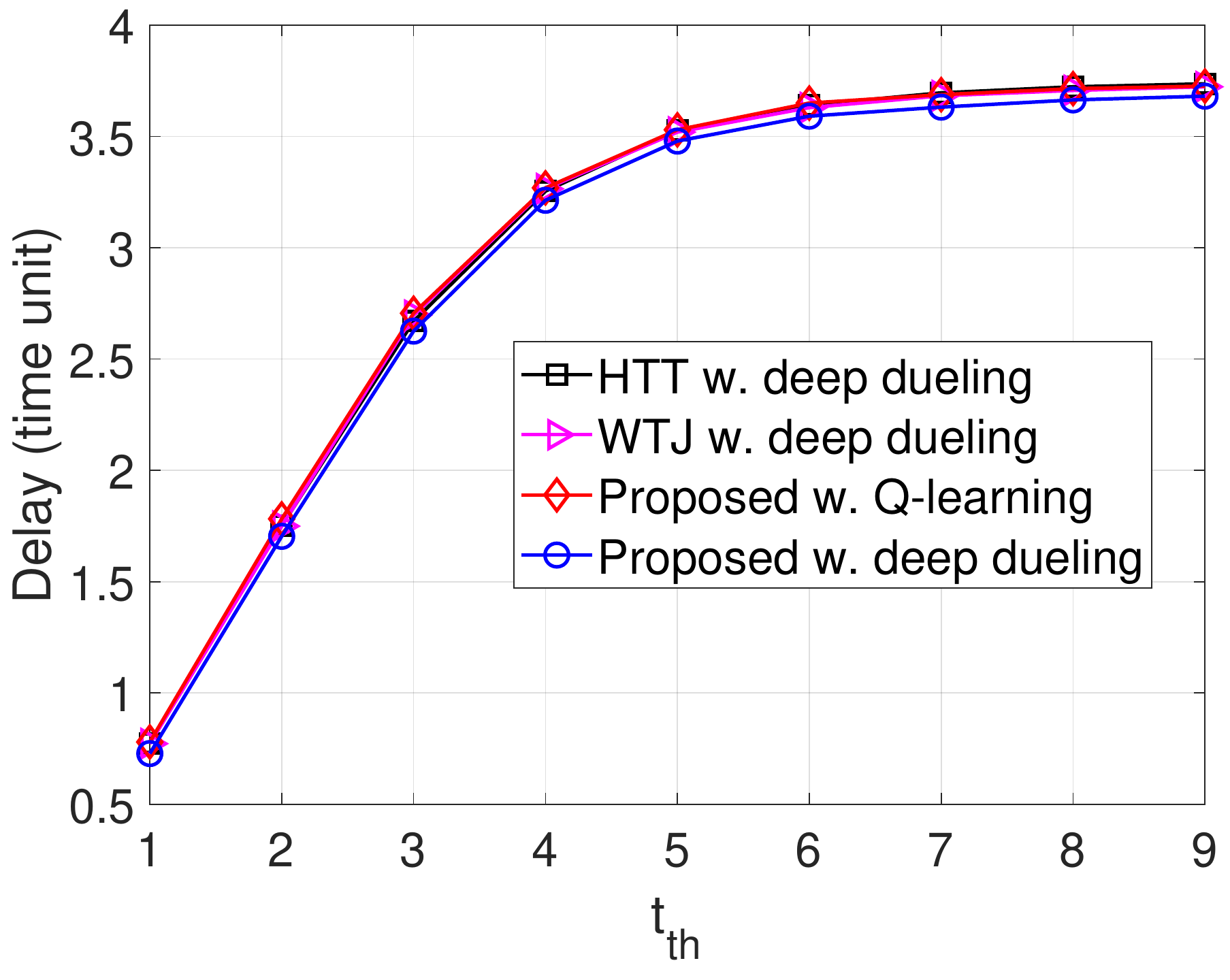}
			\caption{}
		\end{subfigure}%
		\caption{(a) Average throughput (packets/time unit), (b) Packet loss (packets/time unit), (c) Average number of packets in the data queue, (d) PDR, (e) Delay (time/units) vs. $t_{th}$.} 
		\label{fig:varyThreshold}
\end{figure*}

Finally, we vary the latency threshold and investigate the performance of the proposed solution as shown in Fig.~\ref{fig:varyThreshold}. Obviously, when the latency threshold increases from $1$ to $4$ time units, the throughputs and the PDRs obtained by all schemes increase as shown in Fig.~\ref{fig:varyThreshold}(a) and Fig.~\ref{fig:varyThreshold}(d), respectively, and the packet losses decreases as shown in Fig.~\ref{fig:varyThreshold}(b). This is stemmed from the fact that with a very short period of latency, more packets will be discarded from the data queue resulting in lower throughputs. When the latency threshold is large, the throughputs remain the same as the deep dueling algorithm obtains the optimal solution to effectively utilize the ambient signal as well as the jamming signal. As the latency threshold increase, the arrival packets have more time to stay in the data queue. As such, the number of packets in the data queue increases as shown in Fig.~\ref{fig:varyThreshold}(c). Note that the Deep Dueling algorithm always achieves the best performance in all cases. In contrast, the Q-learning algorithm cannot achieve the optimal policy for the transmitter due to the slow-convergence problem.

\begin{figure*}[h]
	\centering
	\begin{subfigure}[b]{0.35\textwidth}
		\centering
		\includegraphics[scale=0.35]{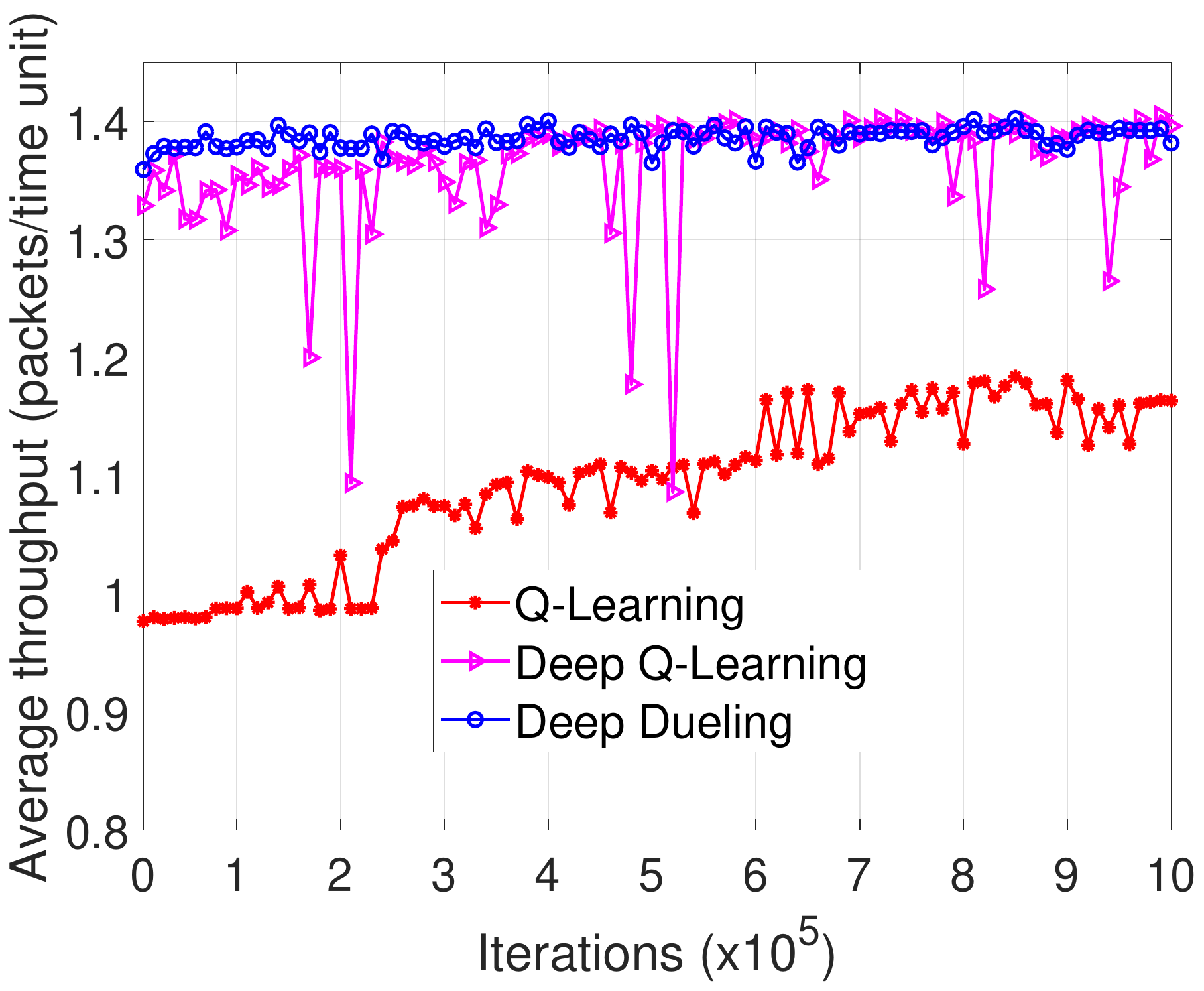}
		\caption{$D=E=10$}
	\end{subfigure}%
	~ 
	\begin{subfigure}[b]{0.35\textwidth}
		\centering
		\includegraphics[scale=0.35]{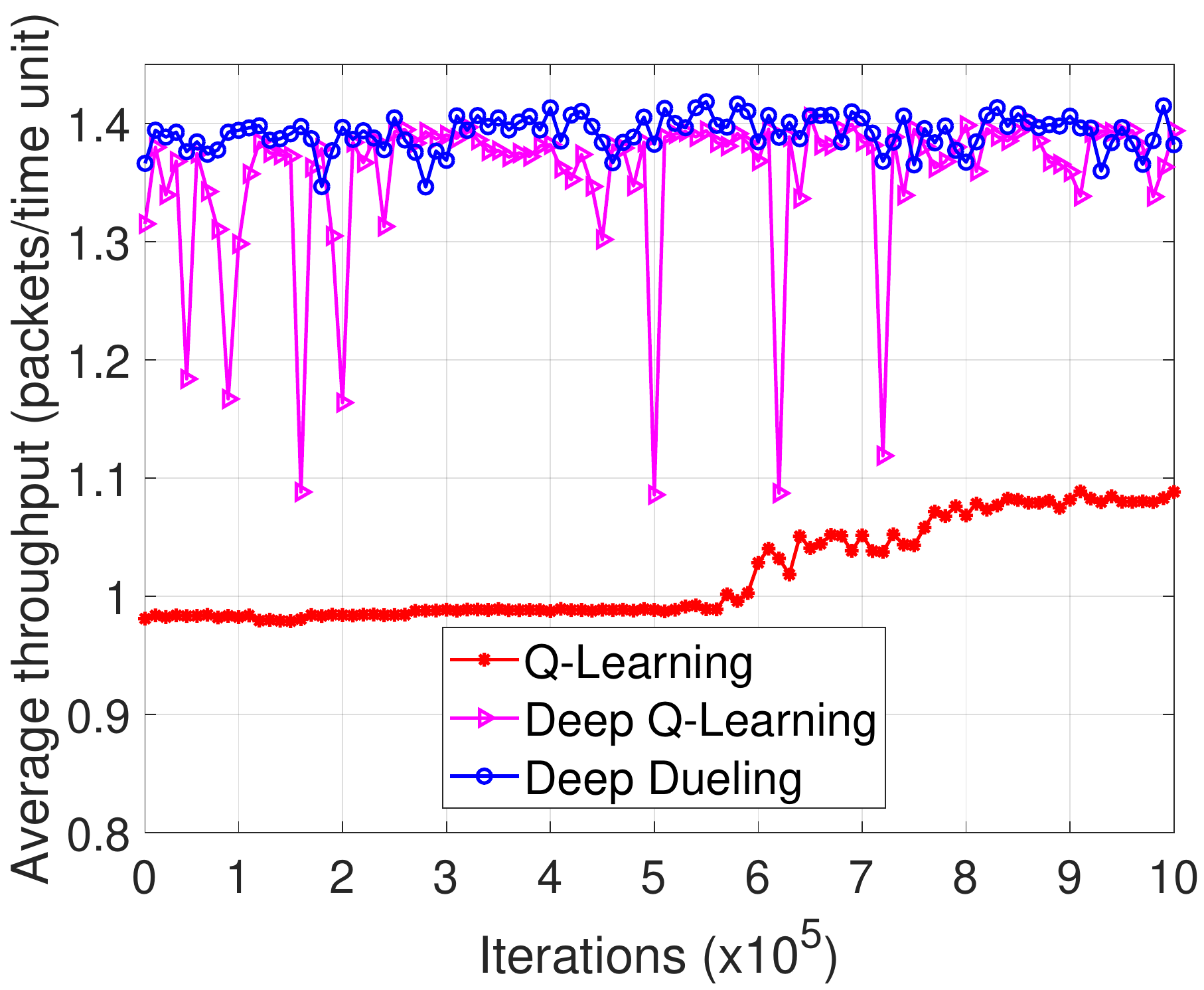}
		\caption{$D=E=20$}
	\end{subfigure}%
	\caption{Convergence rates when (a) $D=E=10$ and (b) $D=E=20$.} 
	\label{fig:convergence1020}
\end{figure*}
	
\paragraph{Convergence of Deep Reinforcement Learning Approaches}
We first show the learning process and the convergence of the proposed deep reinforcement learning algorithms, i.e., deep Q-learning and deep dueling, in several scenarios. As shown in Fig.~\ref{fig:convergence1020}(a) and Fig.~\ref{fig:convergence1020}(b), when the maximum sizes of data and energy queues are set at $10$ and $20$, respectively, after $10^6$ iteration, the average throughput obtained by the Q-learning algorithm is much lower than those of the deep reinforcement learning algorithms, especially in the first $10^5$ iterations. This implies that as the system state space increases, the Q-learning algorithm requires more time to be converged, and thus given a fixed short time period, the system performance obtained by the Q-learning algorithm cannot achieve the results as great as those of deep reinforcement algorithms (i.e., Deep Q-learning and Deep Dueling algorithms). Note that in Fig.~\ref{fig:convergence1020}, the performance obtained by Deep-Q learning algorithm is as close as that of Deep Dueling algorithm, however the average throughput obtained by the Deep-Q learning algorithm is very fluctuated compared with that of the Deep Dueling algorithm. This implies that the Deep-Q learning algorithm requires more time to be converged compared with that of the Deep Dueling algorithm.

\section{Conclusion}
\label{sec:conclusion}
In this paper, we have developed the optimal anti-jamming framework which allows the wireless transceivers to effectively defeat jamming attacks. In particular, with the ambient backscatter capability, while being attacked, the device can either adapt its transmission rate or backscatter its data to the gateway through the jamming signal or harvest energy from the jamming signal to support its operations. To effectively learn about the jamming attacks as well as the channel activities, we have proposed an optimal anti-jamming strategy based on MDP to obtain the optimal defend policy for the transmitter. Then, the reinforcement learning algorithms, i.e., Q-learning, deep Q-learning, and deep dueling, have been developed to maximize the long-term average throughput and minimize the packet loss. Extensive simulations have demonstrated that by using two streams of fully-connected hidden layers, the proposed framework using deep dueling algorithm can improve the average throughput up to 426\% and reduce the packet loss by 24\%. Importantly, with ambient backscatter and energy harvesting technology, jamming signal can be leveraged by the transmitter as the ambient RF signal, thereby effectively eliminating jamming attacks. To the best of our knowledge, this is the first anti-jamming solution that allows wireless transceivers to not only survive jamming attacks without requiring additional resources but also leverage the jamming signal to improve their transmission rate. The proposed ambient backscattering augmented communications framework can be applicable to both civil (e.g., ultra-reliable and low-latency communications or URLLC) and military scenarios (to combat both inadvertent and deliberate jamming).

\appendices

\section{The proof of Theorem~\ref{theo:convergeQ}}
\label{appendix:convergeQ}
Here, we prove that Q-learning converges to the optimum action-values with probability one, i.e., $\mathcal{Q}_t(s,a) \rightarrow \mathcal{Q}^*(s,a)$ as $n \rightarrow \infty$. The main idea of the convergence proof is an artificial controlled Markov process called the action-replay process (ARP)~\cite{Watkins1992QLearning}, which is formulated from the learning rate $\tau_t$ and the episode sequence. In particular, the state space of the ARP is $\{\langle s,t\rangle \}$ together with a special absorbing state, where $s$ is a state of the real process and $t \geq 1$ is the level of the ARP. The action space is $\{a\}$ where $a$ is an action from the real process.

The stochastic reward and state transition consequence when action $a$ is taken at state $s$ are as follows:
\begin{equation}
\mathbf{i_*}	=	\left\{	\begin{array}{ll}
argmax_i\{t^i \le t\},	&	\mbox{if $(s,a)$ is perfomed}\\
	                    &   \mbox{before episode t},	\\
0,	&	\mbox{otherwise},
\end{array}	\right.
\end{equation}
where $i$ is the index of the $i^{th}$ time action $a$ is taken at state $s$. In this way, $t^{i_*}$ is the last time before episode $t$ where $(s, a)$ is executed in the real process. When $i_*=0$, the reward is set as $\mathcal{Q}_0(s,a)$, and the ARP absorbs. Otherwise, let denote
\begin{equation}
\mathbf{i_e}	=	\left\{	\begin{array}{ll}
i_*,	&	\mbox{with probability $\tau_{t^{i_*}}$},	\\
i_*-1,	&	\mbox{with probability $(1-\tau_{t^{i_*}})\tau_{t^{i_*-1}}$},	\\
i_*-2,	&	\mbox{with probability $(1-\tau_{t^{i_*}})(1-\tau_{t^{i_*-1}})\tau_{t^{i_*-2}}$},	\\
 & \vdots\\
0,	&	\mbox{with probability $\prod_{i=1}^{i_*}(1-\tau_{t^i})$},
\end{array}	\right.
\end{equation}
as the index of the episode that is replayed or taken. If $i_e=0$, as above, the reward is set at $\mathcal{Q}_0(s,a)$, and the ARP absorbs. In contrary, when $i_e \# 0$, the reward is $r_{t^{i_e}}$, and a state transition is formed as $\langle {s'}_{t^{i_e}},t^{i_e} - 1\rangle$.

As stated in~\cite{Watkins1992QLearning}, the ARP tends towards the real process. Thus, $\mathcal{Q}_t(s,a)$ tends to $\mathcal{Q}^*(s,a)$, where $\mathcal{Q}_t(s,a) =  \mathcal{Q}^*_{ARP}(\langle s, t\rangle, a), \forall a, s,$ and $t \geq 0$, is the optimal Q-value for the $t^th$ level of the ARP~\cite[Lemma A]{Watkins1992QLearning}. Without loss of generality, we assume that $\mathcal{Q}_t(s,a) < \frac{r^*}{(1-\gamma)}$, where $r* \geq |r_t|$ is the bound of the reward. Given $\chi > 0$, choose $\xi$ such that
\begin{equation}
\gamma^\xi \frac{r*}{1-\gamma} < \frac{\chi}{6}.
\end{equation} 

Based on Lemmas B.2, B.3, B.4 in~\cite{Watkins1992QLearning}, we can compare the value $\bar{\mathcal{Q}}_{ARP}(\langle s,t \rangle, a_1, \ldots, a_\xi)$ of taking actions $a_1, \ldots, a_\xi$ in the ARP with $\bar{\mathcal{Q}}(s, a_1, \ldots, a_\xi)$ of taking them in the real process as follows:
\begin{equation}
\label{eq:difference}
\begin{aligned}
&|\bar{\mathcal{Q}}_{ARP}(<s,t>, a_1, \ldots, a_\xi)-\bar{\mathcal{Q}}(s, a_1, \ldots, a_\xi)| < \\
& \frac{\chi(1-\gamma)}{6\xi r*}\frac{2\xi r*}{1-\gamma} + \frac{2\chi}{3\xi(\xi + 1)}\frac{\xi (\xi + 1)}{2} = \frac{2 \chi}{3}.
\end{aligned}
\end{equation}
Clearly, the effect of taking only $\xi$ actions makes a difference of less than $\frac{\chi}{6}$ for both the ARP and the real processes. As Eq.~(\ref{eq:difference}) can be applied to any set of actions, it applies perforce to a set of actions optimal for either the real process or the ARP. Thus, we have
\begin{equation}
|\mathcal{Q}^*_{ARP}(\langle s, t\rangle,a) - \mathcal{Q}^*(s,a)| < \chi.
\end{equation}
As a result, with probability 1, $\mathcal{Q}_t(s,a) \rightarrow \mathcal{Q}^*(s,a)$ as $n \rightarrow \infty$.

\renewcommand{\baselinestretch}{0.93}

\end{document}